\documentclass[aps, prx, twocolumn, showpacs, superscriptaddress, floatfix, amsmath, amssymb]{revtex4-2}
\pdfoutput=1
\usepackage[utf8]{inputenc}
\usepackage{amstext}

\usepackage{hyperref}

\usepackage{graphicx}
\usepackage{xcolor}

\usepackage[english]{babel}
\usepackage{microtype}

\usepackage{esint}
\usepackage{bm}

\usepackage{hyperref} 
\hypersetup{
	pdftoolbar=true,        		 
	pdfmenubar=true,     		 
	pdffitwindow=false,     	 
	pdfstartview={FitH}, 
	pdftitle={},  
	pdfauthor={},     		 
	pdfsubject={},   			 
	pdfcreator={},   	        	   	
	pdfproducer={}, 		
	pdfnewwindow=true,      	
	colorlinks=true,       		
	linkcolor=black,          	
	citecolor=blue,        		
	filecolor=magenta,    		
	urlcolor=blue          		
}

\usepackage{stmaryrd}
\newcommand*{\lists}[2]{\left\llbracket \begin{matrix} #1 \\ #2 \end{matrix} \right\rrbracket}

\newcommand{\CCQ}{Center for Computational Quantum Physics, Flatiron Institute, 162 5th Avenue, New York, NY 10010, USA}
\newcommand{\CCM}{Center for Computational Mathematics, Flatiron Institute, 162 5th Avenue, New York, NY 10010, USA}
\newcommand{\Grenoble}{Univ. Grenoble Alpes, CEA, Grenoble INP, IRIG, Pheliqs, F-38000 Grenoble, France}
\newcommand{\Neel}{Univ. Grenoble Alpes, CNRS, Institut N\'eel, 38000 Grenoble, France}

\renewcommand\Im{\operatorname{Im}}

\newcommand{\mI}{\mathcal{I}}
\newcommand{\mJ}{\mathcal{J}}

\begin{document}
\title{Learning Feynman Diagrams with Tensor Trains}

\author{Yuriel N\'u\~{n}ez-Fern\'andez }
\email{yurielnf@gmail.com}
\affiliation{\Grenoble}

\author{Matthieu Jeannin}
\affiliation{\Grenoble}

\author{Philipp T. Dumitrescu}
\affiliation{\CCQ}

\author{Thomas Kloss}
\affiliation{\Grenoble}
\affiliation{\Neel}

\author{Jason Kaye}
\affiliation{\CCQ}
\affiliation{\CCM}

\author{Olivier Parcollet}
\affiliation{\CCQ}
\affiliation{Universit\'e Paris-Saclay, CNRS, CEA, Institut de physique th\'eorique, 91191,
   Gif-sur-Yvette, France}

\author{Xavier Waintal}
\email{xavier.waintal@cea.fr}
\affiliation{\Grenoble}

\date{\today}
\begin{abstract}
  
   We use tensor network techniques to obtain high order perturbative
   diagrammatic expansions for the quantum many-body problem at very high
   precision.  The approach is based on a tensor train parsimonious
   representation  of the sum of all Feynman diagrams, obtained in a controlled
   and accurate way with the tensor cross interpolation algorithm. 
   %This representation is an effective separation of variables and therefore enables
   %a direct calculation of the high dimensional integrals.  It also yields, at
   %no additional computational cost, 
   It yields the full time evolution of physical
   quantities in the presence of any arbitrary time dependent interaction.  Our
   benchmarks on the Anderson quantum impurity problem, within the real time
   non-equilibrium Schwinger-Keldysh formalism, demonstrate that this technique
   supersedes diagrammatic Quantum Monte Carlo by orders of magnitude in
   precision and speed, with convergence rates $1/N^2$ or faster, where $N$ is
   the number of function evaluations.  The method also works in parameter
   regimes characterized by strongly oscillatory integrals in high dimension,
   which suffer from a catastrophic sign problem in Quantum Monte-Carlo.
   Finally, we also present two exploratory studies showing that the technique
   generalizes to more complex situations: a double quantum dot and a single
   impurity embedded in a two dimensional lattice. 

\end{abstract}

\maketitle

\section{Introduction}

Many important problems in physics can be formally solved by expressing
physical quantities as sums or integrals in high dimensional spaces, e.g.
equilibrium partition functions in condensed matter and statistical physics, 
or high order perturbative diagrammatic expansions in field theories and in the 
quantum many-body problem.
Calculating integrals in high dimensions is, however, notoriously difficult.
Quantum Monte-Carlo algorithms have emerged as a class of numerical methods of
choice for such problem and have been tremendously successful in many
situations \cite{Sugar1981,Foulkes2001,Sandvik2010,Vanhoucke2010,Carlson2015}.
They have nevertheless well-known major shortcomings.
First, as sampling methods, they can become exponentially inefficient due to massive cancellations, 
a set of related phenomena famously known as the ``sign problem'', which
typically becomes
exponentially more severe at low temperatures and for large systems. % \cite{Pan2022} WHAT IS THIS
Second, as stochastic methods, they have an intrinsically slow convergence 
(as $1/\sqrt{N}$, where $N$ is the number of independent samples),
which can severely limit the accuracy of calculations.
In fact, overcoming the apparent exponential complexity of the fermionic quantum
many-body problem is one of the main motivations for the development
of full scale quantum computers.

Parsimonious (or compressed) representations of high-dimensional functions based on tensor
trains, and more generally on a low rank tensor network (TN)~\cite{oseledets2011ttd, dolgov2020integralRn, tt-integral1,
tt-integral2, tt-example-FP, fiesta5}, offer another route to compute such
large dimensional integrals.  Indeed, they provide an effective separation of
variables that reduces the calculation of high dimensional integrals to the
evaluation of a set of one dimensional integrals, a much simpler
problem~\cite{dolgov2020integralRn}. The \emph{tensor cross interpolation} (TCI)
formula~\cite{tci-dmrg1-2010,tci-dmrg2-2011,ACA-tci-2014} is an
algorithmically efficient way to obtain such a representation, in time
scaling polynomially with the dimension. 
It is a generalization to tensors of cross interpolation for
matrices~\cite{maxvol1997,ACA2000,maxvol2010}, and is closely related to the
interpolative decomposition \cite{matDecomp2017survey}.

The subject of this article is the replacement of Monte Carlo sampling by
tensor network-based algorithms such as TCI in some many-body algorithms, in
particular diagrammatic quantum Monte Carlo. We emphasize that this use of
tensor networks is radically different from their original application in the
density matrix renormalization group algorithm~\cite{dmrg-White-1992} (DMRG)
and its descendants, where is it used as a variational ansatz for the many-body
wavefunction.  Here we use tensor network representations for the many-body
correlation functions arising in the context of Feynman diagram expansions.
Like for tensor train applications in machine learning \cite{Huggins2019}, we
use tensor trains (also known as matrix product states) to {\it learn}, in a
controlled manner, the function representing the sum of Feynman diagrams.

Diagrammatic quantum Monte Carlo methods, 
i.e. high-order diagrammatic perturbation expansions in powers of the interaction strength,
are natural candidates for tensor network techniques.
Despite their perturbative nature, when properly combined
with resummation techniques and judiciously chosen (field-theory) counter-terms
\cite{Prokofev_9804, Prokofev_0801, Mishchenko_9910,
VanHoucke_1110, profumo2015, Wu_1608, Rossi_1612, Chen_1809,
Bertrand_1903_series, Bertrand_1903_kernel, Moutenet_1904,
Rossi_2001, macek2020qqmc, HauleChen2020}, diagrammatic expansions have been successfully used to explore physics
far beyond weak coupling. This includes  the Kondo regime of a quantum
dot~\cite{Bertrand_1903_series, Bertrand_1903_kernel, macek2020qqmc, Bertrand_2021}, the
pseudo-gap regime of the Hubbard model~\cite{SimkovicRossiFerrero2021},
and the low density
electron gas~\cite{Chen_1809,HauleChen2020}. They are particularly useful %for computing %dynamic spectral quantities or 
in non-equilibrium settings~\cite{profumo2015,Bertrand_1903_kernel, macek2020qqmc}, for which there
are very few accurate methods available. 
Computing the expansion coefficient at order $n$ involves, at minimum,
computing
$n$-dimensional integrals over time, as well integrals or sums over other
dimensions, and the different Feynman diagrams themselves. Since the formulation
of perturbation theory as a stochastic sampling over $n!$ Feynman 
diagrams~\cite{Prokofev_9804}, there has been an effort to reformulate the problem
and develop new algorithms for the coefficients in the perturbation
series~\cite{profumo2015, Rossi_1612, macek2020qqmc}. Despite major advances, the
integration techniques used thus far have been variations of sampling from a
non-negative probability distribution. These techniques inevitably suffer
from a sign problem for very oscillatory integrals. Rapidly oscillating
integrals are encountered especially often in the real-time Schwinger-Keldysh
formalism~\cite{profumo2015,Bertrand_1903_kernel,macek2020qqmc}.
We note that among the quantum Monte Carlo algorithms, diagrammatic Monte-Carlo
typically manipulates the integrals with lowest dimensions, 
since the complexity of the calculation of the sum of Feynman diagrams
grows exponentially with $n$ (typically as $O(2^n)$) \cite{profumo2015}.
Hence they are natural first candidates for
a tensor network approach to integration.

In this paper, we explore the use of TCI for real-time
non-equilibrium Schwinger-Keldysh perturbation expansions up to high order
$n\sim 30$ and high precision. We apply the tensor decomposition to the bare
Keldysh $n$-body correlators appearing in Feynman integrals. We demonstrate
very fast convergence, as fast as $O(1/N^2)$ in the number
$N$ of integration points. The final precision is limited in practice only by machine
precision and rounding errors, something usually out of reach in Monte-Carlo.

The main observation underlying our results is that the $n$-body
Keldysh correlators we consider are well approximated by a low-rank tensor train when
viewed as functions of $n$ time differences. We will refer to this property as
``$\epsilon$-factorizability''. The $\epsilon$-factorizability property
yields a separation of variables which reduces the high-dimensional
integrals to a sequence of
one-dimensional integrals which can be computed rapidly.
Crucially, this $\epsilon$-factorizability persists
even in parameter regimes in which the integrands are highly oscillatory. This
renders the approach largely immune to the sign problem, which is reduced to
the problem of integrating oscillatory functions of a single variable.
Finally, the tensor train representation of the $n$-body correlator directly 
provides the full time dependence of the observable for an arbitrary time
dependent interaction coupling strength with a costless post-processing step.

The outline of this paper is as follows. In Section~\ref{sec:overview}, we
summarize our approach and present some illustrative
numerical results showcasing its efficiency.
Section~\ref{sec:TCI} reviews the TCI method, and can be read independently from
the rest of the article.  Section \ref{sec:wick} gives a concise introduction
to the many-body Keldysh formalism, and the notations used to compute high order perturbative
expansions.  In Section~\ref{sec:tdd}, we adapt the TCI method of
Section \ref{sec:TCI} to calculate the high order expansion presented in
Section \ref{sec:wick}.  We refer to this technique as \emph{tensor train
diagrammatics (TTD)}.  Section \ref{sec:siam} presents some numerical results
on TTD for calculating properties of the single impurity Anderson model (SIAM).
Section \ref{sec:beyond} shows results beyond SIAM for an impurity embedded in
a 2D lattice and a double quantum dot. 
Section~\ref{sec:conclusion} contains concluding remarks.  

%%%%%%%%%%%%%%%%%%%%%%%%%%%%%%%%%%%%%%
\section{Overview of the main results}
\label{sec:overview}

Since TCI and the Wick determinant formalism for 
high order expansions might be unfamiliar to some readers, we
begin with a brief motivating overview, including a sample of our main results.
Most technical details are postponed until later sections.

We consider a Hamiltonian of the form
\begin{equation}
  \label{eq:def_H_Hint}
   H = H_0 + U H_\text{int}
\end{equation}
with interaction term $U$,
and a physical observable $Q(U)$, e.g. the charge in a simple quantum
impurity model in steady state. It has a perturbative expansion 
\begin{equation}
  \label{eq:qu_sum}
   Q(U) = \sum_n Q_n U^n
\end{equation}
with
\begin{equation} \label{eq:qnint}
Q_n = \int dv_1 \ldots dv_n\  \tilde Q_n(v_1, \ldots, , v_n).
\end{equation}
where $v$ are time differences. The Schwinger-Keldysh formalism provides
explicit expressions for $\tilde Q_n$ in terms of the propagators of $H_0$.
The difficulty lies in the calculation of the $n$-dimensional integral
\eqref{eq:qnint}.

Our main result is a compressed approximate representation of $\tilde Q_n$ as a
matrix product state (MPS): 
\begin{equation}
   \label{eq:OverviewMPS}
   \tilde Q_n(v_1,\ldots, v_n) \approx M_1(v_1) \cdots M_n(v_n)
\end{equation}
where $M$ are matrices of maximal dimension $\chi$, the so-called bond dimension.
%\times \chi$, except $M_1$ (resp.  %$M_n$) which is $1\times \chi$ (resp. $\chi \times 1$). 
As the variables are now
separated,  we have 
\begin{equation}
  \label{eq:OverviewMPSIntegrated}
  Q_n \approx \left( \int dv_1  M_1(v_1)\right) \cdots \left( \int dv_n
  M_n(v_n)\right).
\end{equation}
The central point of this paper is to demonstrate the existence of a highly accurate tensor interpolation
of the form \eqref{eq:OverviewMPS} for the bare
$n$-body correlators involved in the perturbative expansions at order $n$, with a moderate
bond dimension $\chi$ which does not grow significantly with $n$.
This tensor representation
can be obtained from $O(n \chi^2)$ evaluations of $\tilde Q_n$ using the
TCI algorithm, even though the integration volume
grows exponentially with $n$. Furthermore, the approximation is {\it systematically controlled} by $\chi$.
Using this MPS form, the complexity of computing the $n$-dimensional
integral becomes $O(n d \chi^2)$, rather than $O(d^n)$, where
$d$ is the number of discretization points (or basis functions) in each
dimension. These complexities are expressed in the number of evaluations
of the integrand $\tilde Q_n(v_i)$. To obtain
the total complexity, a factor $2^n$ must be included to account for the
complexity of a single evaluation of $\tilde Q_n(v_i)$ in the Keldysh formalism.

%% results 

\begin{figure}
\includegraphics[width=0.49\textwidth]{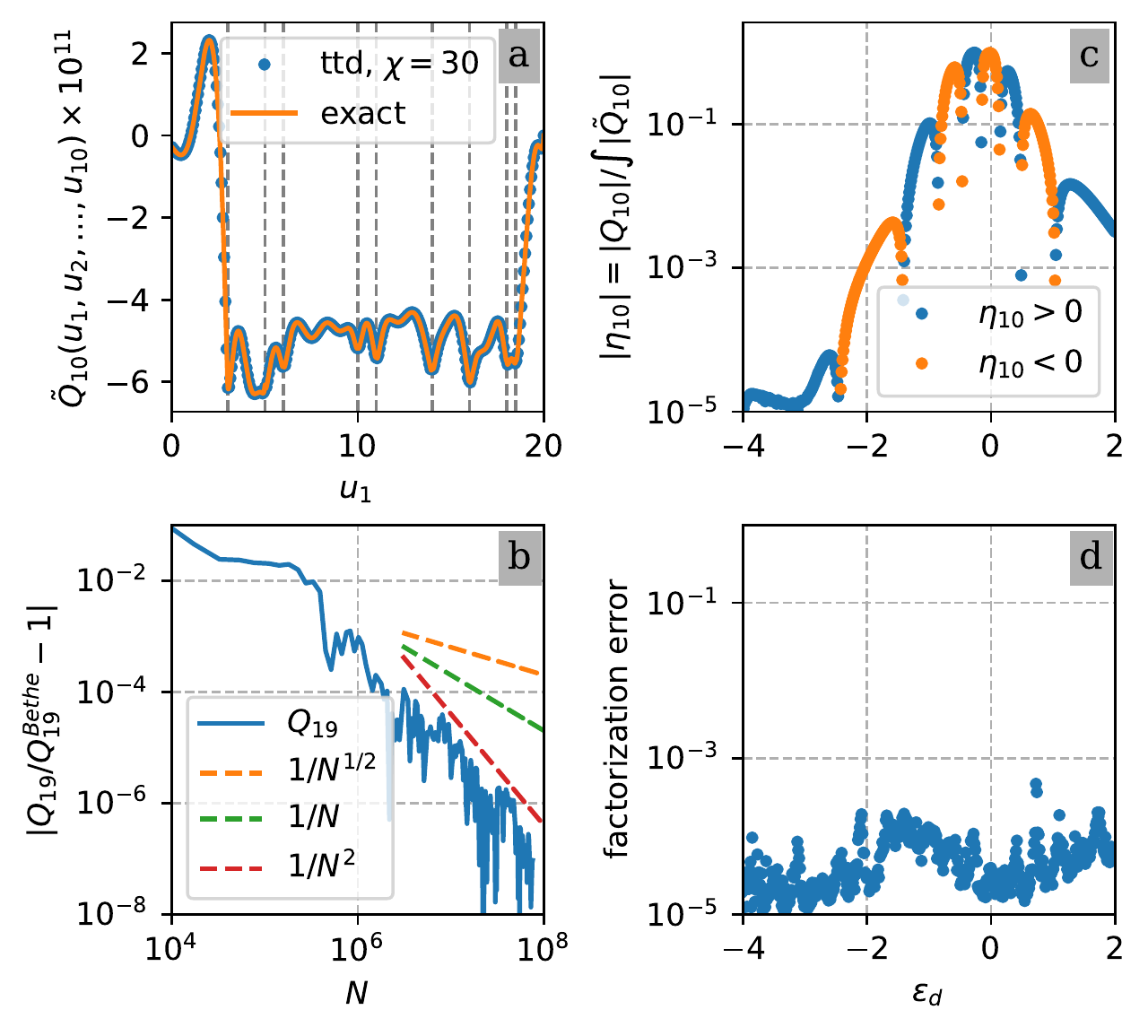}
\caption{%
   \label{fig:Ovw} 
   Overview of the main results.
   {\it a)} Slice of the corresponding integrand $\tilde Q_{10}$ (orange line)
          compared to the MPS approximation (blue dots). The values of $u_2,u_3, \ldots,u_{10}$ were arbitrarily fixed (vertical dashed lines).
  {\it b)} Relative error of the $n$th coefficient (for $n=19$) 
  in the perturbative expansion of the charge $Q$ of the Anderson
  quantum impurity model (compared with the exact Bethe ansatz solution)
  vs. the number $N$
  of evaluation of  $\tilde Q_{19}$.  
   {\it c)} Average sign defined as $\eta_{n} = |Q_{n}|/\int|\tilde Q_{n}|$ for $n=10$,
           versus on-site energy $\epsilon_d$. 
  {\it d)} Relative error of the rank-50 MPS approximation (pivot error,
  as defined in \eqref{eq:errorIsOnA}, divided by the value of the function $\tilde Q_{n}$ at the first pivot) 
           versus $\epsilon_d$.
}
\end{figure}

The quality of the tensor interpolation is illustrated in Fig. \ref{fig:Ovw}a, 
for the coefficient $Q_n$  
of the perturbative expansion of the charge $Q$ of the Anderson quantum impurity model.
We present $\tilde Q_n$ on a path in the $n$-dimensional integration domain (blue line)
and its MPS approximation \eqref{eq:OverviewMPS} (orange dots) for $\chi=30$.
In Fig. \ref{fig:Ovw}b, we show 
the convergence of the integral $Q_{19}$ compared to the exact Bethe ansatz solution, 
as a function of the number $N$ of evaluations of the integrand $\tilde Q_{19}$.
We obtain an unprecedented $O(1/N^2)$ convergence down to a relative
error level of $10^{-7}$.

Since it is based on a full interpolation of the correlators, 
the TTD method allows one to compute, at no extra cost,
{\it i)} the {\it full time dependency of $Q_n(t)$} after the interaction quench at $t=0$,
and 
{\it ii)} the same for any {\it time dependent coupling constant $U(t)$}
(by multiplying by $U(t)$ before integrating). 
This is discussed in detail in Section \ref{sec:SIAMRealTimeDyn}.

The TTD has two fundamental differences with DMRG and its higher
dimensional generalizations.
First, the tensor decomposition applies to
$n$-body correlators instead of many-body wavefunctions.  Second, 
in DMRG the unknown wavefunction is represented by a TN ansatz which 
is variationally optimized. Here, the function $\tilde Q_n$ is known
(it is the input of the problem). We compress it
in order to avoid an exponential integration cost.
TCI belongs to the class of ``active machine learning'' algorithms: the
tensor approximation is constructed by
evaluating an $n$-body correlator and finding the region in its
$n$-dimensional space with the largest approximation error.

The TTD has two major advantages compared to diagrammatic quantum Monte
Carlo.
First, we observe a faster convergence rate of
$O(1/N^2)$ instead of $O(1/\sqrt{N})$.  
Second, the $\epsilon$-factorization is completely unrelated to the average
sign of the integral, as illustrated in Figs. \ref{fig:Ovw}cd. The average sign
$\eta_{10}$ (panel {\it c}) varies over five orders of magnitude as a function of
one parameter of the model (here $\epsilon_d$, the on-site energy of the
Anderson model), while the error of the factorization at fixed tensor rank
$\chi$ (panel {\it d}) is constant with $\epsilon_d$.  A small value of $\eta$
implies a major sign problem for diagrammatic Monte Carlo, cf. Section
\ref{sec:tdd}, whereas the TTD has no such problem. 
The limiting factors of TTD and Monte Carlo are
therefore completely different.

Finally, let us discuss the quasi-Monte Carlo technique which was
recently introduced by some of the authors \cite{macek2020qqmc}.
It represents an intermediate step between Monte Carlo and TTD, 
since it combines a (much weaker) $\epsilon$-factorizability for 
the tails of $\tilde Q_n$ at large $v$
with a quasi-Monte Carlo technique to compute the Feynman integrals.
While it produces convergence as
fast as $1/N$ in good cases, it is, in our benchmarks, much less robust than
the TTD. Furthermore, as a (non-stochastic) sampling technique, it also suffers
from a sign problem when $\tilde Q_n$ is highly oscillatory.

%%%%%%%%%%%%%%%%%%%%%%%%%% Tensor CI %%%%%%%%%%%%%%%%%%%%%%%%%%%%%%%%

\section{Tensor train cross interpolation}\label{sec:TCI}

We start with a review of tensor
cross interpolation.
Most of the material in this section is not original 
(see \cite{dolgov2020integralRn} and 
\cite{ACA-tci-2014,oseledets2011ttd,tci-dmrg2-2011,tci-dmrg1-2010,maxvol2010,ACA2000,maxvol1997})
except, 
to our knowledge, the environment-aware error function of Section \ref{sec:ErrorEnv}.
We present it here in detail so that the article is self-contained. We
also show explicitly that most of the results initially derived for
matrices and tensors are directly generalizable to multi-dimensional
functions. The appendices include explicit proofs of the statements made
here in the main text.
Note that in this class of algorithm, the main difficulty lies in the
bookkeeping of the various slices of the tensor held in memory. Hence,
the choice of notations plays a particularly important role.

\subsection{Matrix cross interpolation}

Given an $M\times N$ matrix $A$,
the {\it cross interpolation} technique (CI) yields an approximate rank
$\chi$ factorization of $A$. It is distinct from the  
truncated singular value decomposition (SVD), in which one approximates
$A$ by its SVD with all but the largest $\chi$ singular values set to zero.
Although the truncated SVD yields an optimal rank $\chi$ approximation of
$A$ in the spectral norm, CI has the advantage that it may be constructed by querying
only a small subset of the entries of $A$. CI is
quasi-optimal in the
sense that its error is at most $O(\chi^2)$ times the optimal one~\cite{matCIerror2010,matCIquasioptimality2011}.

We begin by establishing our notation.
Let $\mI=\{i_1,i_2, \ldots,i_\chi\}$ (respectively
$\mJ=\{j_1,j_2, \ldots,j_\chi\}$) denote a list of the rows (columns)
of $A$, $\mI_a \equiv i_a$ its a$^{\text{th}}$ element, and
$\mathbb{I}=\{1,2, \ldots,M\}$ ($\mathbb{J}=\{1,2, \ldots,N\}$) the list of the indices of
all rows (columns). 
Following the Python/MATLAB
convention, we denote by $A(\mI,\mJ)$ the submatrix
of $A$ comprised of the rows $\mI$ and columns $\mJ$;
$A(\mI,\mJ)_{ab} \equiv A_{\mI_a,\mJ_b}$. 
In particular, $A(\mathbb{I},\mathbb{J}) = A $.

\begin{figure*}[htb]
\includegraphics[width=1.0\textwidth]{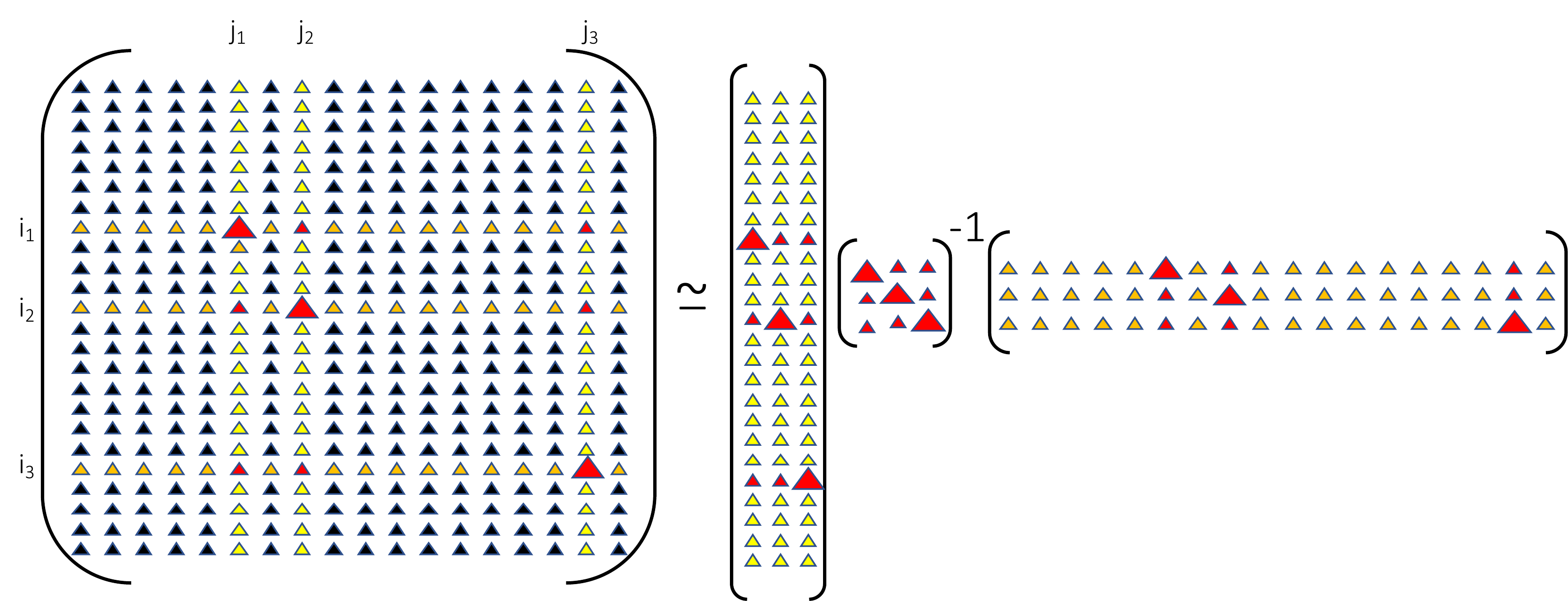}
\caption{\label{fig:ci}
   Illustration of the cross interpolation (CI) of a matrix.
   The large red triangles indicate real pivots and the smaller red
   triangles indicate automatically generated pivots. The right-hand side only contains small subparts of the matrix.}
\end{figure*}

The matrix cross interpolation formula reads
\begin{equation}
\label{eq:matci}
A = A(\mathbb{I},\mathbb{J}) \approx A(\mathbb{I},\mJ) 
A(\mI,\mJ)^{-1}
A(\mI,\mathbb{J}).
\end{equation}
Equation (\ref{eq:matci}) is illustrated graphically in Fig. \ref{fig:ci}. 
It has two main properties
\begin{itemize}
\item[(P1)] It is an interpolation, i.e. it is exact for any $i\in\mI$ or $j\in\mJ$.
   This can be straightforwardly checked from the definition as e.g.
   $A(\mI,\mJ)
   A(\mI,\mJ)^{-1} A(\mI,\mathbb{J}) =
   A(\mI,\mathbb{J})$.
\item[(P2)] It is exact if the matrix $A$ has rank $\chi$ (cf. Appendix \ref{app:ci} for a simple proof).
\end{itemize}

% Pivots
The elements of the non-singular submatrix $A(\mI,\mJ)$ are called the
{\it pivots} and $A(\mI,\mJ)$ the {\it pivot matrix}. 
The pivots should be chosen to minimize the error in the approximation (\ref{eq:matci}).
There is an exponentially large number of possible choices of pivot
matrix, so it is impossible in practice (for a large matrix $A$)
to try all of them. However, well-established
heuristic algorithms exist which provide good quality pivots,
by maximizing the magnitude of the determinant of the pivot matrix.
This is known as the \emph{maxvol principle} (i.e. maximum volume)~\cite{maxvol1997,maxvol2010}. 

% Continuous limit.

In this work, we will need a generalization of the CI to the continuum \cite{matDecomp2017survey,ACA2000,matCIerror2010}.
We refer to a real-valued function $A(x,y)$
as {\it $\epsilon$-factorizable} in the CI sense with {\it finite} rank
$\chi$ if it can be approximated with error $\epsilon$ as
\begin{equation}
A(x,y)\approx  \sum_{ab} A(x,y_a)\left[ A(\mI,\mJ)^{-1}\right]_{ab} A(x_b,y)\text{.}\label{eq:funci}
\end{equation}
Here $\mI = (x_1,x_2, \ldots,x_\chi)$ and $\mJ = (y_1,y_2,
\ldots,y_\chi)$ are {\it finite} sets of $x$ and $y$ values. The CI
\eqref{eq:funci} uses a {\it finite} number $2\chi$ of one-dimensional
functions $A(x,y_a)$ and $A(x_b,y)$. Using
implicit summation, we rewrite \eqref{eq:funci} as
\begin{equation}
A(x,y)\approx A(x,\mJ)A(\mI,\mJ)^{-1} A(\mI,y)\text{.}
\end{equation}
The continuous version of the CI also has the properties (P1) and (P2).

% Separate integrals.

Integrating an {\it $\epsilon$-factorizable} function is greatly
simplified by its approximate 
separability of variables, as only one-dimensional integrals need be
performed:
\begin{multline}
\int dx \, dy \, A(x,y) \approx  
 \left[\int dx \, A(x,\mJ)\right] \\
  \times A(\mI,\mJ)^{-1}
\left[\int dy \, A(\mI,y)\right] \text{.}
\end{multline}
The CI has other similar properties. For instance if the one-dimensional
slices are sufficiently well represented (i.e. a good interpolant has been built for
each of them), then we can also obtain an approximation of the gradient $\vec\nabla
A(x,y)$, from which one may perform optimization:
\begin{equation}
\frac{\partial A}{\partial y}(x,y) \approx  A(x,\mJ)
A(\mI,\mJ)^{-1}
 \frac{\partial A}{\partial y}(\mI,y).
\end{equation}

For practical implementations, it is important to note that evaluating
\eqref{eq:funci} directly may be numerically unstable, since for large values of
$\chi$ the pivot matrix becomes almost singular. An equivalent but stable
evaluation method using the QR decomposition is explained in Appendix \ref{app:ci} for the TCI.

%%%%%%%%%%%%% TENSOR SUBSECTION

\subsection{Tensor train interpolation}

\begin{figure}[ht]
\begin{centering}
\includegraphics[width=0.45\textwidth]{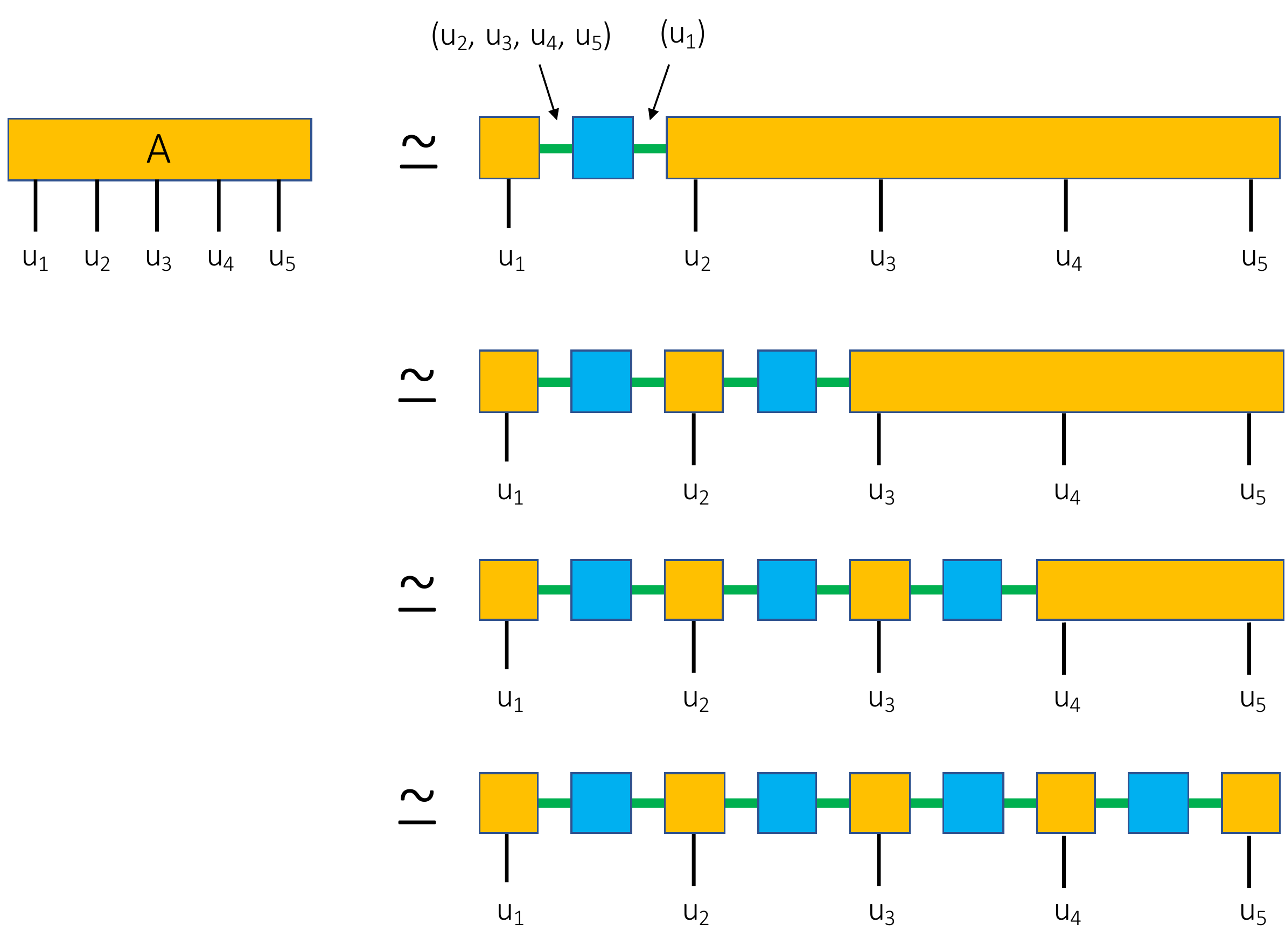}
\par\end{centering}
\caption{
   \label{fig:ttd} 
   Step by step representation of a simple algorithm to factorize a multidimensional tensor into a tensor train.
   The blue squares represent the inverses of the pivot matrices. 
   Summation is implicit over the indices connecting two tensors (green
   lines).
}
\end{figure}

We know turn to the tensor cross interpolation (TCI), as introduced in
\cite{tci-dmrg1-2010}, which is the generalization of the matrix cross
interpolation to $n$-dimensional tensors and functions. TCI is also
quasi-optimal if the {\it maxvol} principle is used~\cite{ACA-tci-2014}. We
consider an $n$-dimensional function $A(u_1, \ldots,u_n)$, where the
$u_i$ are either
discrete or continuous variables. The TCI literature typically deals with the discrete
case, in which $A$ is a tensor, and each index $u_i$ can take $d$ different
values, so that $A$ has $d^n$ entries. In this work, we will also
use a generalization to the continuous case.  Following standard notation,
tensor networks are depicted as a rectangle with n ``legs'' (indices); see
Fig. \ref{fig:ttd}. 

\subsubsection{Naive approach}
\label{sec:naive}
Let us first present a simple algorithm, illustrated in
Fig. \ref{fig:ttd}, to decompose a tensor.
While it is not efficient and not used in practice, it provides a pedagogical introduction to
TCI for the unfamiliar reader.

First, we view the tensor $A$ as a matrix $A_{(u_1),(u_2,u_3, \ldots,u_n)}$ 
by regrouping the indices into $u_1$ and a {\it multi-index}  $(u_2,u_3,u_4 \ldots u_n)$.
Second, we apply the CI to this matrix and decompose it as 
a product of three matrices, as shown in the right hand side of the first line of Fig. \ref{fig:ttd}.
Here, the blue square stands for the inverse of the pivot matrix. 
Crucially, since we keep only a finite number $\chi$ of pivots, 
the summation over the repeated indices (green lines) involves only a small number of
terms, even if the variable $u_i$ is continuous. 
Next, we consider the (orange) tensor on the right side of the first line of Fig. \ref{fig:ttd}. 
We regroup the $\chi$ values of $u_1$ of the pivots and the $d$ values of
$u_2$ into a multi-index $(u_1,u_2)$,
and form the matrix $A_{(u_1,u_2),(u_3,u_4, \ldots,u_n)}$. 
Applying CI to this matrix yields the second line of Fig. \ref{fig:ttd}. 
This process is continued until all the orange tensors have only one black leg, which yields 
the ``tensor train'' represented in the last line of Fig. \ref{fig:ttd}.

From this simple algorithm, we can already observe the extension of property (P2) from matrices to tensors:
if a tensor has rank $\chi$ (which we define as each of the above matrices has rank
$\chi$), then all the steps above are exact for a correct choice of pivots 
and the tensor train is an exact representation of the tensor.
Furthermore, like in the CI, the orange rectangle (respectively blue square) tensors in Fig. \ref{fig:ttd} correspond
to sub-tensors (respectively matrix inverses of sub-tensors) of the initial tensor $A$.

%%%%%%% TENSOR INDEX NOTATION 

\subsubsection{Tensor train interpolation}
%\subsubsection{Multi-index and tensors notations}

\begin{figure*}[ht!]
\begin{centering}
\includegraphics[width=1.0\textwidth]{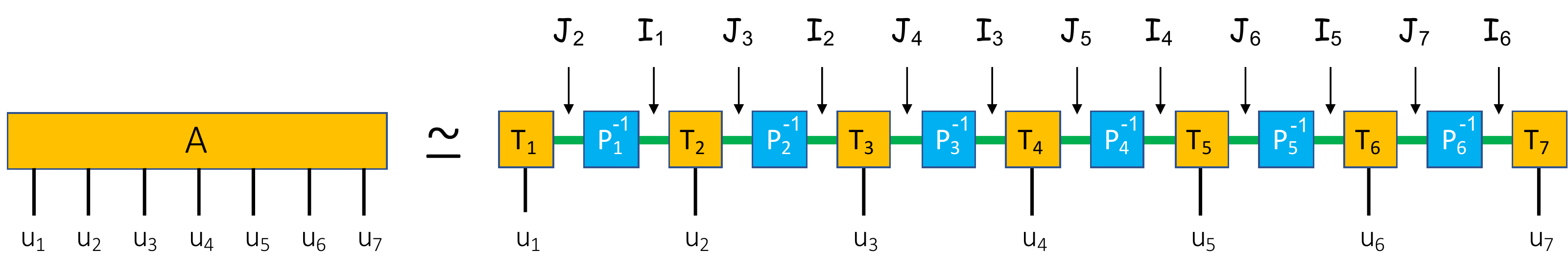}
\par\end{centering}
\caption{\label{fig:ttd_notations} 
   Pictorial representation of TCI formula for $A$ defined in (\ref{eq:defTCI}).
 The $T_\alpha$ tensors and $P^{-1}_\alpha$ pivot
   matrices are represented by orange and blue squares, respectively.
   The green lines correspond to contracted discrete indices $\mI_\alpha$ or $\mJ_{\alpha}$, as indicated by the arrows.
   The thin black lines correspond to the variables $u_\alpha$.
}
\end{figure*}

The goal of TCI is to perform the decomposition of section \ref{sec:naive} using
only few calls to the function $A(u_1,..u_n)$.
Let us now introduce our notation and definitions for TCI, in particular
the multi-index notation. A graphical illustration of the tensor train
notation is shown in Fig. \ref{fig:ttd_notations}.

% FIRST THE NOTATIONs

We consider a tensor $A(u_1, \ldots, u_n)$, with $u_i$ taking a finite
set of $d$ discrete values. The generalization to continuous variables
will be discussed below.
For  any $\alpha$ such that $1 \leq \alpha \leq n$,
we use {\it multi-indices} of the following form:
$i= (u_1,u_2, \ldots, u_{\alpha})$ and $j = (u_{\alpha},u_{\alpha +1},
\ldots,u_n)$.
We let $\mI_\alpha = \{i_1,i_2, \ldots,i_\chi\}$ denote
{\it a set of $\chi$ multi-indices} of size $\alpha$,  
and let $\mJ_\alpha = \{j_1,j_2,\ldots,j_\chi\}$ similarly denote a set of
$\chi$ multi-indices of size $n-\alpha+1$.
Since each of its elements is a multi-index, $\mI$ is a ``list of
lists'' of values of the variables $u_i$. 
For notational convenience, we define $\mI_0$ and $\mJ_{n+1}$ as
singleton sets each comprised of an empty multi-index.
In the following, we reserve the notation $i$, $j$ for such multi-indices,
without emphasizing their dependence on $\alpha$ explicitly.

We use the symbol $\oplus$ to denote concatenation of multi-indices: 
\begin{equation}
  %i \oplus u_\alpha \oplus j = 
  (u_1,u_2, \ldots,u_{\alpha -1}) \oplus (u_\alpha) \oplus (u_{\alpha
  +1}, \ldots,u_n) \equiv (u_1, \ldots,u_n).
\end{equation}
We also define % $d$ the number of possible values of the variable $u$
$\mathbb{K}_\alpha$ as the set of all values of the multi-index $(u_\alpha)$
of size 1, with $1 \leq \alpha \leq n$. 
Finally, we define $\mI \oplus \mJ$ as the set of
all concatenations of the elements of $\mI$ and $\mJ$: 
$\mI \oplus \mJ \equiv \{ i\oplus j \, | \, i \in \mI, j \in \mJ\}$.

To illustrate these notations, let's give a concrete example for $n=4$ and $\chi=2$
and $0\le u_1 <u_2<u_3<u_4 \le 1$. 
A possible choice is $\mI_2 = \{ (0.2,0.45),(0.1,0.6) \}$, 
 $\mI_3 = \{ (0.2,0.45,0.7),(0.1,0.6,0.8) \}$, 
 $\mJ_2 = \{ (0.72,0.92),(0.76,0.92) \}$
and $\mJ_3 = \{ (0.98),(0.92) \}$. Note that this choice respect the
nesting condition defined below. Operations provide e.g. 
$\mI_2\oplus\mJ_3 = \{ (0.2,0.45,0.98),$ $(0.1,0.6,0.98),$ $(0.2,0.45,0.92),$ 
$(0.1,0.6,0.92) \}$.

% NOW WE DEFINE T, P
We now define the tensors $T_\alpha$ and $P_\alpha$ by the expressions
\begin{equation}
   \label{eq:defTP2}
T_\alpha (i,u_\alpha,j) \equiv A(i\oplus (u_\alpha) \oplus j) =
  A(u_1,u_2, \ldots,u_n),
\end{equation}
with $i\in \mI_{\alpha-1}$, $j\in\mJ_{\alpha+1}$, and
\begin{equation}
P_\alpha(i,j) \equiv A(i\oplus j) = A(u_1,u_2, \ldots,u_n),
\end{equation}
with $i\in \mI_{\alpha}$, $j\in\mJ_{\alpha+1}$. Here, $1\leq \alpha \leq n$ and $1\leq \alpha \leq n-1$ for $T_\alpha$
and $P_\alpha$, respectively.
More abstractly, we can write
\begin{subequations}
   \label{eq:defTP}
\begin{align}
T_\alpha &\equiv A(\mI_{\alpha-1}\oplus\mathbb{K}_\alpha\oplus
  \mJ_{\alpha+1}), \\
P_\alpha &\equiv A(\mI_{\alpha}\oplus\mJ_{\alpha+1}).
\end{align}
\end{subequations}
For notational convenience, we define $P_0$ and $P_n$ as the $1\times 1$ unit matrix.
For fixed $\alpha$, $T_\alpha$ is therefore of dimension $\chi \times d \times
\chi$, except $T_1$ and $T_n$, which are of dimension $1\times d \times
\chi$ and $\chi \times d \times 1$, respectively. Similarly, $P_\alpha$
is of dimension $\chi \times \chi$.
$T_\alpha$ is therefore a three-leg tensor (whose name comes from its ``T''
shape), and $P_\alpha$ is a matrix.
From these definitions, we see that if one selects one of the $\chi$
multi-indices $i \in \mI$, and one of the $\chi$ multi-indices $j \in
\mJ$, then $T_\alpha$ defines a one-dimensional slice of the original tensor $A$ along
the variable $u_\alpha$.
We lastly note that the position of the indices in $P^{-1}$ is transposed compared to $P$ due to the inversion. The $T_\alpha$ tensors and the $P_\alpha$ matrices are given a schematic representation shown in the right hand side of Fig.\ref{fig:ttd_notations} as respectively a three legs orange tensor and a blue matrix with the discrete indices $i$ and $j$ in green while
the $u_\alpha$ are in black.

% As matrices.
We will also use matrix notation for $T_\alpha$ by defining
$T_\alpha(u)$ as the matrix of values of the tensor with fixed
$u_\alpha=u$. We have
\begin{subequations}
\begin{align}
   \label{eq:defTPMatrix}
[T_\alpha  (u)]_{ij} &\equiv T_\alpha (i,u,j),
\\
(P_\alpha)_{ij} & \equiv P_\alpha(i,j).
\end{align}
\end{subequations}

Using these notations, the TCI representation of $A$ takes a simple form 
in terms of matrix multiplications. It is a tensor train of the form
\begin{equation}
   \label{eq:defTCI}
   A(u_1, \ldots,u_n) \approx A_\text{TCI}(u_1, \ldots,u_n) \equiv
   \prod_{\alpha=1}^n T_\alpha(u_\alpha) P_\alpha^{-1}.
%A(u_1, \ldots,u_n) \approx \text{Tr} \left[ \prod_{\alpha=1}^n T_\alpha(u_\alpha) P_\alpha^{-1} \right]
%A(u_1, \ldots,u_n) \approx  T_1(u_1) P_1^{-1} T_2(u_2) \ldots P_{n-1}^{-1} T_n(u_n) 
\end{equation} 
Note that given the dimensions of $T_1$, $T_n$, and $P_n$, this product
is a scalar. This TCI representation is illustrated in
Fig. \ref{fig:ttd_notations}. Each green line corresponds to a set of
multi-indices $\mI_\alpha$ (``rows'') or $\mJ_\alpha$ (``columns'').
It is important to notice
that the TCI representation is defined entirely by the selected sets of
``rows'' and ``columns'' $\mI_\alpha$ and $\mJ_\alpha$, so that
constructing an accurate representation of $A$ amounts to optimizing
the selection of $\mI_\alpha$ and $\mJ_\alpha$ for $1 \leq \alpha \leq
n$.

We impose a restriction on the possible choices of $\mI_\alpha$ and $\mJ_\alpha$ called the {\it nesting condition}
~\cite{ACA-tci-2014,dolgov2020integralRn}: 
$\mI_\alpha$ ($\mJ_\alpha$) is constructed from elements of $\mI_{\alpha-1}$ ($\mJ_{\alpha+1}$), except for the last (first) variable, which is taken from $\mathbb{K}_\alpha$.
\begin{subequations}
\begin{align}
   \label{eq:nestingCondition}
   \mI_{\alpha} &\subset \mI_{\alpha-1} \oplus \mathbb{K}_{\alpha}
   \\
   \label{eq:nestingCondition2}
   \mJ_{\alpha} &\subset  \mathbb{K}_\alpha \oplus \mJ_{\alpha+1}.
\end{align}
\end{subequations}
In other words, if
$i \in \mI_\alpha$, then there is an $k \in
\mI_{\alpha-1}$ such that $i  = k \oplus
u_\alpha$ for some $u_\alpha \in \mathbb{K}_\alpha$.
Similarly, if $j \in \mJ_\alpha$, then there is an $k \in
\mJ_{\alpha+1}$ such that $i  = u_\alpha \oplus k$ for some $u_\alpha \in \mathbb{K}_\alpha$.
We show in Appendix \ref{app:proofInterpolationFromNesting} that
imposing the nesting condition guarantees the generalization of the
interpolation property (P1) for the tensor train, namely
\begin{equation}
   \label{eq:TCIinterpolProperty}
   A_\text{TCI} (\mI_{\alpha-1},\mathbb{K}_\alpha, \mJ_{\alpha+1}) = A(\mI_{\alpha-1},\mathbb{K}_\alpha, \mJ_{\alpha+1})
\end{equation} 
for $1 \leq \alpha \leq n$. 
In other words, the approximation is exact for any indices that define one of the 
tensors $T_\alpha$ and {\it a fortiori} to those that define one of the
$P_\alpha$ matrix. 

The TCI approximation of $A(u_1,u_2, \ldots,u_n)$ is built 
from one-dimensional slices (i.e. partial evaluations) of $A$ (the $T_\alpha$ tensors with fixed $\alpha$, $i$, and $j$). Therefore, only $O(n d\chi^2) \ll d^n$ entries of $A$ are used in the
approximation. 

As for matrix cross interpolation, this construction can be directly
generalized to continuous variables.
We call a function $A(u_1,u_2, \ldots,u_n)$ {\it $\epsilon$-factorizable}
if the factorization \eqref{eq:defTCI} 
satisfies $\vert\vert A - A_\text{TCI} \vert\vert_\infty < \epsilon$ 
with $\chi$ finite and increasing ``slowly'' as $\epsilon$ is decreased.
The TCI is particularly useful to compute the $n$-dimensional integral
of $A$, which is our goal in this paper.
Indeed, it separates the variables, reducing the calculation of the
$n$-dimensional integral to that of $O(n\chi^2)$ one-dimensional integrals, 
followed by the tensor contraction \eqref{eq:defTCI}:
\begin{equation}
\int du_1 \cdots du_n \, A(u_1, \ldots,u_n)\approx  \prod_{\alpha=1}^n \int du_\alpha T_\alpha(u_\alpha) P_\alpha^{-1} \text{.}
\end{equation}

%%%%%%%%%%%% Also to obtain TCI 

\subsubsection{Algorithm to construct the TCI} 
\label{sec:TCIalgo}

We now turn to the algorithm used to construct the TCI, and in
particular to find the set of pivots. Our implementation is essentially
equivalent to that described in Ref. \cite{dolgov2020integralRn}.
We start with an initial point $(u_1, \ldots,u_n) = (u_1, \ldots,u_\alpha) \oplus (u_{\alpha+1}, \ldots,u_n)$,
which we split in $n-1$ different ways to obtain 
one element for each of the sets $\mI_\alpha$ and $\mJ_\alpha$.
This yields the initial $\chi=1$ TCI,
which is exact if the function $A(u_1, \ldots,u_n)$ 
factorizes as a product of functions of one variable. 

\begin{figure}[ht!]
\begin{centering}
\includegraphics[width=0.45\textwidth]{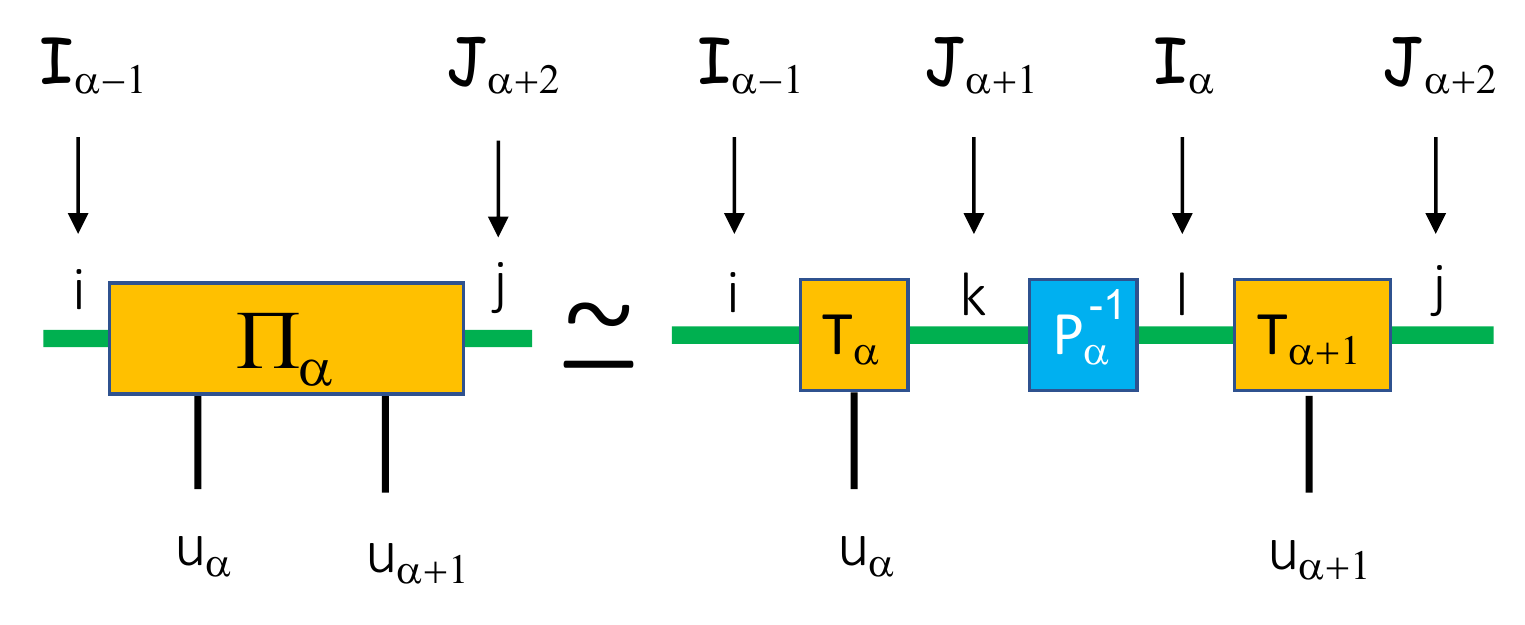}
\par\end{centering}
\caption{\label{fig:PiDecomposition} 
Pictorial representation of the $\Pi_\alpha$ tensor and its cross interpolation.
The notation is the same as in Fig. \ref{fig:ttd_notations}.
}
\end{figure}

Let us now define the tensors $\Pi_\alpha$, named for their four-legged
shapes (see Fig. \ref{fig:PiDecomposition}), by %as in two-site DMRG:
\begin{equation}
\label{eq:piDef}
\Pi_\alpha \equiv A(\mI_{\alpha-1}\oplus \mathbb{K}_\alpha\oplus
  \mathbb{K}_{\alpha+1}\oplus \mJ_{\alpha+2}).
\end{equation}
A pictorial representation of $\Pi_\alpha$ is shown in Fig. \ref{fig:PiDecomposition}.
Considering $\Pi_\alpha$ as a matrix with $(i,u_\alpha)$ being the raw index and 
$(u_{\alpha+1},j)$ the column index, one can build a cross interpolation of $\Pi_\alpha$
using the pivots $\mI_{\alpha}$ and $\mJ_{\alpha+1}$. The resulting approximation of
$\Pi_\alpha$ reads,
\begin{multline}
\label{eq:piTCI2}
\Pi_\alpha (i, u_\alpha,u_{\alpha+1},j) \approx \sum_{kl} T_\alpha (i,u_\alpha,k) P_\alpha^{-1}(k,l) \\
\times T_{\alpha +1}(l,u_{\alpha +1},j),
\end{multline}
or equivalently, using matrix notation, as
\begin{equation}
\label{eq:piTCI}
\Pi_\alpha (u_\alpha,u_{\alpha+1}) \approx T_\alpha (u_\alpha) P_\alpha^{-1} T_{\alpha +1}(u_{\alpha +1}). 
\end{equation} 

We introduce the {\it error function} $\epsilon_\Pi$,
\begin{multline}
\epsilon_\Pi(i,u_\alpha,u_{\alpha+1},j) \equiv 
\Bigl | \Pi_\alpha (i,u_\alpha,u_{\alpha+1},j) -  \\
  \sum_{kl} T_\alpha (i,u_\alpha,k) P_\alpha^{-1}(k,l) T_{\alpha +1}(l,u_{\alpha+1},j)
\Bigr |,
\label{eq:errorFunctionNoEnv}
\end{multline}
where $i\in \mI_{\alpha-1}$, 
$u_\alpha \in \mathbb{K}_\alpha$, 
$u_{\alpha + 1} \in \mathbb{K}_{\alpha + 1}$ and 
$j\in \mJ_{\alpha+2}$. 
We show in Appendix \ref{app:proofErrorEstimate} that as a result of the
nesting condition, the error function satisfies
\begin{equation}
 \epsilon_{\Pi}(i, u_\alpha, u_{\alpha+1}, j) = | A - A_\text{TCI}| (i,
  u_\alpha, u_{\alpha+1}, j).
\label{eq:errorIsOnA}
\end{equation}
In other words, the error of the factorization of $\Pi_\alpha$ is in
fact the error of the interpolation $A_\text{TCI}$ with respect to $A$, 
computed on the two-dimensional slice determined by $i$ and $j$. Hence, improving
the factorization of $\Pi_\alpha$ does indeed improve the overall TCI representation of $A$. 

The algorithm adds more pivots to the sets $\mI_\alpha$ and $\mJ_{\alpha+1}$ in order to improve the approximation of $\Pi_\alpha$, while maintaining the nesting condition.
It finds a local maximum
$\left(i,u_\alpha,u_{\alpha+1},j\right)$ of the error function
$\epsilon_\Pi$ for $i \in \mI_{\alpha-1}$ and $j \in
\mJ_{\alpha+2}$, adds the new pivots
$i \oplus (u_\alpha)$ and $(u_{\alpha +1}) \oplus j$ to  $\mI_{\alpha}$
and $\mJ_{\alpha+1}$, respectively, 
and then updates the pivot matrix $P_\alpha$.
This procedure preserves the nesting condition. 
The rationale for adding the pivot for which the error is {\it maximum}
is that this choice of pivot yields the largest improvement in the accuracy
of the TCI approximation since the corresponding point becomes exact. 
Another way to understand this choice, as shown in Appendix \ref{app:ci:maxvol} is
that this choice gives the largest determinant for the corresponding $P_\alpha$ matrix, 
i.e. follows the max-vol principle.
  
In the  {\it full search} variant of the algorithm, the maximum of $\epsilon_\Pi$
is determined by a brute force search over all $(\chi d)^2$ values of $\Pi_\alpha$.
In the much faster {\it alternate search} variant, 
one searches for a local maximum of $\epsilon_\Pi$ by starting from a 
random point and  scanning $(i,u_\alpha)$ and $(u_{\alpha+1},j)$ alternatively. The search ends when a local maximum is found
or a maximum number of iterations (typically 3 or 4) is reached.
The computational cost of the alternate search variant is only $O(d \chi)$ for adding
a new pivot, hence $O(d \chi^2)$ globally.
In practice, for the cases considered in this paper, we have observed
little difference in the quality of the approximation obtained using the
two variants, and we therefore
used the alternate search variant for all results presented below.

For the case of continuous $(u_\alpha)$, we
have explored two approaches:
{\it i)} search for the pivots on a predefined grid, and
{\it ii)} search for a local maximum directly in the continuum, 
using standard optimization algorithms. Since we did not observe obvious advantages
in using the second approach, we use the first method for the results presented below. 

We perform $n_{\rm sw}$ {\it sweeps} of this procedure, each consisting
of a forward sweep, which
improves all $\Pi$ tensors from $\Pi_1$ to $\Pi_{n-1}$, and a backward
sweep, which improves all $\Pi$ tensors from
$\Pi_{n-1}$ to $\Pi_1$.
Each sweep increases the bond dimension $\chi$ by two (at most, see section \ref{sec:effectof}) so that $\chi \leq 2 n_{\rm sw} + 1$.

%%% PIVOT ERROR FUNCITON AND ENV

\subsubsection{Improved pivoting using an environment-aware error function}
\label{sec:ErrorEnv}

The error function $\epsilon_\Pi$ is quite natural and is used in \cite{ACA-tci-2014,dolgov2020integralRn}.
The standard choice in the literature is to follow {\it maxvol} principle \cite{tci-dmrg1-2010, tci-dmrg2-2011} where one looks for pivots that maximize the determinant of the pivot matrix $P_\alpha$.
Appendix \ref{app:ci:maxvol} shows that the two criteria are closely
related. 
Since our goal is to compute $n$-dimensional integrals,
we have found that another error function, directly associated to the
error of the integral, yields significantly better results in the cases
we have studied.

Let us consider a single $\Pi_\alpha$ tensor in the TCI (\ref{eq:defTCI}), 
and integrate over all variables $u_\beta$ except $u_\alpha$ and $u_{\alpha+1}$.
We have (in matrix notation)
\begin{align}
I &\equiv \int du_1\ldots du_n \, A(u_1, \ldots, u_n) 
\nonumber \\
\approx 
& \int du_\alpha du_{\alpha+1} 
\sum_{i,j} L_i   
\Bigl[  T_\alpha (u_\alpha) P_\alpha^{-1} T_{\alpha +1}(u_{\alpha+1}) \Bigr]_{ij}
R_j
\label{eq:IntegralPartialNotPi}
\end{align}
where 
\def\intT#1{\bigl(\smallint  T_{#1}\bigr)}
\begin{align}
L &\equiv \intT{1} P_1^{-1} \ldots \intT{\alpha-1} P_{\alpha-1}^{-1} 
\\
R &\equiv P_{\alpha+1}^{-1} \intT{\alpha+2} \ldots P_{n-1}^{-1} \intT{n}
\end{align}
are vectors of length $\chi$ and
\begin{equation}
   \label{eq:defIntTTensor}
   \intT{\alpha} \equiv \int du_\alpha \,\, T_\alpha(u_\alpha).
\end{equation}
A better approximation of the $n$-dimensional integral, which replaces the
part of the factorization involving the $u_\alpha$ and $u_{\alpha+1}$
variables with the exact slice $\Pi_\alpha$, is
\begin{equation}
\label{eq:IntegralPartialPi}
I \approx \int du_\alpha du_{\alpha+1} 
\sum_{i,j} L_i   
\Pi_\alpha (u_\alpha,u_{\alpha+1})_{ij}
R_j.
\end{equation}
Here, we see that the $\Pi_\alpha$
tensor in the integral is weighted by the factors $L$, $R$, which we
refer to as the {\it environment} in a manner reminiscent of DMRG.
We therefore propose taking the difference of Eq.\eqref{eq:IntegralPartialNotPi} and 
Eq.\eqref{eq:IntegralPartialPi}, and using the modulus of the resulting
integrand as an error function:
\begin{equation}
   \label{eq:errfunENV}
\epsilon^\text{env}_\Pi(i,u_\alpha,u_{\alpha+1},j) \equiv | L_i R_j | \epsilon_\Pi
  (i,u_\alpha,u_{\alpha+1},j).
\end{equation}
We refer to
this as the {\it env} variant of the algorithm. In practice, we also
multiply this error with another weight $W_a W_b$ defined in the next section (weighted learning variant).

We will show that
the {\it env} variant significantly outperforms the standard algorithm using
$\epsilon_\Pi$ in the cases considered below.
Indeed, it leads to the selection of pivots in regions of large volume in which the integrand is small.
This is illustrated by analogy with the following integral
$\int_0^\infty dx \ (e^{-x} + e^{-x/100}/100) = 2$.
One of the terms in the integrand is rapidly decaying, and the other is
slowly decaying and small, but both contribute equally to the integral.
The ordinary choice of pivots leads to sampling the integrand based on its absolute value, 
ignoring the weighting of the corresponding contribution by its volume.
The corresponding algorithm would focus on improving the description of
the large term, and only start adding points in the tail region $x\gg 1$ once the large term is known very accurately.
Instead, the two terms should be approximated with an error weighted by
their respective contributions to the integral. %not with the same absolute error but with the same relative error, i.e. 
By including the corresponding volumes in the weight, the error function
$\epsilon^\text{env}_\Pi$ implement this idea. 

%%% Computing the integrals.

\subsubsection{Quadrature rules for numerical integration}
\label{sec:integrationTCI}
It remains to specify a method of calculating the one-dimensional integrals Eq.\eqref{eq:defIntTTensor}. The
integration in the case that the underlying domain is a simplex, rather
than a hypercube, is discussed below.
This question is independent of tensor factorizability and the TCI construction.
The behavior of the functions $T_\alpha(u_\alpha)$ varies
from model to model, as does the precise domain of integration, which
will be discussed in the next section on real time computations.

In this work, we use rules based on either Chebyshev polynoms or Gauss Kronrod
quadratures to perform these one-dimensional integrals \cite{kronrod1965nodes}.
The quadrature rule associated to Chebyshev polynoms is known as the Clenshaw–Curtis quadrature.
However, we will also use Chebyshev interpolants to perform integrations on domains smaller than the
initial domain used to construct the Chebyshev interpolant and in that case we get different weights.
We note $CH_x$ (resp. $GK_x$) the rules for Chebyshev polynoms (resp. 
Gauss-Kronrad quadrature) with $x$ points.
In one application, we will encounter highly oscillatory and slowly
decaying integrals. Although these integrals could be calculated with
standard quadratures by brute force, we have found that building a
specialized quadrature yields a significant improvement in efficiency (see Appendix \ref{app:tailored_quad}).

These quadratures specify a set of $d$ points $x_a$ and weights $W_a$ such that
\begin{equation}
\int du \ T_\alpha(u) \approx \sum_{a=1}^d W_a T_\alpha(x_a).
\end{equation} 
The multidimensional integral $I$ then reduces to the full contraction of the tensor with the weights $W_a$, 
i.e. the contraction of $A(x_{i_1}, \ldots,x_{i_n}) W_{i_1} \ldots W_{i_n}$.
In the {\it weighted learning} variant of the algorithm, the TCI is
constructed for this weighted tensor, rather than the original tensor
$A$. It has been argued in the literature~\cite{dolgov2020integralRn}
that weighted learning 
improves the convergence of $I$ with the bond dimension.
We show below that, in the cases considered here, the improvement is
marginal.

%%%%%%%%%   MANY BODY %%%%%%%%%%%%%%%%%%%%

\section{Real time many-body formalism}
\label{sec:manybody}
The formalism used in this article follows the Keldysh approach in real
time that is used in the context of
diagrammatic quantum Monte Carlo \cite{profumo2015}. 
For completeness, we review the main definitions and expressions which will be needed later. We refer to 
\cite{Bertrand_1903_kernel} for proofs and additional details.

\subsection{Perturbation theory with Wick determinants}
\label{sec:wick}
Our starting point is a Hamiltonian $H = H_0 + H_{\rm int}\theta(t)$
consisting of an arbitrary non-interacting
term $H_0$ and an interaction term $H_{\rm int}$ that is switched on at $t=0$. The non-interacting part is arbitrary:
\begin{equation}
H_0 = \sum_{ii',\sigma\sigma'} (H_0)_{ii',\sigma\sigma'}
  c^\dagger_{i\sigma} c_{i'\sigma'}.
\end{equation} 
However, for concretness, we focus on $H_0$ that are diagonal in the spin sector. Here the fermionic operator $c^\dagger_{i\sigma}$ ($c_{i\sigma}$)
creates (destroys) an electron with spin $\sigma$ on site $i$. We
consider systems directly in the thermodynamic limit, i.e. with an
infinite number of sites $i$. The interaction term can in principle be
an arbitrary quartic Hamiltonian. However, since all of the
calculations in this article will be performed using a Hubbard-like
interaction, we concentrate on this specific form for concreteness:
\begin{equation}
H_{\rm int} = U \lambda (t) \sum_{i\in\mathcal{C}} (c^{\dagger}_{i\uparrow} c^{\phantom{\dagger}}_{i\uparrow}
- \bar\alpha) (c^{\dagger}_{i\downarrow} c^{\phantom{\dagger}}_{i\downarrow} - \bar\alpha). 
\end{equation}
Here the sum is taken over a finite subset $\mathcal{C}$ of interacting sites.
The $\bar\alpha$ term shifts a quadratic term between the non-interacting
and the interacting part of the Hamiltonian, and therefore provides a
mathematically different perturbation expansion in powers of $U$\cite{profumo2015, Rubtsov2004}
of the same physical problem. The function $\lambda (t)$ captures the time dependence of the interaction. 
One of the remarkable features of TTD is that $\lambda (t)$ is only
needed after the factorization is performed, in the post-processing step
in which the integration is carried out. Calculating the time evolution
of an observable for different functions $\lambda (t)$ therefore comes
essentially for free. Most of the examples treated in this
article use $\lambda (t)=\theta (t)$, a Heaviside function,
but we also describe a non-trivial example $\lambda (t) = 1 -
\cos(t/t_{\rm sw})$ to illustrate the algorithm's capabilities.

The dynamics of $H_0$ can be formally solved through the introduction of
the corresponding non-interacting Green's functions. The lesser and
greater Green's functions $g^<$ and $g^>$ can be computed explicitly
from $H_0$ and comprise, together with the value of $U$ and $\bar\alpha$, the actual input of the problem. 

We have
\begin{equation}
g_{ii',\sigma\sigma'}^{<}(t)=i\langle c_{i'\sigma'}^{\dagger}(0)c_{i\sigma}(t)\rangle\text{,}
\end{equation}
\begin{equation}
g_{ii',\sigma\sigma'}^{>}(t)=-i\langle c_{i\sigma}(t)c_{i'\sigma'}^{\dagger}(0)\rangle\text{,}
\end{equation}
where the time dependence of an operator $\mathcal{O}$ is given by
$\mathcal{O}(t)=e^{i H_0 t}\mathcal{O}e^{-i H_0 t}$.
These Green's functions can be computed analytically for simple models, or
numerically in more complex scenarios using e.g. Tkwant~\cite{Kloss2021}. 
Their explicit forms for the specific models considered here are given below.

We introduce the general coordinate $X = (i,\sigma, t,a)$ which
describes the site index $i$, the spin $\sigma$, the time $t$, and a
``Keldysh index'' $a$
taking the values $0$ or $1$. The Keldysh Green's function $g(X,X')$ is
defined as
\begin{align}
    &g(X,X') \equiv \\
    \nonumber 
    &\begin{cases}
       g^>_{ii',\sigma\sigma'}(t)\theta(t) + g^<_{ii',\sigma\sigma'}(t) \theta(-t) & \text{for } a=a'=0 \\
       g^>_{ii',\sigma\sigma'}(t)\theta(-t) + g^<_{ii',\sigma\sigma'}(t) \theta(t) & \text{for } a=a'=1 \\
    g^<_{ii',\sigma\sigma'}(t)  & \text{for } a=0, a'=1\\
 g^>_{ii',\sigma\sigma'}(t)  & \text{for } a=1, a'=0,
    \end{cases}
\end{align}
where $\theta(t)$ is the Heaviside function and we have taken $t'=0$ since $g(X,X')$ is a function of $t-t'$. We also introduce the full
interacting Green's functions $G(X,X')$ with an identical definition as
$g(X,X')$ but with the full Hamiltonian $H$ replacing $H_0$.
Observables can be related to $G(X,X')$ in a simple manner. For
instance, the occupation of an
orbital $(i,\sigma)$ at time $t$ is given by
$-i G^<_{ii,\sigma\sigma}(t,t)$.

Using this notation, we can write the perturbative expansion of
$G(X,X')$ in powers of the interaction coupling $U$. We obtain
\begin{equation}
G(X,X') = \sum_{n=0}^\infty G_n(X,X') \ U^n,
\end{equation}
where $G_n(X,X')$ is defined as
\begin{equation}
G_n = \sum_{i_1 i_2 \ldots i_n} \int_{\rm S_u} du_1 du_2  \ldots  du_n    
\,\lambda(u_1)\lambda(u_2) \ldots \lambda(u_n)  \tilde G_n.
\label{eq:quadrature}
\end{equation}
The integration is carried out inside the simplex ${\rm S_u}$ defined by % i.e. the $u_i$ that verify
$0\le u_n \le  \ldots \le u_2 \le u_1 \le t$.
Assuming for simplicity that $H_0$ conserves spin, the integrand $\tilde
G_n(X,X',i_1,i_2 \ldots i_n,u_1,u_2 \ldots .u_n)$ is given explicitly by
\begin{equation}
\label{eq:wick}
\tilde G_n  = i^n \sum_{a_1, \ldots, a_n} (-1)^{\sum a_k} \lists{X, U_1, \ldots, U_n}{X', U_1, \ldots, U_n}
    \lists{U_1, \ldots, U_n}{U_1, \ldots, U_n},
\end{equation}
where $U_k = (i_k, u_k, a_k)$, and the ``Wick determinant'' $\lists{\ldots}{}$ is defined, for $A_1, \ldots, A_m$ and
$B_1, \ldots, B_m$ any collections of points on the Keldysh contour, as
\begin{equation}
    \lists{A_1, \ldots, A_m}{B_1, \ldots, B_m} =
    \begin{vmatrix}
        {g}(A_1, B_1) & \hdots & {g}(A_1, B_m) \\
        \vdots & \ddots & \vdots \\
        {g}(A_m, B_1) &  \hdots & {g}(A_m, B_m)
    \end{vmatrix}.
\end{equation}
For the case in which $\bar\alpha \ne 0$, the diagonal terms of the Wick
determinants must be shifted by $-i\bar\alpha$ \cite{profumo2015}. 

In the following, we will illustrate the method for the calculation of the total charge 
on site $i=0$ at a time $t$ after switching on the interaction:
\begin{equation}
Q(t,U) =  \langle e^{i\int dt H } c^{\dagger}_{0\uparrow}  
c^{\phantom{\dagger}}_{0\uparrow}e^{-i\int dt H }\rangle.
\end{equation}
Here the evolution operator $e^{-i\int dt H }$ is a time-ordered exponential.
This observable admits an expansion $Q(t,U) = \sum_n Q_n(t) U^n$, and we refer
to the corresponding integrand (\ref{eq:wick}) as $\tilde Q_n(i_1,
\ldots,i_n,u_1, \ldots,u_n,t)$,
so that
\begin{equation}
\label{eq:def_Qn}
Q_n(t) =\sum_{i_1 \ldots i_n}\int_{\rm S_u} du_1 du_2  \ldots  du_n   \lambda(u_1)\lambda(u_2) \ldots \lambda(u_n)  
\ \tilde Q_n.
\end{equation}
The $n$th order contribution to the expansion is given by an $n$-dimensional
integral.
The integrand $\tilde Q_n$ is given by a sum of $2^n$ Wick
determinants, which can be computed explicitly from the knowledge of the
non-interacting dynamics. 
The complexity of computing the integrand therefore appears to be $O(n^3 2^n)$, but 
there are known algorithms \cite{Griffin_2006_Pminors,
Simkovic_2022_Pminors} to compute it with $O(2^n)$ complexity. In 
Appendix \ref{app:keldyshfastsumdets}, we present a simpler version of
such an algorithm using only a few lines of codes.
In this way, the computational problem is reduced to that of computing
the high dimensional integrals above.

\subsection{Models}
\label{sec:Models}
We consider three different models in this article:
a single quantum dot weakly coupled to electrodes, a quantum dot strongly coupled to a two-dimensional infinite electrode,
and a double quantum dot weakly coupled to electrodes. 
The inputs to the TTD method are the corresponding non-interacting Green's functions $g$.
Their explicit forms are given in Appendices \ref{app:g0_qd} and \ref{app:g0_2d}. 
Fig.~\ref{fig:g0} shows specific examples of these Green's functions for
the three different problems. Note that all the examples considered here
are invariant with respect to spin rotations, so the Green's functions do not depend on spin.

\begin{figure}
\includegraphics[width=0.49\textwidth]{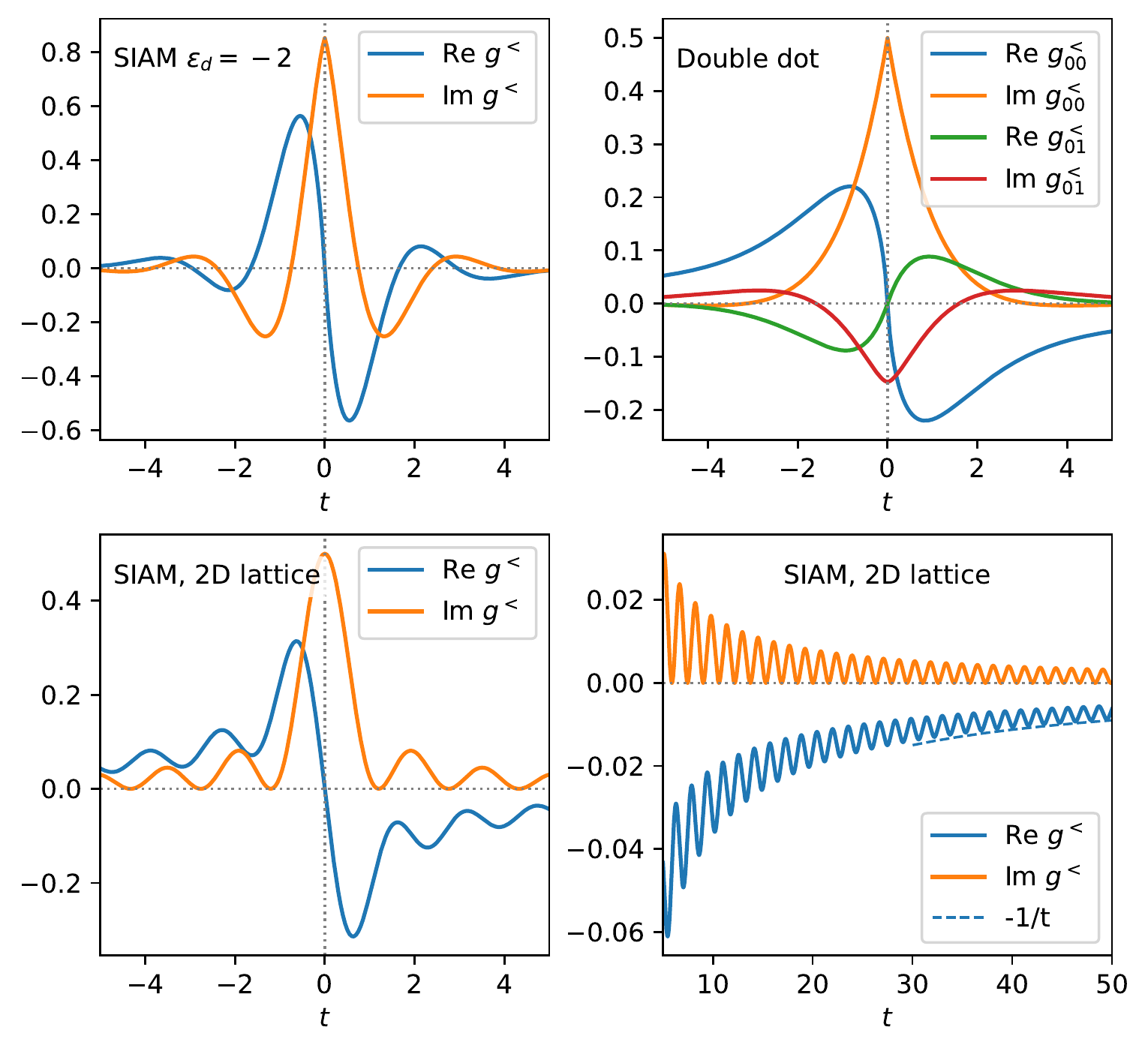}
\caption{\label{fig:g0} Real time non-interacting lesser Green's
  functions for the three models considered in this paper. Upper panel: single quantum dot (SIAM) and double quantum dot connected to leads in the flat-band limit. Lower panel: one interacting site in the 2D lattice, including the tail of $g^<(t)$.
}
\end{figure}

\subsubsection{Single-impurity Anderson model (SIAM)}
The first system is an interacting quantum dot connected to non-interacting
leads: the single-impurity Anderson model (SIAM). The Hamiltonian is
given by $H=H_{0}+H_{\text{int}}\theta(t)$, with
\begin{equation}
H_{0}=\sum_{i \sigma}\gamma_{i}\left(c_{i \sigma}^{\dagger}c_{i+1,\sigma}+h.c.\right)+\epsilon_{d}\sum_{\sigma}c_{0 \sigma}^{\dagger}c_{0 \sigma}\text{,}\label{eq:siam}
\end{equation}
\begin{equation}
H_{\text{int}}=Uc_{0 \uparrow}^{\dagger}c_{0 \uparrow}c_{0 \downarrow}^{\dagger}c_{0 \downarrow}\text{.}
\end{equation}
The hopping parameters are all equal, $\gamma_{i}=\gamma$, except for
the connection of the quantum dot to the leads,
$\gamma_{0}=\gamma_{-1}\ne\gamma$. We work
in the flat-band limit (see Appendix \ref{app:g0_qd}) in which $\gamma_{0},\epsilon_{d}\ll\gamma$
but $\Gamma=2\gamma_{0}^{2}/\gamma$ is finite. $\Gamma$ is used
as our unit of energy. In this limit, the electron-hole symmetric
case $\epsilon_{d}=0$ is taken as a benchmark, since there the exact
expression for the charge $Q(U)$ at equilibrium is given by the Bethe
ansatz \cite{macek2020qqmc}. 

The upper-left panel of Figure \ref{fig:g0} shows an example of the
non-interacting Green's function g on the quantum dot
for $\epsilon_{d}=-2$. In this case, the integrand $\tilde{Q}_{n}$ has many oscillations;
see Figure \ref{fig:Ovw}c (even more oscillations appear in other cases,
for example $\epsilon_{d}=-4$).% , while there is not sign problem for $\epsilon_{d}=0$ or $\epsilon_{d}=-0.3$.

\subsubsection{Single impurity in a 2D lattice}
\label{sec:models:siamlattice}

Our second system is a quantum dot strongly coupled to a more complex electronic bath.
The system is an infinite two-dimensional lattice,
\begin{equation}
H_{0}=\gamma\sum_{\langle ij\rangle \sigma}c_{i \sigma}^{\dagger}c_{j \sigma}\text{,}
\end{equation}
with only one interacting site $i=0$,
\begin{equation}
H_{\text{int}}=Uc_{0 \uparrow}^{\dagger}c_{0 \uparrow}c_{0 \downarrow}^{\dagger}c_{0 \downarrow}.
\end{equation}
Here $\langle ij\rangle$ corresponds to the nearest neighbor indices in
2D, and $\gamma$ is taken as the unit
of energy. The non-interacting correlators are highly oscillatory,
with slow decay $\sim 1/t$; see the lower panel of Figure \ref{fig:g0}. The corresponding integrand $\tilde{Q}_{n}$
decays very slowly, with rapid oscillations.

\subsubsection{Double quantum dot}
\label{sec:models:doubledot}

The last system is a double quantum dot connected to two electrodes,
\begin{equation}
H_{0}=\sum_{i \sigma}\gamma_{i}\left(c_{i \sigma}^{\dagger}c_{i+1,\sigma}+H.c.\right)\text{,}
\end{equation}
\begin{equation}
H_{\text{int}}=U\sum_{i=0,1}c_{i \uparrow}^{\dagger}c_{i \uparrow}c_{i \downarrow}^{\dagger}c_{i \downarrow}\text{,}
\end{equation}
with $\gamma_{i}=\gamma$, except for $\gamma_{-1}=\gamma_{1}\ne\gamma$ and $\gamma_{0}\ne\gamma$.
We work in the flat-band limit in which $\gamma_{0},\gamma_{1}\ll\gamma$,
and $\Gamma=\gamma_{1}^{2}/\gamma$ is taken as the unit of energy. Double quantum dots play an important role in semi-conducting quantum technologies, as qubit systems or detectors.

%%%%%%%%%%%%%%%%%%%%%%%%%%%%%%%%%%%%%%%%%%%%%%%%%%%%%%%%%%%%%%

\section{Tensor Train Diagrammatics}
\label{sec:tdd}

We now turn to the discussion of the {\it tensor train diagrammatics (TTD)}
technique, i.e. the application of tensor train interpolation (TCI) to the
calculation of high order perturbation expansions in the interacting coupling
strength, as in \eqref{eq:def_Qn}.
We use the TCI algorithm to factorize $\tilde Q_n$ and perform
the corresponding $n$-dimensional integral. This section focuses on impurity
models, for which the integral involves $n$ time variables but no
spatial sums. 
The extension to multi-orbital models will be discussed in Section \ref{sec:beyond}.

The calculation can be split into two fundamentally different steps.  In
Section \ref{sec:Method:Factorizability}, we discuss the
factorization of the tensor $\tilde Q_n$ appering in \eqref{eq:def_Qn} 
using a TCI decomposition. In Section
\ref{sec:TTDintegration}, we discuss the computation of the integral, along the
lines outlined above. We discuss the various sources of error in \ref{sec:TTDErrorEstimation}. Detailed benchmarks
and numerical results are then presented in the next section.

\subsection{Factorizability in time differences} 
\label{sec:Method:Factorizability}

The {\it integration domain} in \eqref{eq:def_Qn} is the simplex $S_u$
defined by $0\le
u_n \le  \ldots \le u_2 \le u_1 \le t$.  The TCI
decomposition itself is constructed in a different domain the hypercube $[0,t]^n$.
We could simply integrate over the whole
hypercube and divide the result by $n!$, since the integrand  $\tilde Q_n$ is symmetric in
the $u$ variables as a result of the anticommutativity of fermionic operators under the
time-ordered  product.
However, $\tilde Q_n$ has a cusp whenever two of the $u_i$ are equal,
because of the Heaviside functions introduced by time ordering. Some of these cusps can be seen in Fig.~\ref{fig:Ovw}a (e.g. when $u_1 = u_2$).
Such a function does not factorize well; consider, for example, the
Heaviside function itself, $\theta(u_1 - u_2)$.
We check explicitly in Fig.~\ref{fig:Q10}a that a direct decomposition
in the $u$ variables fails. 

We therefore change to the time difference variables $v_i$,
defined by:
\begin{subequations}
   \label{eq:v_vars}
   \begin{align}
   v_1 &= t - u_1 \\ 
   v_i &= u_{i-1} - u_i \qquad \text{ for } 2 \leq i \leq n.
\end{align}
\end{subequations}
This change of variable has a Jacobian $\bigl| {\rm det} [\partial u_i/\partial v_j] \bigr |= 1$.  
In the $v$ variables, the integration domain becomes 
\begin{subequations}
\begin{align}
   v_i &\geq 0 \label{eq:domainVPositive},\\
   \sum_{i=1}^n v_i & \leq t \leq t_M, \label{eq:domainVTm}
\end{align}
\end{subequations}
where $t_M$ is the maximum time of the calculation (possibly infinite
for a steady state calculation).
The condition \eqref{eq:domainVPositive} enforces the time ordering in
$u$. As a result, the function $\tilde Q_n(v_i)$ 
has no cusps due to time ordering inside the hypercube $[0,t_M]^n$ in the $v$
variables. 

Using the TCI algorithm, we first obtain a factorization of $\tilde Q_n(v_i)$ on $[0,t_M]^n$, 
as in \eqref{eq:defTCI}:
\begin{equation}
   \label{eq:TCIQn}
   Q_n(v_1, \ldots, v_n) \approx Q_n^\text{TCI}(v_1, \ldots, v_n) \equiv
   \prod_{\alpha=1}^n T_\alpha(v_\alpha) P_\alpha^{-1}.
   %Q_n(v_1, \ldots, v_n) \approx  T_1(v_1) P_1^{-1} T_2(v_2) \ldots P_{n-1}^{-1} T_n(v_n) 
\end{equation} 
In practice, we enforce the simplex condition \eqref{eq:domainVTm} not only for
the integration, but also when choosing new pivots: when searching for
pivots which maximize the error $\epsilon_\Pi$ or $\epsilon_\Pi^{\rm env}$, we
only consider candidates that satisfy \eqref{eq:domainVTm}. Indeed, we
only need to improve our approximation in the integration region.
Since the simplex is $n!$ times smaller than the hypercube, this
provides a significant speed up.
We have observed numerically that for $t_M$ sufficiently large, the disregarded candidate pivots
would almost never have been selected anyway. 
We have also observed that the number of evaluations of $\tilde Q_n$ used in the pivot search is only approximately 30\%
of the number of evaluations required to construct the TCI approximation given the pivots. 
In that sense, the algorithm is close to optimal.

Even in the $v$ variables, the $\epsilon$-factorizability of $\tilde Q_n(v_1,\ldots,v_n)$ is far
from obvious. For large $v_i$, some indications can be found.
First, it was shown numerically in \cite{macek2020qqmc} that a rank $\chi=1$
approximation of $\tilde Q_n$ is 
reasonably accurate. This was an essential ingredient in the
construction of the $n$-dimensional change of variables required in
quasi-Monte Carlo methods.
Second, when the Green's function $g$ decays exponentially, 
we expect $\tilde Q_n$ to be dominated by a single Feynman diagram (the nested
tadpole diagram) and hence to factorize with rank $\chi=2$,
\begin{equation}
   \tilde Q_n(v_1, \ldots,v_n) \mathop{\sim}_{v_i \rightarrow \infty}
   {\rm Tr\ } \prod_i M(v_i),
\end{equation}
where the $2\times 2$ matrix $M$ is given by $M_{aa'}(v_i) = (-1)^a \bigl (\hat g_{aa'}(v_i) \bigr)^2 $.

In this paper, we will demonstrate numerically a much stronger property: {\it $\tilde Q_n$ is
$\epsilon$-factorizable in the whole hypercube in $v$}, even for
small time differences.

\subsection{Integration in the simplex} 
\label{sec:TTDintegration}

After obtaining the factorization \eqref{eq:TCIQn}, 
we perform the integral over the times $u$. 
For steady state calculations ($t\rightarrow\infty$), 
we perform the integration in the full $v$-hypercube, see \eqref{eq:domainVTm}. 
For calculations at finite $t$, we need to integrate
over the simplex in $v$ defined by \eqref{eq:domainVTm}.

The integration is carried out as a post-processing step after the
factorization, which is the most time consuming task.
A {\it single} tensor train interpolation is sufficient to
obtain the full curve $Q_n(t)$ for $0\le t\le t_M$, and for any
$\lambda (t)$.
We proceed as follows; Appendix \ref{app:simplex_int} contains further details.
We define the one-dimensional functions $\Psi$ by
\begin{equation}
\label{eq:psi1}
\Psi_{n}(x) =  \int_0^{x} \ dy \ \lambda(y)  T_n(x-y),
\end{equation}
and
\begin{equation}
\label{eq:psi2}
\Psi_{p}(x) =  \int_0^{x} \ dy \ \lambda(y)  T_{p}(x-y) P_{p}^{-1}\Psi_{p+1} (y)
\end{equation}
for $p < n$. We then have
\begin{align}
\label{eq:simplex}
Q_n(t) \approx Q_n^\text{TCI}(t) = \Psi_1(t).
\end{align}

An alternative method using a Fourier transform to perform the integration in $v$ is 
presented in Appendix \ref{app:fourier_int}.
However, the direct approach above is more general [arbitrary $\lambda
(t)$] and numerically faster so it is preferred in practice. 

Calculations in the steady state ($t\rightarrow\infty$) limit or at large $t_M$ present 
a specific difficulty. % due to the large volume of the integration domain.
The tail of $\tilde Q_n(v_i)$ at large $v_i$ is small, so it 
has little effect on the factorization, 
but can make a significant contribution to the integral because of its large volume.
This problem is addressed by the {\it env} error function $\epsilon_\Pi^\text{env}$.
Alternatively, this problem could be solved manually in simple cases 
using a second change of variables mapping $[0,t_M]$ onto $[0,1]$:
\begin{equation}
   \label{eq:definitionWvariable}
   w_i(v_i)\equiv 2 v_i/( t_M + v_i).
\end{equation}
Since this change of variables is diagonal, it does not effect
the factorizability of the tensor. 
However, decomposing the function in $w_i$ introduces a large weight 
$dv_i/dw_i =2t_M/(2-w_i)^2$ in the tail region from the Jacobian,
which affects the choice of pivots.
These techniques will be illustrated in section \ref{sec:effectof}.
\begin{figure}[ht]
\includegraphics[width=0.49\textwidth]{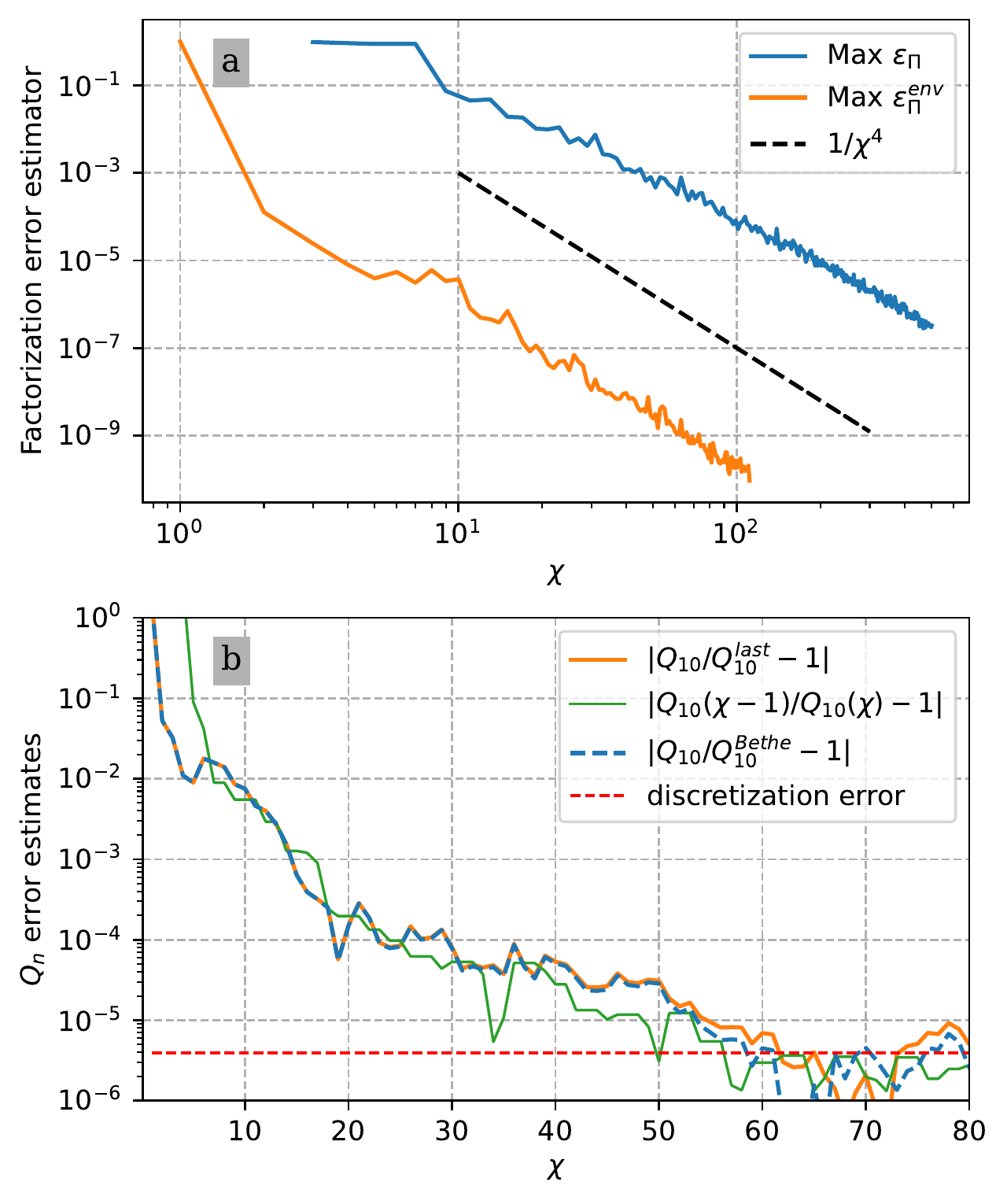}
\caption{\label{fig:MethodTTD} 
Upper panel: factorization error estimator as a function of the tensor rank $\chi$, 
for the SIAM model, for $n=10$, $\epsilon_d=0$, $t_M= 15$, Gauss-Kronrod
  points with $63$ points,
using the error functions $\epsilon_\Pi^{env}$ (orange curve) and $\epsilon_\Pi$ (blue curve), see text.
Lower panel: error of the integral $Q_n$ vs. $\chi$, measured using different estimators.
In this case $Q_{10}^{last}$ is $Q_{10}$ for $\chi=100$.
  The red line is the error due to the integral discretization obtained by varying the number of integration points.
}
%\caption{\label{fig:figErrorEst} Error estimation. The parameters are: gt,wt="gk","V" and tmax,nChe,dim=30,127,10.}
\end{figure}

\begin{figure*}[ht]
\includegraphics[width=1.0\textwidth]{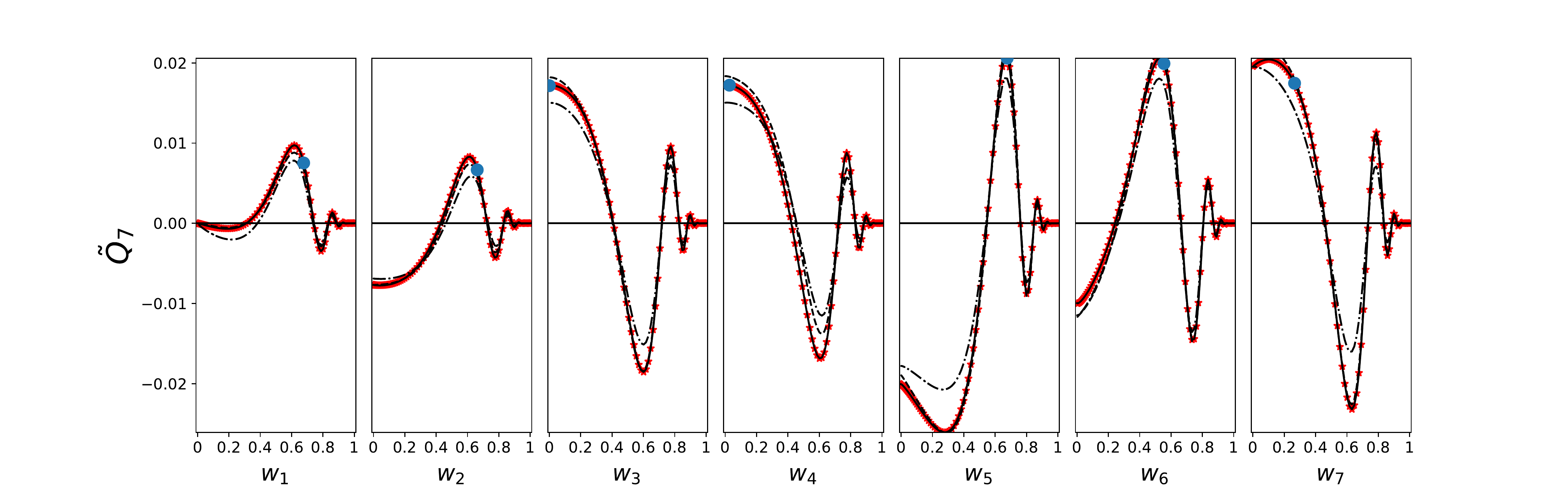}
\caption{\label{fig:errorMax} Comparison of the integrand $\tilde Q_7$ 
for $\epsilon_d=-2$ (red stars) with the TCI of rank $\chi = 2,4, 40$
  (dash-dotted, dashed, and solid black lines, respectively). The function is plotted in the $w$ variable \eqref{eq:definitionWvariable} with
   $t_{M}=5$.
   The  maximum error point is $(0.71, 0.62, 0.0, 0.07, 0.45, 0.60,
   0.25)$ (blue circles).
   Each panel corresponds to the variation of one variable $w_i$
   starting from this point.
}
\end{figure*}

%%%%%%%%%%%%%%% ERROR 

\subsection{Error estimation}
\label{sec:TTDErrorEstimation}

In this section, we present a practical error monitoring scheme for TCI calculations. 
The TTD method has three main sources of errors, which are controlled by
$\chi$ (factorization error), $d$ (discretization error) and $t_M$ (time
truncation error; only for steady state calculations), respectively.

The {\it factorization error}, which comes from the
approximation of the integrand by a tensor train \eqref{eq:TCIQn}, is
the most important source of error. 
We use an estimator of this factorization error defined, for a given value of $\chi$, as the maximum over the sweep at rank $\chi$ of the maximum error $\epsilon_\Pi$ (resp.
$\epsilon_\Pi^{env}$) in the regular (resp. {\it env}-variant) of the TCI
algorithm. 
In machine learning terminology, this is an {\it in-sample
error}, since it is
computed solely from the data used in the construction of the approximation.
It is nevertheless a conservative in-sample error, as we actively seek pivots
with large values of the error. 
In Fig. \ref{fig:MethodTTD}a, we show this estimator for the SIAM model with
$\epsilon_d=0$, for which the Bethe ansatz solution gives high accuracy
benchmarks. In this case, the error decays quickly as $\sim 1/\chi^4$.
 This fast decay is the signature of $\epsilon$-factorizability.

The error on the actual physical quantity $Q_{10}$ in the SIAM model with $\epsilon_d=0$ is shown in Fig. \ref{fig:MethodTTD}b. 
We compare different estimates of the error: the exact error (known here from the Bethe ansatz
solution, but unavailable in general), the {\it running relative error} between $Q_n(\chi-1)$
and $Q_n(\chi)$, where $Q_n(\chi)$ denotes the approximation of $Q_n$ at rank
$\chi$, and the error with respect to the largest value of $\chi$ used,
available only at the end of the calculation. We find that these estimators
are in excellent agreement. 
In practice, the running relative error yields a satisfactory estimate of the true error.
The saturation observed for large
$\chi$ is the result of discretization error.
 
Apart from the factorization error, three more sources of error need
to be controlled. First, the {\it 
discretization error} stems from the 
one-dimensional integration, and is determined by the number of
integration points $d$. In highly oscillatory cases, one may require a
specialized integration technique, as discussed in Section \ref{sec:lattice}.
Second, for calculations at infinite time in the steady state $t_M = \infty$,
the convergence with $t_M$ must be verified as well.
Third, in some cases {\it rounding errors} can become significant, in particular when very high
precision ($\sim 10^{-8}$ or smaller) is sought with large expansion
orders $n$. These result from cancellations in the
summation over Keldysh indices in the calculation of $\tilde Q_n$ (see
Appendix \ref{app:keldyshfastsumdets}).

In practice, monitoring the different errors discussed above is sufficient. 
Nevertheless, in order to illustrate the quality of the TCI approximation, we now also present 
an {\it out-of-sample error} estimate, obtained as follows.
Starting from a random point $(v_1,\ldots, v_n)$, we sweep over the
variables $v_1$, $v_2, \ldots, v_n$ several times. At the $i$th step of
each sweep, we update $v_i$ to maximize the error, with all other variables
fixed. This eventually yields a point $(v_1^*,\ldots, v_n^*)$ that maximizes 
the error locally. We have found that the value of the error obtained by
this procedure is robust with respect to the starting point.
The results are illustrated in Fig.~\ref{fig:errorMax}, which shows one dimensional
slices of $\tilde Q_7$ along every possible direction starting from $v_i^*$. The
blue circles indicate the position of $v_i^*$. The TCI approximation is already
qualitatively correct for $\chi=2$, quantitatively correct for
$\chi=4$, and indistinguishable from the exact solution for 
$\chi=40$. Despite the presence of strong oscillations (the average sign is $10^{-3}$
in this case), the maximum relative error that we observe for this
calculation is of the order of $1\%$, only one order of magnitude larger than the in-sample factorization error.

%%%%%%%%%%%%%%%%%%%%%%%%%%%%%%%%%%%%%%%%%%%%%%%%%%%%%%%%%%%%%%%%%%%%55

\section{Results for the SIAM model}
\label{sec:siam}

In this section, we present comprehensive numerical results for the 
SIAM model in various regimes, in order to illustrate the practical
performance of TTD.

\subsection{High precision benchmarks using the Bethe ansatz}

\begin{figure*}[th]
\includegraphics[width=1.0\textwidth]{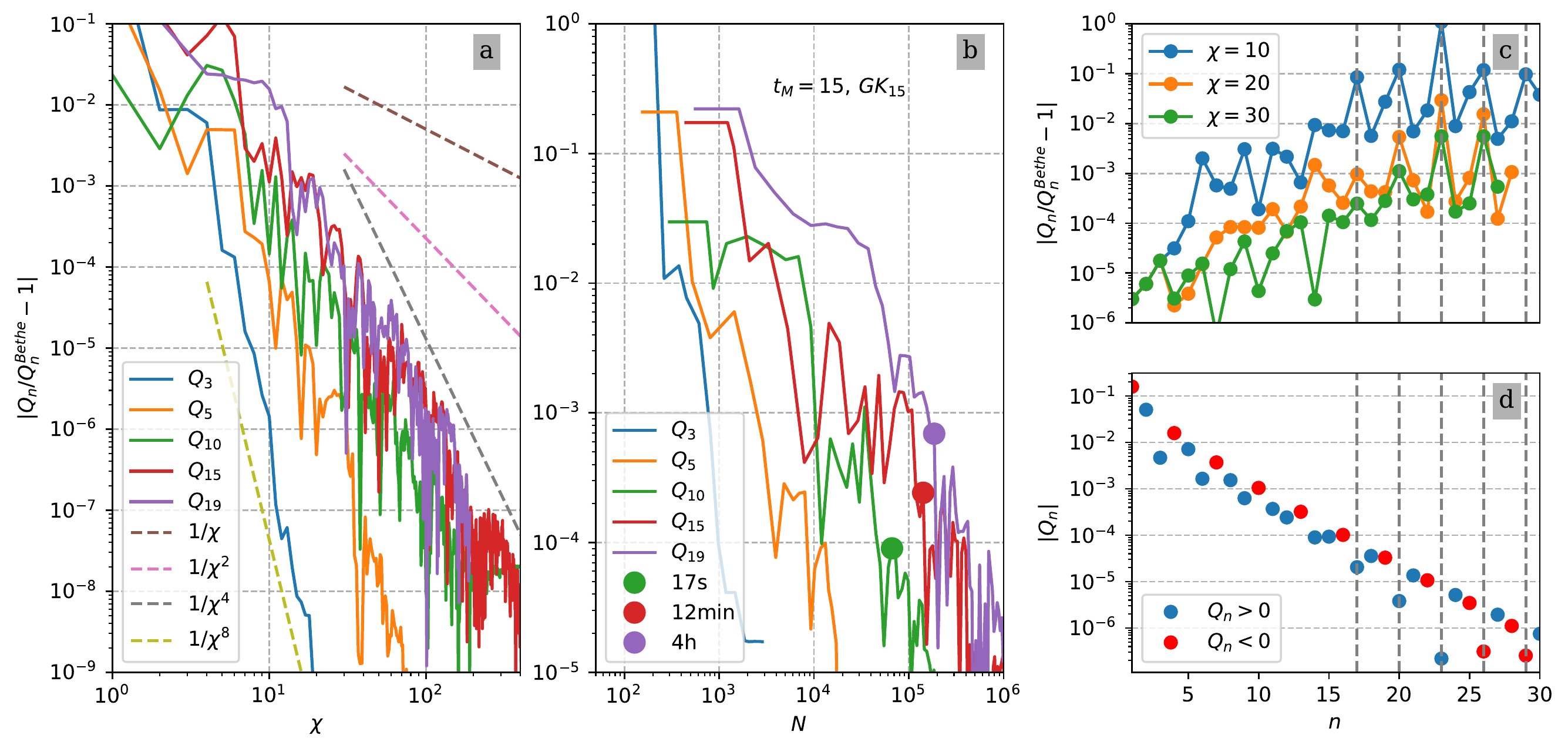}
\caption{
   \label{fig:Qn} 
Relative error of the coefficients $Q_n$ with respect to the exact
  solution for $n=3$ (blue), $n=5$ (orange), $n=10$ (green), $n=15$
  (red), and $n=19$
(purple). All calculations were carried out using the {\it env} and 
alternate search variants of the algorithm.
{\it a)} Error versus
tensor rank $\chi$. Large values of $t_M$ and $d$ were used so
that the accuracy is limited only by $\chi$: $t_M=20$, $CH_{255}$ ($n=3$),
  $t_M=25$, $CH_{255}$ ($n=5$) and $t_M=30$, $GK_{63}$ ($n=10,n=15,n=19$)
The dashed lines  are guides to the eye for $1/\chi^p$ scaling which
  corresponds to $1/N^{p/2}$ in terms of the number $N$ of function
  calls. $p=1$ corresponds to Monte Carlo scaling. 
{\it b)} Error versus $N$ in a low precision calculation with $t_M=15$ and $d=15$ ($GK_{15}$). 
The colored circles indicate the CPU time of the calculation 
  when performed on a single core. 
{\it c)} Error versus $n$ for $\chi=10$ (blue),
  $\chi=20$ (orange) and $\chi=30$ (green) ($t_M=30$ and $GK_{15}$).
{\it d)} $|Q_n|$ versus $n$, distinguishing positive and negative coefficients.
The vertical lines indicate values of $n$ where $|Q_n|$ is very small,
  leading to larger relative errors.
}
\end{figure*}

We begin with calculations obtained using TTD on the SIAM model at
$\epsilon_d=\bar\alpha=0$ and $t_M\rightarrow\infty$.
One primary motivation for studying this particular regime is the existence of an analytic Bethe ansatz solution
\cite{Wiegmann_1983_a, macek2020qqmc}, from which we can extract the perturbative expansion with arbitrary accuracy.
Such a high precision benchmark is essential to study convergence and
the scaling of errors using different parameters and variants of the
algorithm.
In most figures in this section, we show the relative error 
$\epsilon^Q_n\equiv |Q_n - Q_n^{\rm Bethe}|/|Q_n^{\rm Bethe}|$ 
of $Q_n$ calculated by TTD, compared with the exact Bethe solution.

The error $\epsilon^Q_n$ is presented in Fig.~\ref{fig:Qn} as a function of the
rank $\chi$ of the tensor train.  The total number $N$ of  $\tilde Q_n$ evaluations
scales as $N \sim n d \chi^2$, so the different dashed lines correspond
to different scalings ranging from $1/\chi$ ($\sim 1/\sqrt{N}$, the
scaling of Monte-Carlo calculations) to $1/\chi^8$ ($\sim 1/N^4$). Remarkably,
one can reach very high precision--better than $10^{-8}$--even for large
pertubation orders $n\sim 20$, using a moderate value of $\chi$. We observe an effective
scaling of the error as $1/N^2$ for $Q_{10}, Q_{15}, Q_{19}$, and a faster
scaling for lower orders. The exact asymptotic scaling is unknown. 
In \cite{dolgov2020integralRn} a stretched exponential has been observed at very high precision.
This behaviour cannot be excluded by our data.
In any case, the convergence we
observe is dramatically faster than that of Monte Carlo methods, which is
$\sim 1/\sqrt{N}$, or even of
quasi-Monte Carlo techniques \cite{macek2020qqmc}, which are at best
$\sim 1/N$ in favorable cases but much less robust.

The error is presented as a function of $N$ in Fig.~\ref{fig:Qn}b for
lower precision calculations, $\epsilon^Q_n \ge 10^{-5}$, for $t_M = 15$.
Here we use a coarser discretization (smaller $d$) than that used in 
Fig.~\ref{fig:Qn}a, 
which is less costly but limits the accuracy.
We see that $10^6$ evaluations of $\tilde Q_n$ is sufficient to obtain
error below $10^{-4}$ in all cases. We
include the total computation time for three points, using a single core
on a recent modern workstation, e.g. $17$
seconds for $Q_{10}$ with four-digit accuracy.  Reaching the same
accuracy using our previous Monte Carlo implementation \cite{profumo2015} 
would require thousands of CPU hours. 

The error of $Q_n$ is plotted against $n$ in Fig.~\ref{fig:Qn}c, up to
$n=30$, and $|Q_n|$ is plotted in Fig.~\ref{fig:Qn}d.
We note that we
can obtain $Q_{30}$ with better than two-digit accuracy in approximately
$10^4$ CPU hours, a significant improvement over our previous works
($n=10-15$ using Monte Carlo \cite{profumo2015} and $n=20-22$ using
quasi-Monte Carlo \cite{macek2020qqmc}).

Crucially, we observe that the $\epsilon$-factorizability does not
deteriorate significantly with increasing $n$: in Fig. \ref{fig:Qn}a, 
we observe similar error for $n = 10$, $15$ and $19$; in Fig.
\ref{fig:Qn}c, the error stabilizes after $n=15$ (note that the
indicated values
at which the relative error is peaked simply correspond to values of
$n$ for which $Q_n$ is particularly small).

%%%%%%%%% Real time dyn ..........

\subsection{Real time dynamics}
\label{sec:SIAMRealTimeDyn}

\begin{figure}
\includegraphics[width=0.49\textwidth]{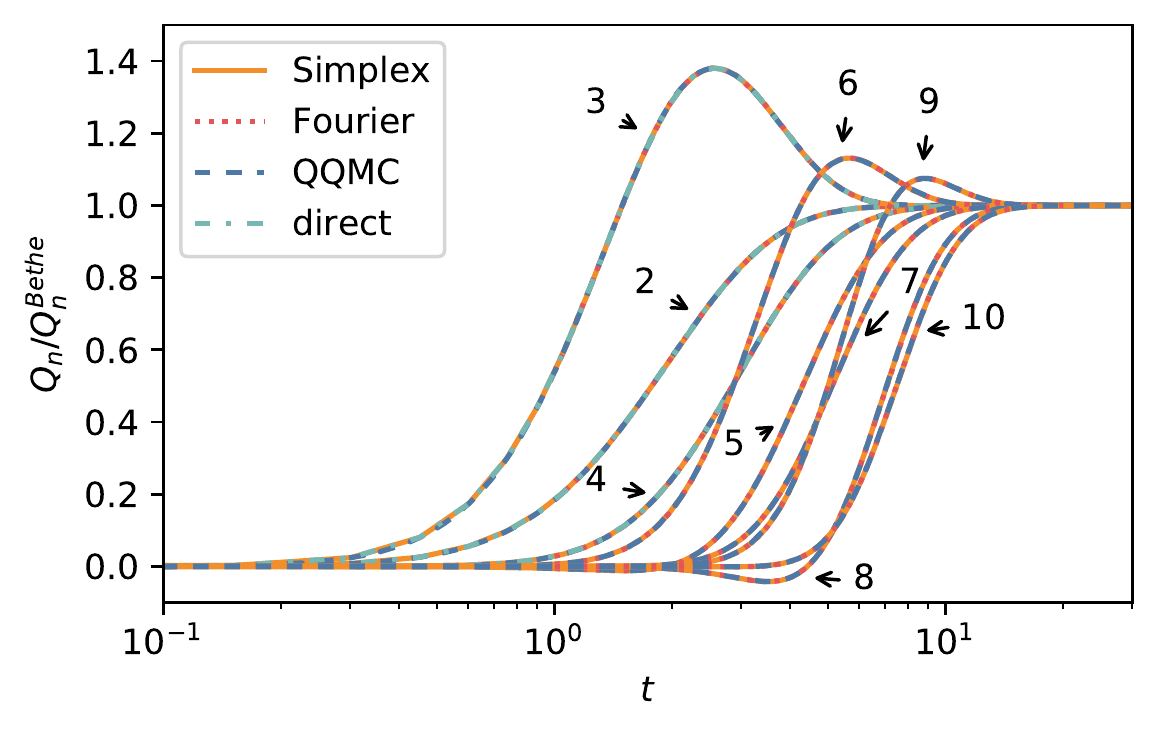}
\includegraphics[width=0.49\textwidth]{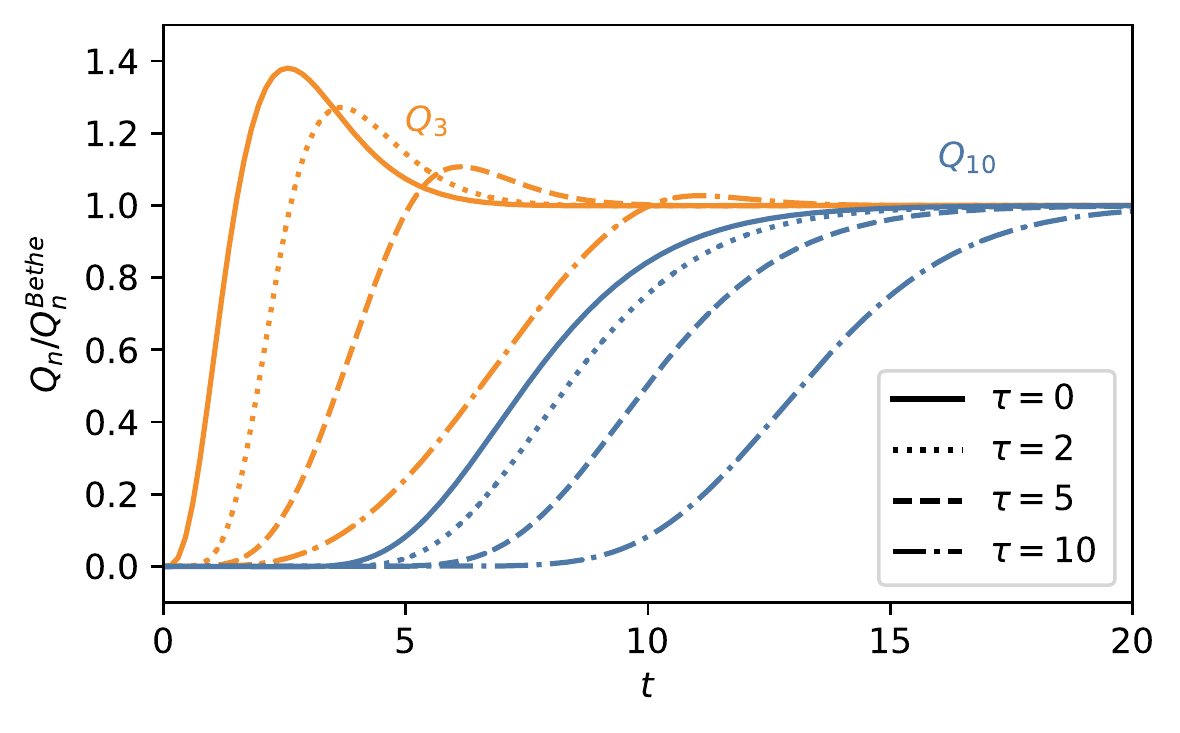}
\caption{\label{fig:qn_vs_time}
   $Q_n(t)$ after a quench at $t=0$ for the SIAM model at
   $\epsilon_d=\alpha=0$, in equilibrium. The coefficients are normalized by their exact
   asymptotic values $Q_n(t=\infty) = Q_n^{\textrm{Bethe}}$.  
   Upper panel:
   Abrupt quench $\lambda(t) = \theta(t)$. Different curves correspond
   to different integration techniques, with indistinguishable results: 
   simplex integration \eqref{eq:simplex} (yellow continuous line), 
   Fourier technique \eqref{eq:def_Qn_v_num} (red dotted line), 
   quantum quasi-Monte Carlo (QQMC) \cite{macek2020qqmc} (blue dashed
   line), and direct numerical integration of \eqref{eq:def_Qn} for $n = 2, 3, 4$ (green dashed-dotted line). 
   The arrows indicate the value of $n$.  
   Lower panel: 
   $Q_n(t)$ for $n = 3$ (orange) and $n = 10$ (blue) using a continuous
   increase of the interaction strength
   $\lambda(t) = \sin(\pi t / 2 \tau)$ for $t < \tau$, and  $\lambda(t) = 1$, for $t\ge \tau$
   (with simplex integration).
}
\end{figure}

\begin{figure}
\includegraphics[width=0.49\textwidth]{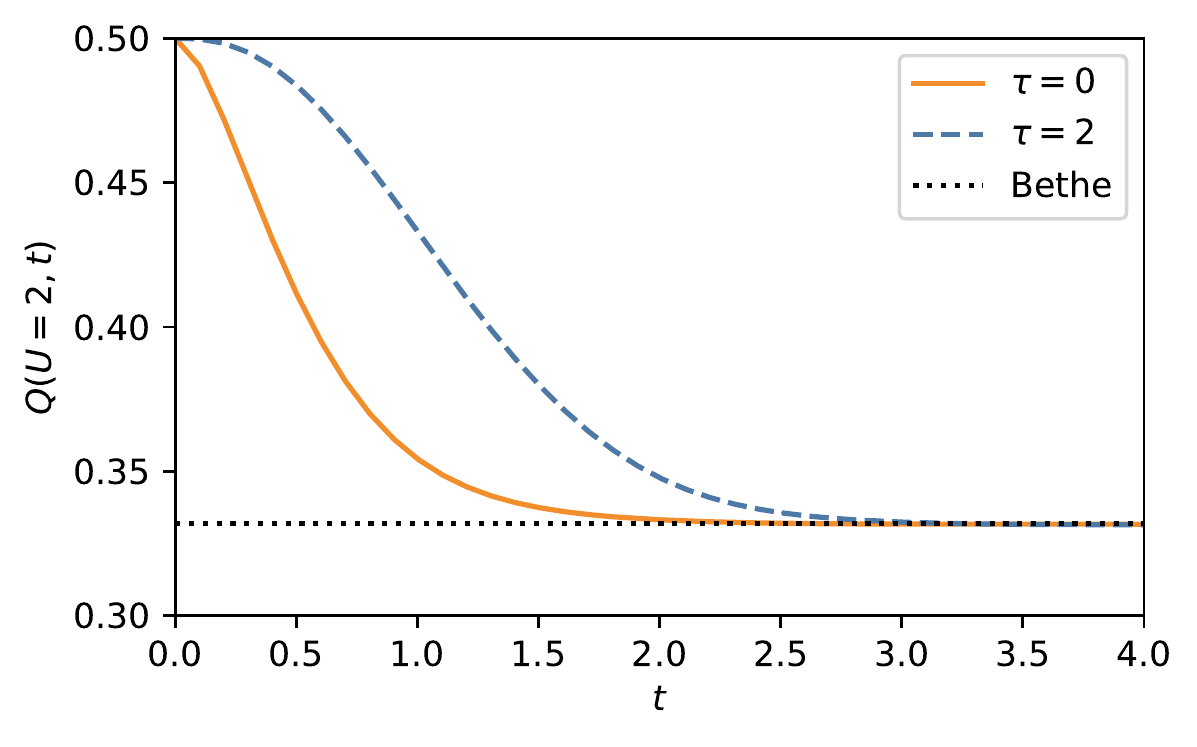}
\caption{\label{fig:qu_lambda}
Charge $Q(U, t)$ on the dot for the SIAM model after switching
on the interaction abruptly (solid line, $\tau = 0$), or smoothly (dashed
line, $\tau = 2$) with $\lambda(t\le\tau) = \sin(\pi t / 2 \tau)$ and
$\lambda(t\ge\tau) = 1$. Both curves include terms
  $Q_n(t)$ up to $n = 14$. The black dotted line corresponds to the
  resummed series using the Euler transform of the first 20
  $Q_n^{\textrm{Bethe}}$ coefficients, as in Ref.\ \cite{profumo2015}.
In all
  calculations we use $\epsilon_d=\alpha=0$ with $U = 2, t_{M} = 20$.
}
\end{figure}

As explained in Sec. \ref{sec:TTDintegration},
the full time dependency of the charge $Q(t)$ and its expansion coefficient $Q_n(t)$, after switching on the
interaction at $t=0$, can be obtained from a {\it single} factorization of $\tilde
Q_n$ at negligible additional cost. 
Figure \ref{fig:qn_vs_time} (upper panel) 
shows an example of $Q_n(t)$ curves for
different orders $n$. 
At large $t$, each $Q_n(t)$ converges towards the
Bethe ansatz equilibrium value.

The time integration (\ref{eq:psi1}, \ref{eq:psi2})
can be performed with any time-dependent coupling constant $U \lambda
(t)$ at a negligible increase in cost.
In Fig. \ref{fig:qn_vs_time}b, we show $Q_n(t)$ in two cases: 
{\it i)} an abrupt turning on of the interaction, $\lambda(t) = \theta(t)$ 
and
{\it ii)} a continuously differentiable $\lambda(t)$ given by  
\begin{equation}
\label{eq:lambda_ex}
\lambda(t) = \left\{
\begin{array}{ll}
\sin(\pi t / 2 \tau)   & 0 \le t\le \tau \\
1  & t> \tau.
\end{array}
\right.
\end{equation}

The ability to quickly calculate the effect of any time-dependent coupling
constant suggests interesting possibilities for studying the effect of various types of quenches.
It might also be used to optimize the convergence to the asymptotic value. 
Indeed, although the series \eqref{eq:qu_sum} has an
infinite radius of convergence for any {\it finite} time \cite{Bertrand_1903_series}, 
a very high-order expansion may still be required at intermediate and
long times.
An interesting open question is whether a smooth adiabatic turning on of the interaction 
could lead to an easier resummation of the perturbative series (i.e.
using fewer terms) at 
intermediate times than an abrupt quench, which puts the system far out of equilibrium.

In Fig.~\ref{fig:qu_lambda} we show an example of the actual physical observable
$Q(U,t)$ for two values of the switching time $\tau$, and $U=2$. 
These results are
obtained by truncating the series \eqref{eq:qu_sum} to
a finite number of terms, varying the expansion order to check
convergence. Both curves converge to the asymptotic Bethe ansatz value
with high accuracy. Note that $U=2$ is beyond the radius of convergence
of the series for $t\rightarrow \infty$, so the two curves can only be
obtained up to a finite time without resummation \cite{Bertrand_1903_series}.

\subsection{Factorizing the ``sign problem''}

\begin{figure}
\includegraphics[width=0.49\textwidth]{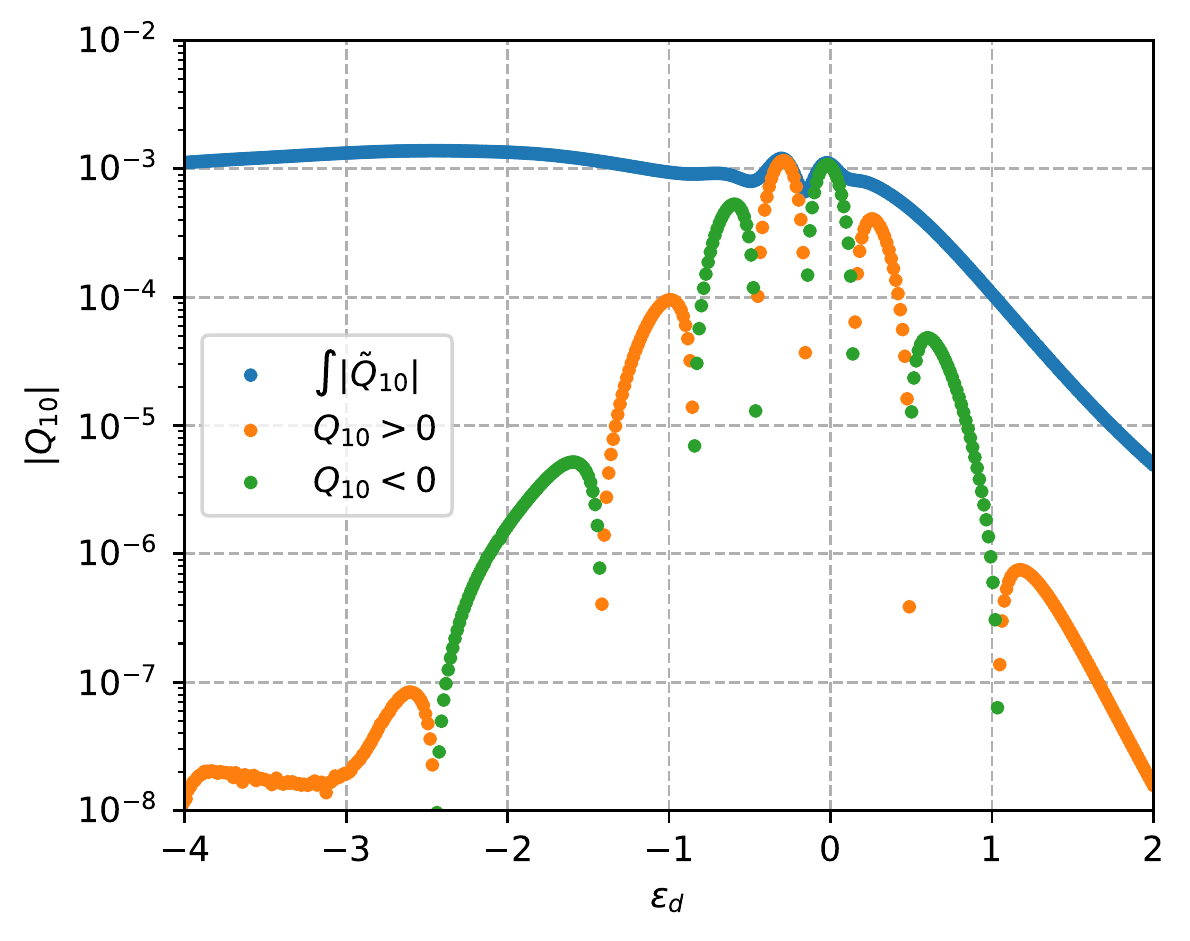}
\caption{\label{fig:SignPb} 
 Integrals $Q_n = \int\tilde Q_n$ and $\int|\tilde Q_n|$ for $n=10$
 versus $\epsilon_d$ for the SIAM model.
 The green and orange colors correspond $Q_{10} < 0$ and $Q_{10} > 0$,
  respectively. 
}
\end{figure}

\begin{figure}
\includegraphics[width=0.49\textwidth]{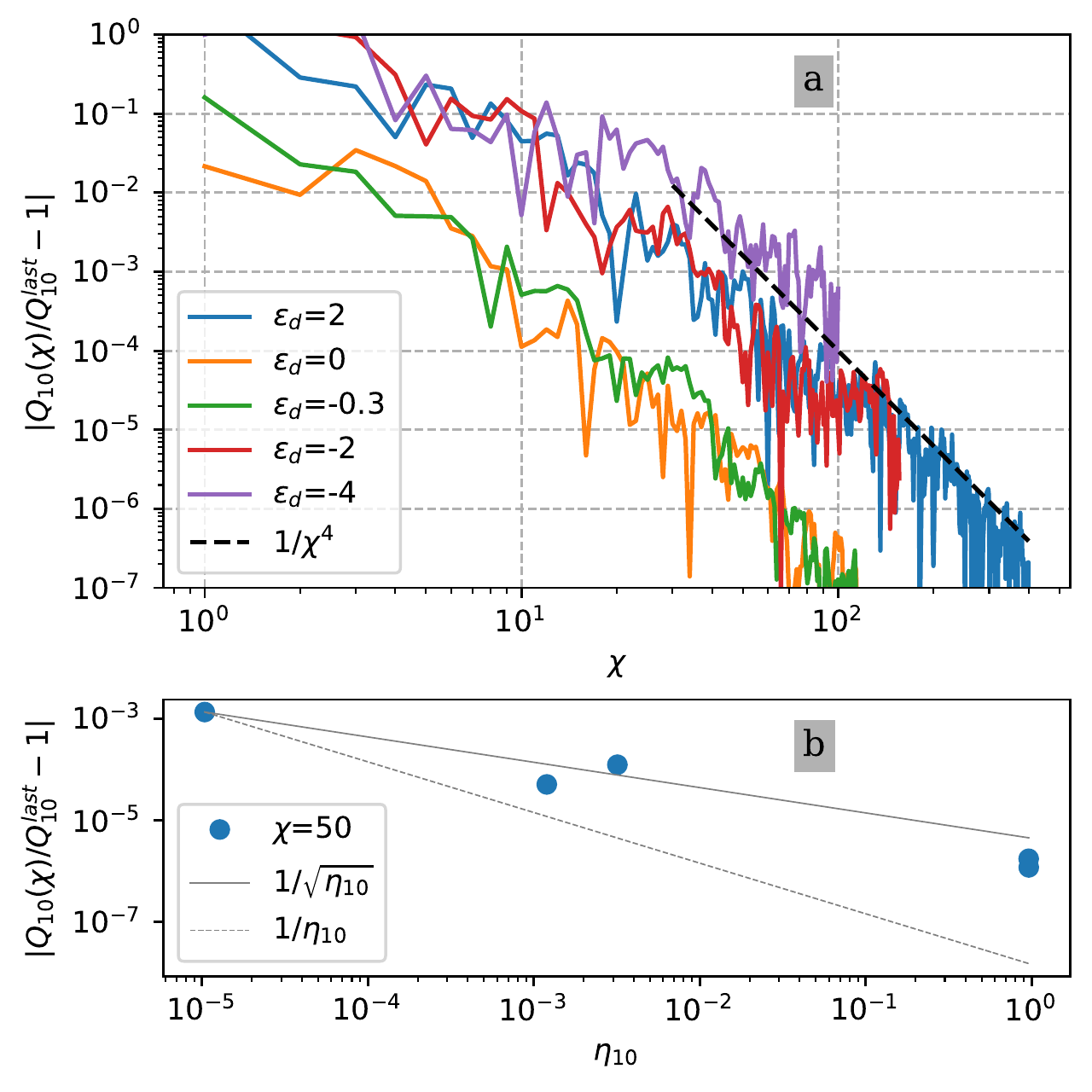}
\caption{\label{fig:SignPb2} 
  Upper panel: error of $Q_{10}$ versus $\chi$ in the SIAM model, for
  different $\epsilon_d$. Gauss-Kronrod integration is used with 
$d=31$ points for $\epsilon_d= 2,0, -0.3$, and $d=255$ points for
  $\epsilon_d= -2, -4$.
  Lower panel: the same error, but for fixed $\chi = 50$, varying
  $\eta_{10}$. 
}
\end{figure}

In the previous benchmark, the integral is fairly non-oscillatory.
When $\epsilon_d$ is non-zero, $\tilde
Q_n$ oscillates much more, a challenge for high-dimensional integration
techniques like Monte Carlo and quasi-Monte Carlo.
We illustrate the issue using a simple toy function
\begin{equation}
A(v_1, \dots, v_n) = \prod_{i=1}^n \bigl( 1 + a \cos 2\pi v_i \bigr)
\end{equation}
defined on the hypercube $[0,1]^n$, which is completely factorizable.
A direct Monte Carlo estimator of the integral
of $A$ using $N_\text{MC}$ random points is given by $\bar A =
(1/N_\text{MC}) \sum_{\alpha=1}^{N_\text{MC}} A(v^\alpha)$.
The variance of $\bar A$ is exponentially large in $n$:
\begin{equation}
   \text{var}\ \bar A = \frac{1}{N_\text{MC}} \left [ \left( 1 +
   \frac{a^2}{2}\right)^n - 1 \right]. 
\end{equation}
By contrast, using the TTD of $A$ simply requires computing $n$ one-dimensional integrals.
The central question is then the robustness of $\epsilon$-factorizability in
strongly-oscillating cases.

A consequence of the presence of oscillations is the well-known ``sign
problem'':  the {\it average sign} at order $n$,
\begin{equation}
   \label{eq:signdef}
\eta_n = \frac{\int du_1\ldots du_n \ \tilde Q_n(u_1, \ldots, u_n)}{\int
  du_1\ldots du_n \ |\tilde Q_n(u_1 , \ldots,  u_n)|},
\end{equation}
may be small as a consequence of cancellations in the integral. If one is
interested in maintaining {\it relative} accuracy, then a small $\eta_n$
poses an additional challenge, as the absolute
precision of the factorization of $\tilde Q_n$ must be increased. In our
calculations, however, small $\eta_n$ usually means that the corresponding contribution to
the observable is small, and we are interested in {\it absolute}
accuracy (or more precisely, in relative accuracy with respect to the largest
contribution usually found at low order).

Let us now review our empirical observations of the behavior of TTD
in the presence of strong oscillations in the integral and a small
average sign.  
The numerator and denominator in the definition \eqref{eq:signdef} of $\eta_{10}$ are
presented in Fig. \ref{fig:SignPb}, as a function of $\epsilon_d$, for
the SIAM model. The sign
$\eta_{10}$ is of magnitude approximately $1$ for $\epsilon_d\approx 0$.
Away from $\epsilon_d = 0$, $\eta_{10}$ decreases, reaching $\eta_{10} \approx
10^{-5}$ for $\epsilon_d=-4$. A Monte Carlo simulation in this regime would
be prohibitively expensive. On the other hand, as illustrated in Fig. \ref{fig:Ovw}d, the TCI decomposition was
performed with an approximately constant computational time for every value of $\epsilon_d$, 
indicating its insensitivity to small average sign.

Since the $\epsilon$-factorizability is
independent of the oscillatory character of the integrand,
the difficulty is reduced to that of integrating 
oscillatory one-dimensional functions, a much less formidable problem. In practice, this may still be a
difficult task, and we discuss our approach for specific cases in Section \ref{sec:lattice}
and Appendix \ref{app:tailored_quad}.
%(e.g. $d=255$ for $\epsilon_d=-4$, while $d=31$ is sufficient for $\epsilon_d=0$). % in fig 13 caption already

However, a small average sign leads to a very general and simple  
issue: the relative error involves division by a small number.
In order to keep it at a given level when varying $\epsilon_d$, 
a smaller absolute error is required by at most $1/\eta_n$.
This effect is illustrated in Fig.~\ref{fig:SignPb2}a, where the
relative error of $Q_{10}$ is presented as a function of $\chi$ for various values of $\epsilon_d$.
The convergence rate is the same ($1/N^2$) for every $\epsilon_d$, reflecting again that a small average sign 
does not affect the quality of the $\epsilon-$factorization.
However, we observe that a small $\eta_{10}$ implies a larger relative error.
This effect is difficult to predict quantitatively, as it depends on cancellations in integrating the error in 
\eqref{eq:TCIQn}, {\it i.e.} $\tilde Q_n - \tilde Q_n^\text{TCI}$.
In Fig.~\ref{fig:SignPb2}b, the relative error is plotted as a function of $\eta_{10}$, for fixed $\chi = 50$.
We observe that it increases approximately like $1/\sqrt{\eta_{10}}$, slower than $1/\eta_{10}$.
Since $Q_{10} = O(\eta_{10})$, it follows that the absolute error actually {\it decreases} when $\eta_{10}$ gets smaller (not shown).
Due to the $1/N^2$ convergence rate, keeping the same relative error when varying $\epsilon_d$
therefore corresponds to a moderate increase in computing time $\sim 1/\eta^{1/4}$ in this model.
Keeping a constant absolute error is actually easier in the presence of a
sign problem.

Since our implementation uses double precision arithmetic, we cannot go
beyond an absolute precision of $10^{-9}$ in the integrand due to rounding
errors in the Keldysh sum of determinants (see Appendix \ref{app:keldyshfastsumdets}).
This translates to a relative error in $Q_{10}$ of $10^{-4}$ for the worst case $\epsilon_d=-4$. 
Beyond this point, the error saturates. 

In sharp contrast with Monte Carlo, computing $\int |\tilde Q_n|$ with TTD is in fact significantly {\it
harder} than computing $\int \tilde Q_n$. Indeed, the function $|\tilde Q_n|$  is not
$\epsilon$-factorizable with a low rank, most likely as a result of the cusps introduced
by taking the absolute value. 

This section illustrates a central point of this paper: the
property of the integrand $\tilde Q_n(v_1,\ldots, v_n)$  that makes the problem
amenable to integration with TTD ($\epsilon$-factorizability) is orthogonal to the property that would make
it amenable to a solution with Monte Carlo sampling (positivity). 
In particular, TTD works seamlessly in some situations in which
Monte Carlo fails. 
This is a strong incentive to revisit problems that suffer from a strong 
sign problem in Monte Carlo algorithms using the TCI algorithm.

% -------------------------------------------------------------------

\subsection{Extrapolation vs. interpolation}

\begin{figure}
\includegraphics[width=0.49\textwidth]{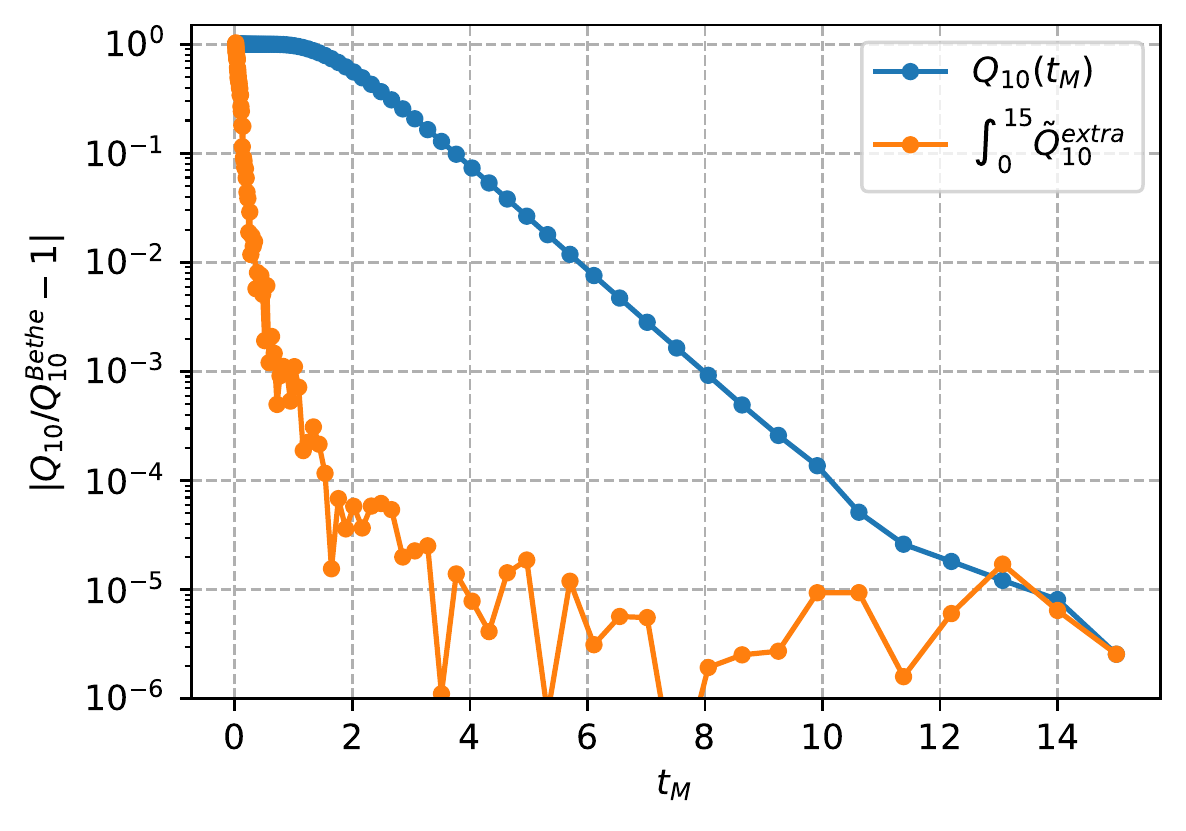}
\caption{\label{fig:Extrapol} 
   Relative error with respect to the Bethe ansatz for 
   $Q_{10}(t_M)= \int^{t_M}_0\tilde Q_{10}$
   and $\int^{15}_0\tilde Q_{10}^{\rm extra}$, versus $t_M$.
   $\tilde Q_{10}^{\rm extra}$ is the extrapolation of the TCI of $\tilde Q_{10}$ obtained for $v_i \le t_M$.
   We used Gauss-Kronrod rule with $63$ points for the integration.
    The error saturates at $10^{-5}$ due to the discretization error.
   }  %: gt,wt="gk","V" and tmax,nChe,dim=15,63,10.}
\end{figure}

In this section, we discuss a remarkable feature of our integrand discovered by the TCI decomposition. 
As mentioned earlier, the crucial and non-trivial property of $\tilde Q_n$ is the 
$\epsilon-$factorizability of the {\it core} of the function (e.g. with all variables confined to
a small pocket $v_i\in [0,1]$), while the factorizability at large-$v$ is easier to understand.
It turns out that this core factorization is also an excellent {\it extrapolation} at large-$v$
(e.g. $v_i\in [0,8]$). 
The factorization of $\tilde Q_n$ at short times (difference) is not only possible, 
it is in fact sufficient to approximate the whole function.

We illustrate this observation with Fig.~\ref{fig:Extrapol}, where 
two calculations of $Q_{10}$ are presented.
The first one (blue curve) is simply $Q_{10}(t_M)= \int^{t_M}_0\tilde Q_{10}$, the direct hypercube integral.
At large $t_M$, as expected, the error with respect to the stationary value decreases quickly.
The second calculation consists in computing the integral for the hypercube $v_i \le 15$, 
but with a TCI approximation computed for smaller hypercube $v_i \le t_M\le15$, {\it i.e.}
with the pivots confined to $[0, t_M]^n$. 
The second computation converges to the equilibrium value much faster than the first one. 
In summary, in order to obtain a precision of three digits, 
the integration must be done on a large volume $[0,8]^{n}$, but it is
sufficient to perform the learning part inside a volume $[0,1]^{n}$, which is 
exponentially smaller with $n$.  

%--------------------------------------------

\subsection{Effect of various parameters and variants on the convergence}
\label{sec:effectof}

\begin{figure*}
\includegraphics[width=1.0\textwidth]{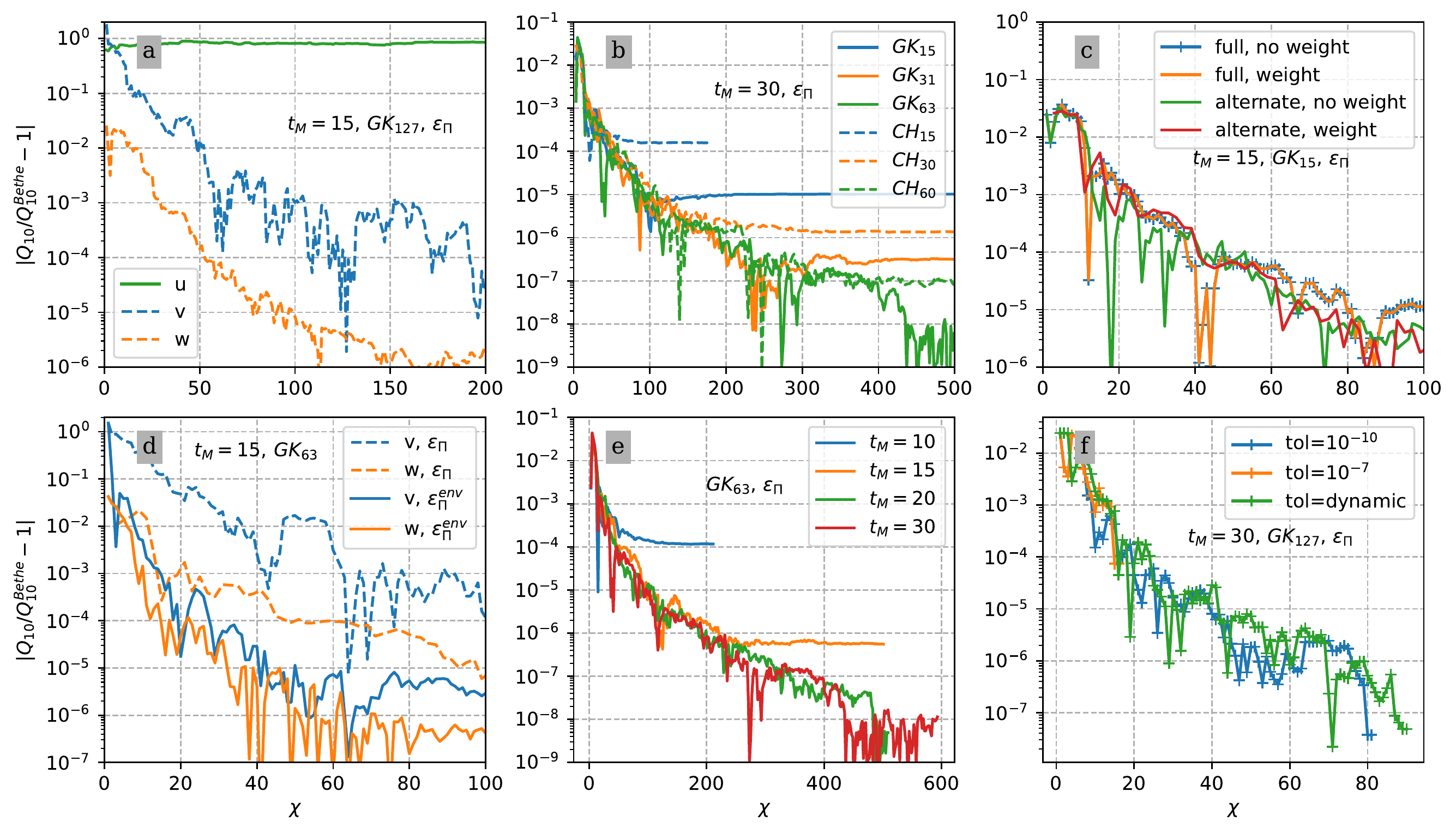}
\caption{\label{fig:Q10}
Convergence with respect to the tensor rank $\chi$ of the 10-dimensional integral
$Q_{10}$, for the SIAM model. All panels show the relative error $\epsilon^Q_{10} = |Q_{10}(\chi)/Q_{10}^{\rm Bethe}-1|$ 
measured against the Bethe ansatz solution versus the bond dimension $\chi$. 
(a) Comparison of the choices of variables $u$, $v$, and $w$. %, see (\ref{eq:v_vars}, \ref{eq:definitionWvariable}).
(b) Comparison of different one-dimensional quadrature rules of $d$
  nodes: Gauss Kronrod (GK$_d$) and Chebyshev (CH$_d$). 
(c) Comparison of different pivot selection algorithms, 
{\it full search} and {\it alternate search}, defined in Sec. \ref{sec:TCIalgo}, 
and the {\it weighted learning} variant defined in Sec. \ref{sec:integrationTCI}.
(d) Effect of using $\epsilon_\Pi$ \eqref{eq:errorFunctionNoEnv} or $\epsilon_\Pi^\text{env}$ \eqref{eq:errfunENV} as the error function in the pivot selection.
(e) Saturation of the error due to the maximum time cutoff $t_M$. 
(f) Absence of a dependence on the pivot acceptance condition, see text.
}
\end{figure*}

We next examine the convergence of TTD for
 different parameter choices and variants of the algorithm.
Results for the error $\epsilon^Q_{10}$ are summarized in Fig. \ref{fig:Q10}.

We first compare the factorizability in the $u$, $v$, and $w$
variables given in (\ref{eq:v_vars}, \ref{eq:definitionWvariable}).
Fig. \ref{fig:Q10}a shows the error in the factorization of $\tilde
Q_{10}$ for the three choices of variables, using the pivot error function $\epsilon_\Pi$ of \eqref{eq:errorFunctionNoEnv}.
We observe that $\tilde Q_{10}(u_1, \ldots,u_n)$ {\it is not} $\epsilon$-factorizable (green curve).
As discussed above, this is likely a consequence of the cusps on the
boundaries between the $n!$ different smooth components of the function,
corresponding to different orderings of the $u_i$.
By contrast, the error decreases quickly when the $v$ variables are
used, and using the
$w$ variables gives a further reduction by two orders of magnitude. The
observed saturation around $10^{-6}$ is a consequence of making the
cutoff $t_M=15$, as shown below. 
The factorizability is the same in $v$ and $w$, and it
would provide the same approximation if the same pivots had been
selected. 
The use of $w$ clearly produces better pivots.

Fig. \ref{fig:Q10}d establishes that using the error function $\epsilon_\Pi^\text{env}$ defined in \eqref{eq:errfunENV} removes the need for introducing a problem-dependent change of variable $w$. 
We use this function for all computations in this paper, unless
otherwise specified. 

Fig.\ref{fig:Q10}b illustrates that the discretization of
the one-dimensional integrals limits the overall accuracy.
The different curves are essentially on top of each other until the number of
points become a limitation in the precision.
The closeness of the curves before this limit is reached suggests a
robustness relative to the precise position of the pivots, which are
different for the different curves, since they are chosen from different
grids.
We also see that, for this model, Gauss-Kronrod integration has slightly
better convergence properties than Chebyshev integration,  and has the additional advantage of
providing a built-in estimate of the integration error. 
We have also
tried Gauss-Legendre quadrature rules (not shown), with similar convergence
to Gauss-Kronrod. 

In Fig. \ref{fig:Q10}e, we show the error of the steady state
value with respect to the maximum time $t_M$ since the interaction is switched
on in the Keldysh formalism. A large choice of $t_M$ is required
for high accuracy. We note that increasing $t_M$ may also require
increasing $d$ to maintain the accuracy of one-dimensional integrals.

In Fig. \ref{fig:Q10}c, we test the effect of using the {\it alternate search}
and {\it full search} pivot selection methods, defined in Sec. \ref{sec:TCIalgo}.  
We find no significant difference in the result, so for all calculations
in this paper we use the
{\it alternate search} approach, which has computational complexity
$\propto n \chi^2 d$ rather than $\propto n \chi^2 d^2$.
 
Fig. \ref{fig:Q10}f illustrates the robustness of the TCI algorithm with respect to
different criteria to accept pivots. We introduce a pivot acceptance condition:
we accept a new pivot only if the
 error $\epsilon_\Pi$ is above a given threshold, i.e. not adding pivot that
 only improve the error marginally. Fig.\ref{fig:Q10}f shows the convergence for 
 two different levels for this threshold as well as a dynamical algorithm where the threshold
 is fixed to 1\% of the current typical pivot error $\epsilon_\Pi$. 
 We observe no significant effect.
 For very large bond dimensions ($\chi > 10^3$) where the
 contraction of the tensor train might require a significant computing time,
 using such a condition might become useful. However for the rather small
 values of $\chi$ used in this article, the gain is marginal. 

\section{Preliminary studies beyond the single impurity model}
\label{sec:beyond}

The next step, after the benchmarks on a single impurity model, is to
generalize the TTD method to more complex systems like
multi-site and lattice models, and to other
perturbative expansions.
In this section, we take the first steps in this
direction with two preliminary studies which indicate that the
$\epsilon$-factorizability property is robust beyond a single site model
in a flat bath.

\subsection{Single impurity embedded in a two-dimensional lattice}
\label{sec:lattice}

\begin{figure}
\includegraphics[width=0.49\textwidth]{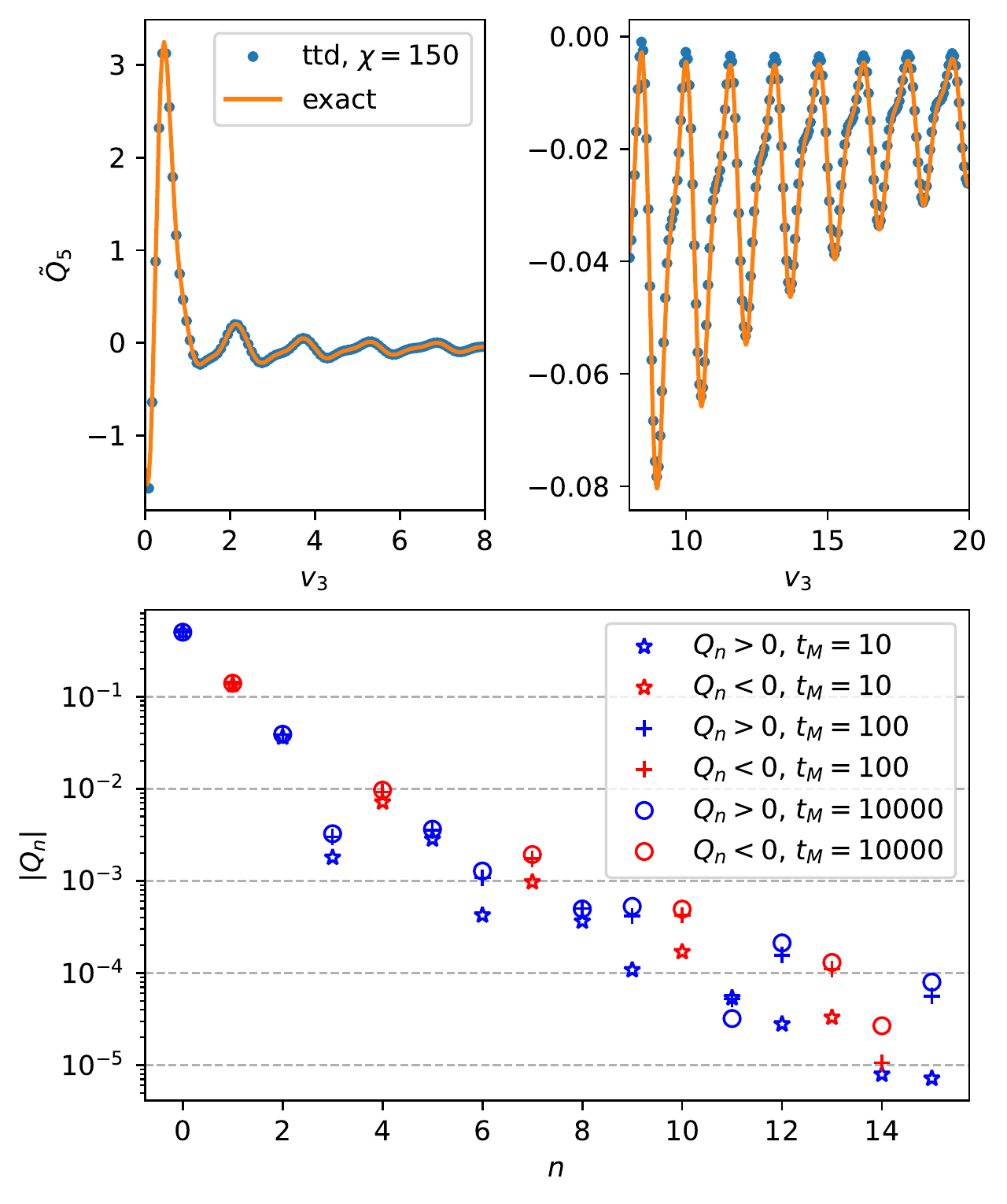}
\caption{\label{fig:HubSiam} One interacting site in an infinite two-dimensional lattice.
Upper panels: integrand $\tilde Q_5(v_1,v_2,v_3,v_4,v_5)$ versus $v_3$ for a random choice
of the values $v_1,v_2,v_4,v_5$. The right panel zooms in on the tail.
  Actual integrand (orange curve) and
TTD approximation at $\chi=150$ (blue circles) are both shown.
  Lower panel: $|Q_n|$ versus $n$ for three values of the maximum
  time, $t_M=10$ (stars), $t_M=100$ (pluses), and $t_M=10000$ (circles). Blue (red) symbols correspond to positive (negative) values of $Q_n$.
   }  
\end{figure}

We first consider a single site model with a more complex bath than the SIAM:
an infinite 2D lattice in which a single site is interacting, defined in Sec. \ref{sec:models:siamlattice}.
%The most important difference with the SIAM is
%that the coupling of the impurity to the rest of the lattice is {\it large}. 
As a result of the band edges, the non-interacting Green's functions have strong
oscillations at a frequency set by the bandwidth (see lower panels of
Fig. \ref{fig:g0}), so $g^<(t) \sim \cos(4t)/t$ is both highly oscillatory and
slowly decaying.

The upper panels of Fig. \ref{fig:HubSiam} show a one-dimensional slice of the
integrand $\tilde Q_5$, demonstrating strong
oscillations with several harmonics of $\omega_0=4$ present.
Such a calculation would be very challenging for Monte Carlo approaches \cite{profumo2015}.
Nevertheless, the TCI approximation, also shown in Fig. \ref{fig:HubSiam}, works well, 
indicating $\epsilon$-factorizability despite
the strong oscillations (note however that $\chi = 150$).

In the lower panels of Fig. \ref{fig:HubSiam}, we show the first $15$
coefficients of the interaction expansion of the charge. The rapid
oscillation and slow decay of the integrand is so severe that even the calculation
of the one-dimensional integrals in TTD is non-trivial. While for the SIAM a small
cutoff $t_M=15$ is sufficient to obtain several digits of accuracy, we have found
in this case that $t_M = 10000$ is required to go beyond two digits, or even a single
digit at large $n$. To perform these one-dimensional integrals efficiently, we
use a specifically tailored quadrature rule, described in Appendix
\ref{app:tailored_quad}, which makes use of an asymptotic expansion of
the integrand. The interval of integration is broken into a short time
region $[0,10]$, on which we use a $63$-point
Clenshaw–Curtis rule, and a large time region $[10,t_M]$, on which the
custom quadrature rule is used.
We note that the slow decay with $t_M$ is specific to the $T=0$ case.
We have checked that at a higher temperature $T=0.1$, convergence is reached for a much smaller
$t_M\approx 10$ (not shown). However, the factorizability appears to be
independent of temperature.

%----------------------------------------------------------------
\subsection{Double quantum dot}

\begin{figure}
\includegraphics[width=0.48\textwidth]{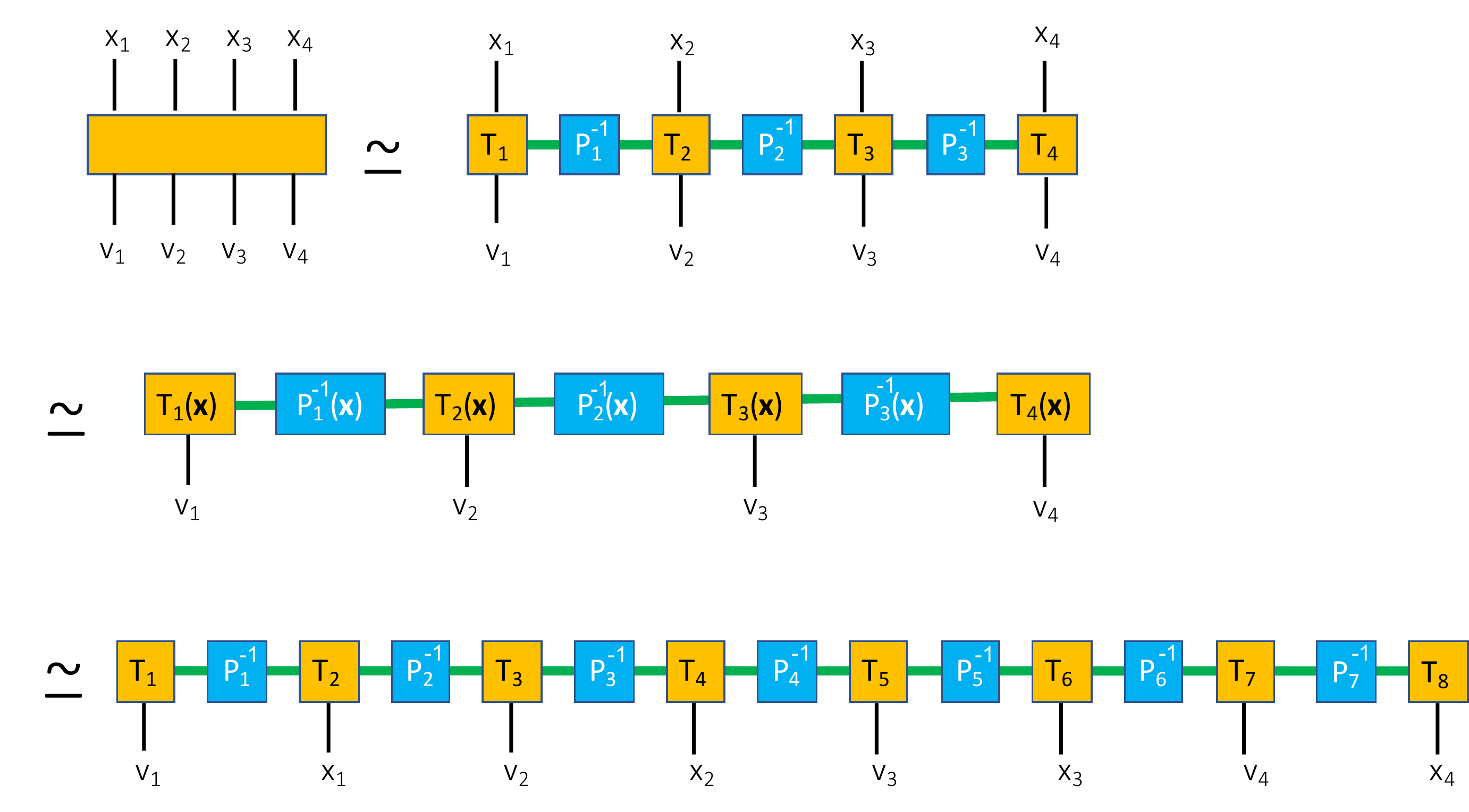}
\caption{\label{fig:2Qd_schematic} 
Three different factorizations used for the double quantum dot problem.
From top to bottom: vertex factorization, time factorization, and full factorization.   
} 
\end{figure}

\begin{figure}
\includegraphics[width=0.49\textwidth]{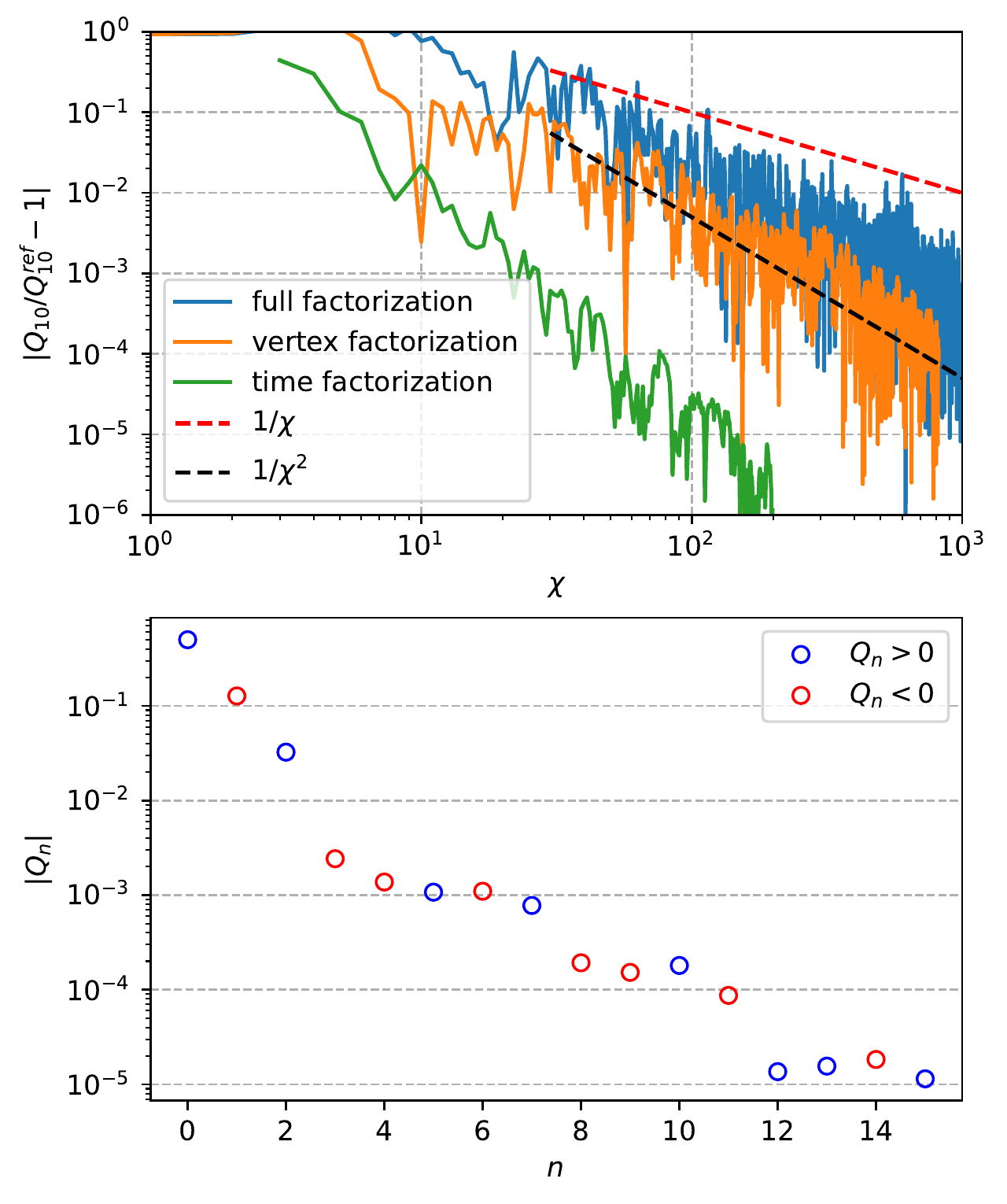}
\caption{\label{fig:2Qd} 
   Double quantum dot. Upper panel: error of $Q_{10}$ versus $\chi$ for the
   three algorithms introduced in the text. The reference value
   $Q_{10}^\text{ref}$ was
   obtained using the time factorization algorithm. Lower panel: $|Q_n|$ versus
   $n$, obtained using the vertex factorization algorithm. Blue (red) symbols
   correspond to positive (negative) values of $Q_n$. We use $\gamma_0 =
   0.5\Gamma$ and $t_M=15$. 
}  
\end{figure}

We next consider a double quantum dot, i.e. with {\it two interacting sites}.
This system plays a central role in various approaches to semi-conducting qubits.
Apart from its intrinsic importance in mesoscopic physics, it is the
simplest case in which the perturbative expansion involves sums over both spatial
indices $x_i \in\{0,1\}$ and time differences $v_i$, since $\tilde Q_n(x_1, \ldots,x_n,v_1, \ldots,v_n)$ now depends on both.
It is therefore a good
starting point from which to extent the TTD to a function of space and time.

We first emphasize that there are multiple ways to include spatial
indices in the tensor network form.
Our goal is to find the one with the lowest rank $\chi$, and the best 
convergence rate with $N$.
We study three possibilities, depicted in Fig. \ref{fig:2Qd_schematic}.
Corresponding results are presented in Fig. \ref{fig:2Qd}.

First, we can take $T$ tensors that depend on both a spatial and a time variable,
$T_\alpha (x_\alpha,v_\alpha)$ (first line of Fig. \ref{fig:2Qd_schematic}). 
We refer to this approach as {\it vertex factorization}, since the
factorization is done vertex by
vertex. The computational cost is only increased by a factor of two
compared to the SIAM, and would scale linearly with the number $L$ of dots ($L=2$ here), as 
$\propto Ld$. 
In  Fig.~\ref{fig:2Qd} (orange curve), we observe that this method
converges quite quickly, as $1/N$, 
but more slowly than the SIAM, which was $1/N^2$.

Second, we can fix the spatial indices $x_1, \ldots, x_n$  
and use TTD for the times (second line of Fig. \ref{fig:2Qd_schematic}).
We refer to this approach as {\it time factorization}.
After integrating over times, we obtain an intermediate function $\hat Q_n(x_1, \ldots,x_n)$ given by 
\begin{equation}
\hat Q_n(x_1, \ldots,x_n) \equiv \int \prod_i dv_i \ \tilde Q_n(x_1, \ldots,x_n,v_1, \ldots,v_n).
\end{equation}
The summation over the $x_i$ can be carried out in two ways.
We can explicitly simply sum over the $L^n$ combinations, 
with an exponential computational scaling $\propto L^n d$, which is manageable for
$L=2$. 
In Fig.~\ref{fig:2Qd} (green curve), we observe that this method
converges as $1/N^2$, like the SIAM. We also observe this convergence
rate for each fixed set of spatial indices $x_i$.
Alternatively, we can use TCI again on the spatial variables to factorize
$\hat Q_n$. We found (not shown) that this approach also converges.
However, for the small value $L=2$, the $\hat Q$ tensor is not large
enough to draw a definite conclusion on the performance 
of this technique for large $L$.

Third, we can use an MPS form, with alternating $v$ and $x$ variables
(third line of Fig. \ref{fig:2Qd_schematic}).
The computational cost of this approach is essentially the same as that
for the SIAM, with $d$ replaced by $L+d$.
In Fig.~\ref{fig:2Qd} (blue curve), we observe that this method converges more 
slowly with $N$ --- only slightly faster than $1/\sqrt{N}$ ---
indicating that the ``entanglement'' between space and time variables has
a non-trivial structure, which is not captured efficiently by this simple tensor train. 

Thus, various tensor forms can be used to apply TTD to the double or multiple dots.
The three methods presented here are all convergent, but
with different rates, and for this example vertex factorization is the most efficient.
However, many further possibilities could be explored, e.g. using spatial position differences, 
or different orderings of the variables in the MPS, or a PEPS generalization of the tensor form
in space-time.
The search for an optimal tensor form for the lattice case is an
interesting open question, which we leave for future work.

\section{Conclusion}\label{sec:conclusion}

Tensor network methods offer a new approach to high-dimensional
integration, and in particular to computing high-order diagrammatic
perturbative expansions. The $n$-body (bare) correlation functions have a
mathematical structure that allows a parsimonious representation in term of a
tensor network, which can be efficiently obtained using the Tensor Cross
Interpolation (TCI) algorithm.  While a naive direct integration in $n$
dimensions would scale exponentially with $n$, the TCI algorithm can reveal the
underlying structure and perform the sum in a number of calls of the integrand
that scales linearly with $n$.
We have illustrated this approach for quantum impurity models (single and double
dots) within the real-time Schwinger-Keldysh formalism, with high-precision
benchmarks. It significantly outperforms previous Monte Carlo and
quasi-Monte Carlo methods.  In particular, it is insensitive to the
infamous ``sign
problem'' appearing in parameter regimes in which the integrals are highly
oscillatory.
Furthermore, it allows calculations of the full time dependency, and of the effect of a time dependent
coupling constant, at negligible additional cost.

The main open question at this stage is the generality of the
$\epsilon$-factorizability property and its potential application to other
diagrammatic techniques, e.g. for multi-orbital or lattice models, imaginary
time perturbative expansions, and inchworm algorithm in real or imaginary 
time ~\cite{Cohen2013, Cohen2014a, Cohen2014b, Cohen2015, Eidelstein_2020, Li_2022}.  
For example, it is necessary to investigate whether a simple MPS is
sufficient to handle the lattice case (with spatial and time indices), or
whether a more sophisticated tensor network like PEPS is needed.

We point out, more generally, that the limiting factor of the TCI approach ({\it i.e.} the rank of
the $\epsilon$-factorization) is entirely orthogonal to that
of sampling methods like Monte Carlo (the sign problem). This suggests
reexamining various cases (e.g. partition function
calculations) which are known to be limited by the sign problem when Monte Carlo
methods are used.

% %%%%%%%%%%%%%%%%%%%% APPENDIX %%%%%%%%%%%%%%%%%%%%%%%%%%

\begin{acknowledgments}
   O. P., X. W. and P. D. thank Miles Stoudenmire for numerous enlightening discussions on tensor
   network techniques. We thank Fedor \v{S}imkovic and Michel Ferrero for sharing
   the results of \cite{Simkovic_2022_Pminors} prior to publication. 
   The Flatiron Institute is a division of the Simons Foundation. 
   X. W. thanks the Plan France 2030 ANR-22-PETQ-0007 and the French-Japanese ANR QCONTROL for funding.
\end{acknowledgments}

\appendix

\section{Schur complement}
\label{app:Schur}
Two important components of this article (the cross interpolation formula and the principal minor algorithm)
are based on the concept of Schur complement \cite{GoluVanl96} that we recall here briefly for completeness.
We consider an arbitrary matrix $A$ that we put in a $2\times 2$ block form,
\begin{equation}
A =
\begin{pmatrix}
A_{11} & A_{12} \\
A_{21} & A_{22}
\end{pmatrix}
\end{equation} 
It is straightforward to show that provided the $A_{11}$ block is invertible one has,
\begin{eqnarray}
\begin{pmatrix}
1 & 0 \\
- A_{21} A_{11}^{-1} & 1
\end{pmatrix}
\begin{pmatrix}
A_{11} & A_{12} \\
A_{21} & A_{22}
\end{pmatrix}
\begin{pmatrix}
1 & -A_{11}^{-1} A_{12} \\
0 & 1
\end{pmatrix} \nonumber \\
=
\begin{pmatrix}
A_{11} & 0 \\
0 & A_{22} - A_{21}A_{11}^{-1} A_{12}
\end{pmatrix}.
\end{eqnarray}
From which we obtain that,
\begin{equation}
\label{eq:Schur}
{\rm det} A =  {\rm det}  [A_{11}]  \ {\rm det} [A_{22} - A_{21}A_{11}^{-1} A_{12}]
\end{equation}
The matrix $A_{22} - A_{21}A_{11}^{-1} A_{12}$ is called the Schur complement of $A$
with respect to the $11$ block. We refer to \eqref{eq:Schur} as the Schur complement theorem. The $11$ block is referred to as the "pivot".

\section{Properties of the Cross Interpolation}
\label{app:ci}

\subsection{Proof of property (P2)}
We begin with the proof of the property (P2) introduced in the main text, i.e. that if
a matrix $A$ is of rank $r$ then a cross interpolation with $\chi=r$ is exact.
Let's consider an arbitrary point $(x_0,y_0)$ and form the $(r+1)\times (r+1)$ block matrix by adding
one row and one column to the pivot matrix $A(\mI,\mJ)$,
\begin{equation}
\begin{pmatrix}
A(\mI,\mJ) & A(\mI,y_0) \\
A(x_0,\mJ) & A(x_0,y_0)
\end{pmatrix}.
\label{eq:proof_p2}
\end{equation} 
This sub matrix of $A$ has a vanishing determinant. Since the determinant of the pivot matrix is non zero, applying \eqref{eq:Schur} to \eqref{eq:proof_p2} gives,
\begin{equation}
A(x_0,y_0) - A(x_0,\mJ) A(\mI,\mJ)^{-1} A(\mI,y_0)=0
\end{equation} 
which proves property (P2) using (P1).

\subsection{Link between the pivot error and the volume of the pivot matrix}
\label{app:ci:maxvol}

The construction of the previous subsection can also be used to show that when adding a new pivot to a cross interpolation, looking for the pivot that maximizes the error of the approximant is equivalent to trying to maximize the volume of the new pivot matrix. Indeed, suppose that we have a pivot matrix $A(\mI,\mJ)$
and we want to enlarge it with a new pivot $(x_0,y_0)$. Using \eqref{eq:Schur}, the determinant of the new pivot matrix reads,
\begin{eqnarray}
&\left| {\rm det }
\begin{pmatrix}
A(\mI,\mJ) & A(\mI,y_0) \\
A(x_0,\mJ) & A(x_0,y_0)
\end{pmatrix}
\right| =  \\
&| {\rm det} A(\mI,\mJ)   | \times |A(x_0,y_0) - A(x_0,\mJ) A(\mI,\mJ)^{-1} A(\mI,y_0)|.\nonumber
\label{eq:proof_maxvol}
\end{eqnarray} 
Since ${\rm det} A(\mI,\mJ)$ is fixed, it follows that maximizing the volume of the pivot matrix
(left hand side of the above equation) is equivalent to finding the pivot $(x_0,y_0)$ where the error of the approximant $|A(x_0,y_0) - A(x_0,\mJ) A(\mI,\mJ)^{-1} A(\mI,y_0)|$ is the largest.

\subsection{Stable QR decomposition for tensor train contractions}
During the evaluation of the tensor train approximant, one needs to evaluate expressions of the form
$T_\alpha (u_\alpha) P^{-1}_\alpha$. As the tensor train approximation becomes better, the volume of the pivot matrices become
smaller so that expression of this type, although mathematically well defined, eventually become
numerically unstable. Let's consider the $T_\alpha (i,u_\alpha,j)$ tensor as a matrix $T_\alpha$ of $\chi\times d$ rows indexed by $(i,u_\alpha)$ and $\chi$ columns indexed by $j$. 
The nesting condition guarantees that the pivot matrix $P_\alpha$ is in fact a sub-matrix of $T_\alpha$.  
Using this structure, we perform a QR decomposition of the $T_\alpha$ matrix, we get,
\begin{equation}
T_\alpha= 
\begin{pmatrix}
P_\alpha \\
T_\alpha'
\end{pmatrix}
=
\begin{pmatrix}
Q \\
Q'
\end{pmatrix}
R
\end{equation}
where $T'_\alpha$ contains all the rows of $T_\alpha$ that are not in $P_\alpha$. The diagonal of the triangular $R$ matrix contains potentially very small values while the matrices $Q$ and $Q'$ (which together form a unitary matrix) are well conditioned. Using this decomposition the product $T_\alpha P^{-1}_\alpha$ can be computed explicitly without usage of the $R$ matrix,
\begin{equation}
T_\alpha P^{-1}_\alpha
=
\begin{pmatrix}
1 \\
Q'Q^{-1}
\end{pmatrix}
\text{.}
\end{equation}

\section{Role of the nesting condition in TCI}

\subsection{Proof of the interpolation property}
\label{app:proofInterpolationFromNesting}

In this appendix, we show that the nesting condition (\ref{eq:nestingCondition}, \ref{eq:nestingCondition2}) implies 
that the TCI form is a proper interpolation of the tensor $A$ as given by \eqref{eq:TCIinterpolProperty}.
The proof is done in four steps (I)-(IV).
% repeat it here.   
%\begin{equation}
%   \label{eq:TCIinterpolPropertyAppendix}   
%   A_\text{TCI} (\mI_{\alpha-1},\mathbb{K}_\alpha, \mJ_{\alpha+1}) = A(\mI_{\alpha-1},\mathbb{K}_\alpha, \mJ_{\alpha+1})
%\end{equation} 

(I). We note that the nesting property (\ref{eq:nestingCondition}) implies
\begin{align}
   \mI_{\alpha} &\subset \mI_{\alpha-1} \oplus \mathbb{K}_{\alpha} \\
   \mI_{\alpha} &\subset \mI_{\alpha-2} \oplus \mathbb{K}_{\alpha-1}\oplus \mathbb{K}_{\alpha} \\
   \mI_{\alpha} &\subset \mI_{0} \oplus \mathbb{K}_{1}\oplus \ldots \oplus \mathbb{K}_{\alpha}
\end{align}
In other words, one can see an element of $\mI_{\alpha}$ as an element of $\mI_{p}$ for $p <\alpha$ concatenated with some $u$ values.
Similar relations apply for $\mJ$.

(II). We reinterpret the three indices tensor $T_\alpha$ by regrouping the left and $u$ index,
to obtain a matrix $T_\alpha^{(L)}$ of indices $\mI_{\alpha-1} \times \mathbb{K}_{\alpha}$ and $\mJ_{\alpha+1}$.
This matrix is in general rectangular.
Because of the nesting condition, a subset of its row indices is in fact $\mI_{\alpha}$, and the restriction 
of $T_\alpha^{(L)}$ to these rows is $P_\alpha$ from the definition of $T$ and $P$ (\ref{eq:defTP}).
Similarly, we introduce $T_\alpha^{(R)}$ by regrouping the $u$ index and the right index, to obtain 
a matrix of indices $\mI_{\alpha-1} $ and $\mathbb{K}_{\alpha} \times  \mJ_{\alpha+1}$, and we have 
\begin{align}
   \label{app:eq:TLR}
   T_\alpha^{(L)}(\mI_{\alpha}, \mJ_{\alpha+1}) P_\alpha^{-1}(\mJ_{\alpha+1}, \mI_{\alpha}') & = \delta(\mI_{\alpha}, \mI_{\alpha}')
   \\
   P_{\alpha-1}^{-1}(\mJ_{\alpha}, \mI_{\alpha-1}) T_\alpha^{(R)}(\mI_{\alpha-1}, \mJ_{\alpha}') &= \delta(\mJ_{\alpha}, \mJ_{\alpha}')
\end{align} 
where $\delta(\mI_{\alpha}, \mI_{\alpha}')$ and $\delta(\mJ_{\alpha}, \mJ_{\alpha}')$ are identity matrices.

(III). We write the TCI in the following form, with implicit contraction over repeated indices $\mI_\alpha$ and $\mJ_\alpha$, which highlights the role of the different sets of indices:
\begin{align}
   \label{eq:defTCIWithIndices}
   A_\text{TCI}(u_1, \ldots,u_n) \approx  \,
                          &T_1(\mI_{0}, u_1, \mJ_{2}) P_1^{-1}(\mJ_{2}, \mI_{1}) \times \nonumber\\
			  &T_2(\mI_{1}, u_2, \mJ_{3}) P_2^{-1}(\mJ_{3}, \mI_{2})   \times \nonumber \\
			  &T_3(\mI_{2}, u_3, \mJ_{4}) P_3^{-1}(\mJ_{4}, \mI_{3}) \ldots  
\end{align} 

(IV). We now fix one value of $\alpha$, and evaluate the TCI form on the
pivot indices and $u_\alpha$, as in (\ref{eq:TCIinterpolProperty}). Our goal is
to show that the $T$ and $P$ on the left and on the right of $T_\alpha$ cancel.
For any multi-index $(u^*_1, \ldots, u^*_{\alpha-1}) \in \mI_{\alpha-1}$ and 
$(u^*_{\alpha+1}, \ldots, u^*_n) \in \mJ_{\alpha+1}$, 
we evaluate $A_\text{TCI}$:
\begin{align}
\label{eq:proof_interpolation}
   %A_TCI(\mI_{\alpha-1},\mathbb{K}_\alpha, \mJ_{\alpha+1}) 
   A_\text{TCI}&(u^*_1, \ldots, u^*_{\alpha-1}, u_\alpha, u^*_{\alpha+1}, \ldots, u^*_n) \nonumber  \\
   =\,  &T_1^{(L)}((u^*_1), \mJ_{2}) P_1^{-1}(\mJ_{2}, \mI_{1}) \times \nonumber \\
			  &T_2^{(L)}(\mI_{1} \oplus (u^*_2), \mJ_{3}) P_2^{-1}(\mJ_{3}, \mI_{2})   \times \nonumber\\
			  & \ldots \nonumber \\
			  &T^{(L)}_{\alpha-1}(\mI_{\alpha-2} \oplus (u^*_{\alpha-1}), \mJ_{\alpha}) P_{\alpha-1}^{-1}(\mJ_{\alpha}, \mI_{\alpha-1}) \times \ldots \nonumber \\
			  &T_\alpha(\mI_{\alpha-1}, \mathbb{K}_\alpha, \mJ_{\alpha+1})  \times \nonumber \\
			  &P_{\alpha}^{-1}(\mJ_{\alpha+1}, \mI_{\alpha}) T^{(R)}_{\alpha+1}(\mI_{\alpha}, (u^*_{\alpha+1}) \oplus \mJ_{\alpha+2}) \nonumber \\
			  & \ldots \nonumber \\
			  &P_{n-2}^{-1}(\mJ_{n-1}, \mI_{n-2})  T_{n-1}^{(R)}(\mI_{n-2},  (u^*_{n-1}) \oplus \mJ_{n}) \times \nonumber \\
			  &P_{n-1}^{-1}(\mJ_{n}, \mI_{n-1}) T_n^{(R)}(\mI_{n-1},  (u^*_{n}))  
\end{align} 
Using (\ref{app:eq:TLR}), the first line reduces to $\delta\bigl (\mI_{1}, {(u^*_1)} \bigr)$, 
hence the second line becomes 
$T_2^{(L)}((u^*_1, u^*_2), \mJ_{3}) P_2^{-1}(\mJ_{3}, \mI_{2})  = \delta\bigl (\mI_{2}, {(u^*_1, u^*_2)} \bigr)$
since $(u^*_1, u^*_2) \in \mI_{2}$.  
The $T^{(L)}$ cancel telescopically from the left until $T_\alpha$. 
The same happens from the right, and we obtain finally
\begin{align}
   A_\text{TCI} &(u^*_1, \ldots, u^*_{\alpha-1}, u_\alpha, u^*_{\alpha+1}, \ldots, u^*_n)\nonumber \\
   = &T_\alpha(u^*_1, \ldots, u^*_{\alpha-1}, u_\alpha, u^*_{\alpha+1}, \ldots, u^*_n) \nonumber \\
   =&A(u^*_1, \ldots, u^*_{\alpha-1}, u_\alpha, u^*_{\alpha+1}, \ldots, u^*_n)
\end{align}
where we used the definition of $T$ \eqref{eq:defTP} in the last line.
This is exactly \eqref{eq:TCIinterpolProperty}.

%%%%%%%%%%%%%%
\subsection{Proof of Eq.\eqref{eq:errorIsOnA} }
\label{app:proofErrorEstimate}
Here we prove \eqref{eq:errorIsOnA}, i.e. that the error between the $\Pi$ tensor and its cross-interpolation is equal to the global error of the TCI of $A$ on the corresponding subset of points.
Let us define $A_\text{TCI}^{\Pi_\alpha}$ by the $TCI$ form in which the product $T_\alpha P^{-1}_\alpha T_{\alpha+1}$ is 
replaced by $\Pi_\alpha$ in \eqref{eq:defTCIWithIndices} and (\ref{eq:proof_interpolation}). In other words
we keep $\Pi_\alpha$ whole and only factorize the other degrees of freedom. 
The proof of Appendix \ref{app:proofInterpolationFromNesting} can be straightforwardly extended to show that, 
$ \forall (u_\alpha, u_{\alpha+1}) \in \mathbb{K}_\alpha \times \mathbb{K}_{\alpha+1} $%
% \mathbb{K}_\alpha, \mathbb{K}_{\alpha+1}
\begin{align*}
   A_\text{TCI} (\mI_{\alpha-1}, u_\alpha, u_{\alpha+1},\mJ_{\alpha+2}) 
   &= T_\alpha (u_\alpha) P_\alpha^{-1} T_{\alpha +1}(u_{\alpha +1}) 
   \\
   A_\text{TCI}^{\Pi_\alpha} (\mI_{\alpha-1},u_\alpha, u_{\alpha+1},\mJ_{\alpha+2}) 
   &= \Pi_\alpha(\mI_{\alpha-1},u_\alpha, u_{\alpha+1},\mJ_{\alpha+2})
   \\
   &= A(\mI_{\alpha-1},u_\alpha, u_{\alpha+1},\mJ_{\alpha+2})
\end{align*}
where the last line is due to (\ref{eq:piDef}).
From the definition of the error function (\ref{eq:errorFunctionNoEnv}), we get
\begin{equation}
 %\epsilon_{\Pi}(\mI_{\alpha-1},u_\alpha, u_{\alpha+1},\mJ_{\alpha+2}) &= | A - A_\text{TCI}| (\mI_{\alpha-1},u_\alpha, u_{\alpha+1},\mJ_{\alpha+2})
 \epsilon_{\Pi}(i, u_\alpha, u_{\alpha+1}, j) = | A - A_\text{TCI}| (i, u_\alpha, u_{\alpha+1}, j)
\end{equation}
for $i\in \mI_{\alpha-1}$, 
$u_\alpha \in \mathbb{K}_\alpha$, 
$u_{\alpha + 1} \in \mathbb{K}_{\alpha + 1}$ and 
$j\in \mJ_{\alpha+2}$.

\subsection{Canonical form and the nested condition}
We end this appendix with a short remark. 
In analogy with the same standard notations in the DMRG literature, we introduce the
mixed {\it canonical } forms of the TCI approximation. Noting
\begin{align}
\overrightarrow{B}_\alpha(u) &= T_\alpha(u) P_\alpha^{-1}  \\
\overleftarrow{B}_\alpha(u) &= P_{\alpha-1}^{-1} T_\alpha(u),
\end{align}
the TCI approximation can be written in the mixed canonical form centered around $T_\beta$,
\begin{equation}
\label{eq:canonical}
A_{\rm TCI}(u_1, \ldots,u_n) = \prod_{\alpha=1}^{\beta-1} \overrightarrow{B}_\alpha(u_\alpha) \ T_\beta(u_\beta)\prod_{\alpha=\beta+1}^{n} \overleftarrow{B}_\alpha(u_\alpha).
\end{equation}
In this form, the interpolation property (P1) is the direct analogue of the norm computation of
a canonical MPS \cite{Schollwoeck_1008}: 
when (\ref{eq:canonical}) is evaluated on the elements of the $T$ tensor, the product of $B$ matrices 
telescopically reduces to identity.

%%%%%%%%%%%%%%%%%%%%%%%%%%%%%%%%%%%%%%%%%%55

\section{Fast summation over Keldysh indices}
\label{app:keldyshfastsumdets}

The calculation of the integrand $\tilde Q_n(u_i)$ amounts to summing
up $2^n$ determinants of size $(2n+1)\times (2n+1)$ \cite{profumo2015}. These determinants
factorize into products of a $n\times n$ with a $(n+1)\times (n+1)$ determinant in the case considered in this article. A naive calculation of a determinant requires a computing time $\propto n^3$ so that the overall computational price of one call to the integrand is $2^n n^3$. Ref. \cite{Griffin_2006_Pminors} proposes an algorithm to calculate all the principal minors of a $n\times n$ matrix $M$ (the determinants of all the submatrices of $M$) at a much smaller cost of $2^n$. This algorithm was later adapted to speed up both imaginary and real time diagrammatic quantum Monte-Carlo calculations~\cite{Simkovic_2022_Pminors}.
Here we propose an algorithm that is equivalent to the one developed in \cite{Simkovic_2022_Pminors}, yet does not require the use of nilpotent polynomials and as a result is perhaps more transparent. We also discuss the techniques used to avoid numerical instabilities or loss of precision.

\subsection{Algorithm}
The problem can be formulated as follows. Let $g(a,\alpha;a',\alpha')$ be a (Green) function that depends on the Keldysh indices $a,a'\in\{0,1\}$ and on all the other degrees of freedom (time, space and  possibly spin, orbitals,...) labeled collectively as the $\alpha,\alpha'$ variables. Let $a_i$ ($\alpha_i$) be a list of $n$ values of the Keldysh (other variables).  
Calculating the integrand $\tilde Q (\alpha_1 \ldots \alpha_n)$ amounts to performing sums of the form
\begin{equation}
\label{eq:minor1}
\tilde Q = \sum_{a_1 \ldots a_n} (-1)^{\sum_p a_p} {\rm det }\ M{\{a_p\}}  
\end{equation}
where the matrix $M{\{a_i\}}$ is defined as,
\begin{equation}
\label{eq:minor1p}
M{\{a_1a_2 \ldots a_n\}}_{ij} = g(a_i,\alpha_i; a_j,\alpha_j).
\end{equation}
Note that \eqref{eq:minor1p} defines only the first $n$ rows and columns. The matrix can be completed by adding more columns and rows of arbitrary value depending on which observables is computed. 
The first step of the algorithm is to introduce a matrix ${\cal M}$ that contains all the matrices $M{\{a_p\}}$ defined in 
\eqref{eq:minor1p} as submatrices. ${\cal M}$ is obtained by stacking the two values of the Keldysh indices one after the other. More precisely, using the "C" convention where matrix indices start from zero, we write $p= 2 i  + a$ and $p'=2i' +a'$ with $a,a' \in \{0,1\}$ and $i,i' \in \{0,1,2, \ldots,n-1\}$ and define
\begin{equation}
\label{eq:mat2n}
{\cal M}_{pp'} = g(a,\alpha_i; a',\alpha_{i'}).
\end{equation}
The first column and row of ${\cal M}$ contains the elements of $g$ corresponding to $a_1=0$,
the second column and row corresponds to $a_1=1$, the third to $a_2=0$ and so on. 

The principal minor algorithm uses Schur complement (see Appendix \ref{app:Schur}) to iteratively "remove" Keldysh indices. One starts with $a_1$. For a given value of $a_1 \in \{0,1\}$, one first
remove the row and column corresponding to the other value $1-a_1$. Then one uses Schur complement to "integrate out" the row and column associated with $a_1$ and define the matrix ${\cal M}^{a_1}$ with the following elements
\begin{equation}
[{\cal M}^{a_1}]_{pp'} = [{\cal M}]_{pp'} - [{\cal M}]_{p,a_1}
\frac{1}{z_1(a_1)} [{\cal M}]_{a_1, p'}
\end{equation}
with 
\begin{equation}
z_1(a_1) = [{\cal M}]_{a_1,a_1}
\end{equation}
and $p,p' \in\{2,3 \ldots 2n-2\}$.
One can continue and define ${\cal M}^{a_1a_2}$, ${\cal M}^{a_1a_2a_3}$... by integrating out $a_2$, then $a_3$...
We define ${\cal M}^{a_1 \ldots a_{k}}$ iteratively by,
\begin{eqnarray}
\label{eq:minor5}
&[ {\cal M}^{a_1 \ldots a_{k+1}} ]_{pp'} = [ {\cal M}^{a_1 \ldots a_{k}} ]_{pp'} - \\
&[ {\cal M}^{a_1 \ldots a_k} ]_{p,2k+a_{k+1}} \
\frac{1}{z_{k+1}} \
[ {\cal M}^{a_1 \ldots a_{k}} ]_{2k+a_{k+1},p'} \nonumber
\end{eqnarray}
with $p,p'\in\{2k+2, \ldots,2n\}$ and
\begin{equation}
\label{eq:minor6}
z_{k+1}(a_1, \dots, a_{k+1}) = [{\cal M}^{a_1 \dots a_{k}}]_{2k+a_{k+1},2k+a_{k+1}}.
\end{equation}
The coefficients $z_i(a_i)$ are directly linked to our target determinant:
\begin{equation}
\label{eq:minor4}
{\rm det} \  M\{a_1a_2 \ldots a_n\} = \prod_{i=1}^{n} z_i(a_1,\dots, a_i).  
\end{equation}
The key remark to prove  Eq.\eqref{eq:minor4} is a property of the Schur complement \eqref{eq:Schur}: 
if one is interested in the determinant of a sub-matrix $M$ of $\cal M$, one can equivalently either apply the Schur complement before or after deleting the corresponding rows or columns, 
i.e. Schur complement commutes with row and matrix selection as long as the Schur pivot belongs to the sub-matrix. 
Noting ${\cal M} /  a_0 a_1 \ldots a_p$ the sub-matrix of
$\cal M$ where one has deleted the rows and columns corresponding to 
$\bar a_0=1-a_0,\ldots, \bar a_p=1-a_p$, one has $M = {\cal M} / a_0 a_1  \ldots a_n$. Using the Schur complement theorem \eqref{eq:Schur}, one can prove iteratively that
\begin{equation}
%\text{det} [{\cal M} /  a_0 a_1...a_p] = \left(\prod_{i=1}^{p} z_i\right) \ 
%\text{det} {\cal M}^{a_1...a_p}
\text{det} [{\cal M} /  a_0 a_1 \dots a_p] = \left (\prod_{i=1}^{p} z_i(a_1,\dots, a_i) \right)\ 
\text{det} {\cal M}^{a_1 \ldots a_p}
\end{equation}
from which Eq.\eqref{eq:minor4} follows.

With these notations, the algorithm reads as follow. One initializes the algorithm with $a_1a_2 \ldots a_n= 00...0$
and construct the list of matrices ${\cal M}^0, {\cal M}^{00}, \dots, {\cal M}^{00...0}$ as well as the associated list
of weights $z_1(0)z_2(0)\dots z_n(0)$. Then, one iterates over the different values of $a_1 a_2 \dots a_n$ sequentially
with the inner loop on $a_n$.
%by considering the list of bits as an integer in binary format. 
At each stage, we keep the list of matrices
$\bigl({\cal M}^{a_1}, {\cal M}^{a_1a_2}, \dots , {\cal M}^{a_1a_2...a_{n}}\bigr)$ and the weights $\bigl(z_1(a_1), z_2(a_2), \dots, z_n(a_n)\bigr)$. Upon going from one
set of Keldysh indices to the next, one uses \eqref{eq:minor5} and \eqref{eq:minor6} to update the matrices and weights that have changed. The result of \eqref{eq:minor4} gives the contribution of the
set $\bigl(a_1, a_2, \dots, a_n \bigr)$ to the integrand. One can check that the overall computational cost is $\propto 2^n$.

The algorithm can also be extended straightforwardly to compute integrands of the form
\begin{equation}
\label{eq:minor2}
\tilde Q = \sum_{a_1 \ldots a_n} (-1)^{\sum_p a_p} {\rm det }\ M{\{a_p\}}  {\rm det }\ M'{\{a_p\}}
\end{equation}
where $M{\{a_p\}}$ and $M'{\{a_p\}}$ are two matrices of the form defined by \eqref{eq:minor1p}, possibly with two different functions $g$ and $g'$. One simply perform the algorithm simultaneously
on the two matrices ${\cal M}$ and ${\cal M}'$. The product of the result of \eqref{eq:minor4} for the two matrices gives the contribution of the set $a_1a_2 \ldots a_n$ to the integrand.

\subsection{Technical implementation}
The above algorithm can be implemented in a straightforward way. Below we show a simple \verb|c++| implementation using "armadillo" library \cite{armaSanderson2016}. The input of the function \verb|EvalSum| is the matrix $\cal M$ in \eqref{eq:mat2n}.  We have found that the speed up of the simple implementation below against a direct sum of determinants is a factor $15$ for $n=12$. A more optimized (but less transparent) version can be obtained by preallocating the matrices or using an iterative implementation instead of a recursive one. In the implementation used in this article (using two matrices as input $\cal M$ and $\cal M'$), we have observed a typical speed up of a factor $40$ compared to the direct sum for $n=12$.

\begin{verbatim}
cx_mat SchurComplement(cx_mat const& M, bool a)
{
    int s=M.n_rows;
    cx_mat Mc(s-2, s-2);
    for(int j=2;j<s;j++)
    {
        auto f=M(a,j)/M(a,a);
        for(int i=2;i<s;i++)
            Mc(i-2,j-2)=M(i,j)-M(i,a)*f;
    }
    return Mc;
}

cx_double EvalSum(cx_mat const& M,
                   cx_double r=1.0,bool sg=0)
{
    if (M.n_rows<2) return sg ? -r*det(M) 
                              :  r*det(M);
    cx_double sum=0;
    for(int a=0;a<2;a++)
    {
        cx_mat Mc=SchurComplement(M,a);
        sum+=EvalSumR(Mc,r*M(a,a), sg!=a);
    }
    return sum;
}
\end{verbatim}

Lastly, we would like to mention two practical issues. 

First, one call to the integrand is a summation over $2^n$ terms $\sum_{a_i}
f(a_i)$ and there is a possibility of large cancellation between these terms
resulting in a loss of precision.  To detect this problem we compute both
$\sum_{a_i} f(a_i)$ and the sum of absolute values $\sum_{a_i} |f(a_i)|$. When
these two quantities differ by many orders of magnitude, we recompute
$\sum_{a_i} f(a_i)$ using the higher precision "long double" mode. 

Second, the above algorithm is not applicable if the diagonal element ${\cal
M}_{00}$ vanishes as the corresponding $1\times1$ Schur complement is ill
defined (or ill conditioned if ${\cal M}_{00}$ is non-zero but very small). To
address this issue for  ${\cal M}_{00}\ll \left\Vert {\cal M} \right\Vert $ we
switch to a $2\times2$ Schur complement and use  "partial pivoting" to maximize
the determinant of the $2\times 2$ matrix on which we perform the Schur
complement (i.e. we reorder the matrix to maximize the magnitude of the
incoming $2\times2$ determinant). The practical implementation of this
$2\times2$ variant is only a factor 2 slower than the $1 \times 1$ version.

\section{ Non-interacting Green's functions in the flat-band limit}
\label{app:g0_qd}
In this appendix, we discuss how to obtain the non-interacting Green's
functions that form the input of the TTD algorithm. These Green's functions can
be calculated for arbitrary tight-binding models using approaches developed
e.g. in the Tkwant package \cite{Kloss2021}.  For systems weakly coupled to a
environment, such as the quantum dots or double quantum dots studied in this
article, an excellent approximation of these Green's function is given by the
flat band limit. This is the limit considered in this article. It is very
suitable for benchmarks as (i) it corresponds to the limit for which we have
the Bethe Ansatz analytical solution at $\epsilon_d=0$ and (ii) the Green's
function can be written in terms of the exponential integral special function
for which there exists machine precision implementations.

We partition our system into the "system" \emph{S} (a set of quantum dots) and an "environment"
$E$ (typically the infinite leads).
To compute the correlators of a given non-interacting Hamiltonian $H_0$, we need the retarded Green's function 
\begin{equation}
g^{R}(\omega)=(\omega-H_0)^{-1}\text{.}
\end{equation}
The one-particle Hamiltonian $H_0$ has a $2\times 2$ block structure
\begin{equation}
H_0=\begin{pmatrix}H_{SS} & H_{SE}\\
H_{ES} & H_{EE}
\end{pmatrix}.
\end{equation}
Since we are interested only in the correlator $g_{SS}^R$ in the $SS$ sub-block, we can write (using the inverse-by-block of a matrix, see \cite{GoluVanl96}),
\begin{equation}
g_{SS}^{R}(\omega)=\left[ \omega-H_{SS}-\Delta(\omega) \right]^{-1}\label{eq:g_block} \text{,}
\end{equation}
where the \emph{hybridization function} 
\begin{equation}
\Delta(\omega)=\lim_{\eta\rightarrow 0^+} H_{SE}\frac{1}{\omega-H_{EE}+i\eta}H_{ES}
\end{equation}
contains all the effect of the bath $E$.

In many practical situations, the coupling of the system to the bath is sufficiently weak that the hybridization 
matrix $\Delta(\omega)$ can be considered as constant in the energy range of interest for the system.
Neglecting the frequency dependence of the hybridization we arrive at i.e. $\Delta(\omega)\approx\Gamma_{1}-i\Gamma_{2}=\text{constant matrix}$ which is known as the \emph{flat-band limit}.  
In this limit, the local Green's function above is given by
\begin{equation}
\label{eq:gR_qd}
g_{SS}^{R}(\omega)=U\left(\omega-D\right)^{-1}U^{-1} \text{,}
\end{equation}
where $U$(resp. $D$) are the eigenvectors (resp. eigenvalues) of
the effective Hamiltonian
\begin{equation}
H_{\text{eff}}=H_{SS}+\Gamma_{1}-i\Gamma_{2}=U\cdot D\cdot U^{-1}\label{eq:Heff_qd} \text{.}
\end{equation}
Note that $H_{\text{eff}}$ is not Hermitian and the eigenvalues
$D$ are complex in general. 

Once the retarded Green's function is known in the energy domain, we can obtain the
lesser and greater Green's functions in real time. At thermal equilibrium and zero temperature 
the lesser and greater Green's functions are given by,
\begin{equation}
g_{SS}^{\lessgtr}(t)=\frac{\mp i}{\pi}\int d\omega\exp(-i\omega t) 
\theta (\mp \omega)
\Im g_{SS}^R
\end{equation}
where $\Im$ stands for imaginary part.

Since in \eqref{eq:gR_qd} the eigenvectors $U$ don't depend on $\omega$ and $D$ is
a diagonal matrix, these integrals can be computed explicitly:
\begin{align}
\begin{aligned}g_{SS}^{\lessgtr}(t) & =\frac{1}{2\pi}\left[UI^{\lessgtr}(D,t)U^{-1}-U^{*}I^{\lessgtr}(D^{*},t)\left(U^{-1}\right)^{*}\right]\\
g_{SS}^{<}(0) & =\frac{-i}{\pi}\Im\left\{ U\left[\log(-D)+i\pi\text{sg}\left(\Im D\right)\right]U^{-1}\right\} \\
g_{SS}^{>}(0) & =-\frac{i}{\pi}\Im\left\{ U\log(-D)U^{-1}\right\} 
\end{aligned}
\label{eq:g_qd}
\end{align}
with 
\begin{equation}
\begin{aligned}I^{\lessgtr}(a,t)=\exp(-iat) & \left\{ E_{1}(-iat)\pm\right.\\
 & \left.2\pi i\text{sg}(t)\theta\left[-\Im(at)\right]\theta\left[\mp\Re(a)\right]\right\} 
\end{aligned}
\end{equation}
where $E_{1}(z)=\int_{z}^{\infty}dt\exp(-t)/t$ is the exponential-integral
function $E_{1}$ (see \cite{NIST:DLMF}), $\text{sg}(x)$ is the sign function
and $\theta(x)$ is the Heaviside step function with $\theta(0)=1/2$.

\subsection{Single quantum dot (SIAM)}

For a single quantum dot, the effective Hamiltonian matrix becomes a scalar
$H_{\text{eff}}=\epsilon_{d}-i\Gamma$ yielding to:
\begin{align}
\begin{aligned}
g_{SS}^{\lessgtr}(t) & =\frac{1}{2\pi}\left[I^{\lessgtr}(\epsilon_{d}-i\Gamma,t)-I^{\lessgtr}(\epsilon_{d}+i\Gamma,t)\right]\\
g_{SS}^{<}(0) & =\frac{-i}{\pi}\Im\left[\log(-\epsilon_{d}+i\Gamma)+i\pi\text{sg}\left(-\Gamma\right)\right]\\
g_{SS}^{>}(0) & =-\frac{i}{\pi}\Im\log(-\epsilon_{d}+i\Gamma).
\end{aligned}
\end{align}
This expressions form the inputs for our SIAM benchmark. All the energies (times) are measured in unit of $\Gamma$ ($1/\Gamma$).

\subsection{Double quantum dot}

We also consider a double quantum dot with local Hamiltonian matrix
$H_{SS}=\begin{pmatrix}0 & \gamma_{0}\\
\gamma_{0} & 0
\end{pmatrix}$ and hybridization function $\Delta(\omega)\approx-i\begin{pmatrix}\Gamma & 0\\
0 & \Gamma
\end{pmatrix}$, leading to 
\begin{equation}
H_{\text{eff}}=\begin{pmatrix}-i\Gamma & \gamma_{0}\\
\gamma_{0} & -i\Gamma
\end{pmatrix}=U\cdot D\cdot U^{-1}
\end{equation}
\begin{equation}
U=\begin{pmatrix}-1 & 1\\
1 & 1
\end{pmatrix}\text{, }D=\begin{pmatrix}-i\Gamma-\gamma_{0} & 0\\
0 & -i\Gamma+\gamma_{0}
\end{pmatrix}
\end{equation}
In our example, we use $\gamma_{0}=0.5\Gamma$ and apply \eqref{eq:g_qd}
to compute the non-interacting Green's functions. All the energies (times) are measured in unit of $\Gamma$ ($1/\Gamma$).

%----------------------------------

\section{ Non-interacting Green's functions in 2D lattice}
\label{app:g0_2d}
Here we calculate the non-interacting Green's function for a particle in an infinite
two-dimensional lattice used in Section \ref{sec:lattice}. The non-interacting Hamiltonian
reads (omitting the spin index since the problem is diagonal in spin),
\begin{equation}
H_{0}=\sum_{<ij>}c_{i}^{\dagger}c_{j}
\end{equation}
with sum over nearest neighbors.

\subsection{Explicit summation in momentum space}
The dispersion relation of $H_0$ is $E_{\boldsymbol{k}}=2\cos k_{x}+2\cos k_{y}$.
Since the corresponding velocities $\vec v = \partial E/\partial \boldsymbol{k}$ are
bounded by $2$ in both spatial directions, it follows that it is enough to consider a
finite lattice of length $L > 2t$ to calculate the Green's function without finite size effects. Hence, we consider a system of $L\times L$ sites with periodic boundary conditions.  It can be diagonalized using the operators  $\{d_{\boldsymbol{k}}\}$
in the momentum basis
\begin{equation}
c_{i}=\frac{1}{L}\sum_{\boldsymbol{k}}e^{i\boldsymbol{k}\cdot\boldsymbol{r}_{i}}d_{\boldsymbol{k}}\text{,}
\end{equation}
where $\boldsymbol{r}_{i}$ is the lattice position of site $i$ and
$\boldsymbol{k}=(k_{x},k_{y})$ with $k_{x,y}=\frac{2\pi}{L}\kappa_{x,y}$,
and $\kappa_{x,y}=0,1, \ldots,L-1$. 
In Heisenberg representation, we simply have $d_{\boldsymbol{k}}(t)=e^{-i E_{\boldsymbol{k}}t}d_{\boldsymbol{k}}\text{.}$ It follows that  the lesser and greater Green's functions in real time at $i,j$ are given by
\begin{equation}
g_{ij}^{<}(t)=\frac{i}{L^{2}}\sum_{\boldsymbol{k}}e^{i\left(\boldsymbol{k}\cdot\boldsymbol{r}_{ij}-tE_{\boldsymbol{k}}\right)}f_{\text{FD}}(E_{\boldsymbol{k}})\text{,}
\end{equation}
\begin{equation}
g_{ij}^{>}(t)=\frac{-i}{L^{2}}\sum_{\boldsymbol{k}}e^{i\left(\boldsymbol{k}\cdot\boldsymbol{r}_{ij}-tE_{\boldsymbol{k}}\right)}\bar{f}_{\text{FD}}(E_{\boldsymbol{k}})\text{,}
\end{equation}
where $\boldsymbol{r}_{ij}=\boldsymbol{r}_{i}-\boldsymbol{r}_{j}=(x,y)$ is the position difference with $x,y \in \mathbb{Z}$,
$f_{\text{FD}}(E)=1/\left[e^{\beta(E-\mu)}+1\right]$ is the Fermi-Dirac
distribution, $\bar{f}_{\text{FD}}(E)=1-f_{\text{FD}}(E)$, and
$\beta$ is the inverse of the temperate and $\mu$ is the chemical
potential. In practice we compute the sum above for $L=500$ and $i=j=0$ for the single impurity in a lattice problem.

\subsection{Thermodynamic limit }

For $L\rightarrow\infty$ the previous expression for $g_{ij}^{<}(t)$ can be written as
\begin{equation}
\label{eq:g0_hubb}
g_{xy}^{<}(t)=\frac{i}{(2\pi)^{2}}\int_{-\pi}^{\pi}dk_{x}\int_{-\pi}^{\pi}dk_{y}e^{i\left(\boldsymbol{k}\cdot\boldsymbol{r}_{ij}-tE_{\boldsymbol{k}}\right)}f_{\text{FD}}(E_{\boldsymbol{k}})\text{.}
\end{equation}
At zero temperature the Fermi function can be expanded as a Fourier integral
\begin{equation}
\lim_{\beta\rightarrow\infty}f_{\text{FD}}(E)=\theta(-E)=\frac{i}{2\pi}\int\frac{dw}{w+i0^{+}}e^{i wE}\text{.}
\end{equation}
which allows one to decouple the two integrals on $k_x$ and $k_y$ in
\eqref{eq:g0_hubb}. Using the definition of the Bessel's functions (for
$n\in\mathbb{Z}$) \cite[Eq.~10.9.2]{NIST:DLMF}:
\begin{equation}
\int_{-\pi}^{\pi}dke^{i(kn-x\cos k)}=(-i)^{n}2\pi J_{n}(x)\text{,}
\end{equation}
we arrive at,
\begin{equation}
\begin{aligned}g_{xy}^{\lessgtr}(t)=\frac{\mp(-i)^{x+y}}{2\pi} & \left[\fint\frac{dw}{w}J_{x}(2t\mp2w)J_{y}(2t\mp2w)+\right.\\
 & \left.+i\pi J_{x}(2t)J_{y}(2t)\right]\text{,}
\end{aligned}
\end{equation}
where the last integral represents its Cauchy principal value.

\section{Calculation of the integral in the simplex domain}
\label{app:simplex_int}
 The calculation of the integral in the simplex domain S, $0 \leq u_n \leq u_{n-1} \ldots u_1 \leq t$
 is not as straightforward as the hypercube integration and requires an iterative algorithm that we now explain.
The multi-dimensional integral \eqref{eq:def_Qn} over the simplex in $u$-variables has the explicit form,
\begin{eqnarray}
\label{app:eq:def_Qn_simplex}
  Q_n(t) = \int_0^t \lambda(u_1) d u_1 \ldots \int_{0}^{u_{n-2}} \lambda(u_{n-1}) d u_{n-1}
   \nonumber \\
  \int_{0}^{u_{n-1}} \!\!\! \lambda(u_{n})du_n   \tilde Q_n(u_1, \ldots, u_n)
\end{eqnarray}
Since the TTD approximation is performed in the $ v$ variables, our approximation reads,
\begin{align}
  \tilde Q_n(u_1, u_2, \ldots, u_n) \approx T_1(t-u_1) P_1^{-1}   \nonumber \\
 \times  T_2(u_1-u_2) P_2^{-1}  \ldots P_{n-1}^{-1} T_{n}(u_{n-1}-u_n) 
\end{align}
The integrals over the $u$ variables are performed one by one starting with $u_n$ and ending with
$u_1$. The $u_n$ variable is only present in the last tensor $T_n$. We perform the corresponding one-dimensional integral. Defining
\begin{equation}
\label{app:eq:psi1}
\Psi_{n}(x) \equiv  \int_0^{x} \ dy \ \lambda(y)  T_n(x-y),
\end{equation}
we find that
\begin{align}
Q_n(t) \approx \int_0^t \lambda(u_1) d u_1 \ldots \int_{0}^{u_{n-2}} \!\!\! \lambda(u_{n-1}) d u_{n-1}
T_1(t-u_1) P_1^{-1}  \nonumber \\
T_2(u_1-u_2) P_2^{-1}  \ldots T_{n-1} (u_{n-2}-u_{n-1}) P_{n-1}^{-1} \Psi_{n} (u_{n-1})
\end{align}
We continue with the one-dimensional integral over $u_{n-1}$ which is only present in the terms
$T_{n-1} (u_{n-2}-u_{n-1})  \Psi_{n-1} (u_{n-1})$. Defining for $p<n$
\begin{equation}
\label{app:eq:psi2}
\Psi_{p}(x) \equiv  \int_0^{x} \ dy \ \lambda(y)  T_{p}(x-y) P_{p}^{-1}\Psi_{p+1} (y),
\end{equation}
we find that
\begin{align}
Q_n(t) \approx \int_0^t \lambda(u_1) d u_1 \ldots \int_{0}^{u_{n-3}} \!\!\! \lambda(u_{n-2}) d u_{n-2}
T_1(t-u_1) P_1^{-1}  \nonumber \\
T_2(u_1-u_2) P_2^{-1}  \ldots T_{n-2} (u_{n-3}-u_{n-2}) P_{p-2}^{-1} \Psi_{n-1} (u_{n-2})
\end{align}
We continue to perform the integrations one by one until we arrive at the final integration
\begin{align}
\label{app:eq:simplex}
Q_n(t) \approx \int_0^t \lambda(u_{1}) d u_1 T_1(t-u_1) P_{1}^{-1}\Psi_2(u_1) = \Psi_1(t)
\end{align}
In practice the above algorithm requires the precise knowledge of the
$\Psi_p(u_{p-1})$ functions.  We use precise Chebyshev interpolants of the
$T_n(v_n)$ matrices and $\Psi_p(u_{p-1})$ vectors to define the right hand side
of \eqref{eq:psi1} and (\ref{eq:psi2}) in terms of large order polynomials
whose primitive is known exactly. The result is projected again on Chebyshev
polynomials. It is important to note that the last integral
\eqref{eq:simplex} provides the entire $t$ dependence of $Q_n(t)$ and that
it is a post-treatment calculation that can be performed for any time dependent
switching on of the interaction $\lambda(t)$.

%---------------------------------------------

\section{Calculation of the simplex integral using Fourier transform}
\label{app:fourier_int}
As an alternative to the integration in $u$-variables in the simplex domain
discussed in Sec.\ \ref{app:simplex_int}, the multi-dimensional integral can be
as well calculated in the variables $v$ [defined in \eqref{eq:v_vars}],
together with the domain condition (\ref{eq:domainVPositive},
\ref{eq:domainVTm}). Note that this alternative route is defined only for the
abrupt switching of the interaction $\lambda (t)=\theta(t)$ and cannot be
generalized to arbitrary functions $\lambda (t)$.  In the $v$ variables, the
integral in \eqref{eq:def_Qn} is essentially a multidimensional convolution,
\begin{align}
\label{eq:def_Qn_v}
  Q_n(t) = \prod_{\alpha=1}^n \int_0^t d v_\alpha \theta \left[t - \sum_{i=1}^n v_i\right] \tilde Q_n(v_1, \ldots, v_n),
\end{align}
where $\theta(x)$ is the Heaviside step function. The Fourier representation of the Heaviside function,
\begin{align}
 \theta(t) = \lim_{\epsilon \rightarrow 0^+ }\int_{-\infty}^{\infty} \frac{d \omega}{2 \pi i} \frac{e^{i \omega t}}{\omega - i \epsilon} ,
\end{align}
can be used to remove the constraints on the $v$ variables. Using the tensor train factorization approximation in \eqref{eq:TCIQn} for $\tilde Q_n(v_1, \ldots, v_n)$,
one can write \eqref{eq:def_Qn_v} as
\begin{align}
  Q_n(t) &\approx  \int_{-\infty}^{\infty}  \frac{d \omega}{2 \pi i} \frac{e^{i \omega t}}{\omega - i 0^+}
  \prod_{\alpha=1}^n \int_0^t d v_\alpha e^{- i \omega v_\alpha} T_\alpha(v_\alpha) P_\alpha^{-1} .
\end{align}
With $(x \pm i 0^+)^{-1} = \mp i \pi \delta(x) + \text{p.v.}(1/x)$, where p.v. stands for the principal value.
One arrives at
\begin{subequations}
\label{eq:def_Qn_v_num}
\begin{align}
  Q_n(t) &=  \frac{\tilde{q}_n(0)}{2} + \fint_{-\infty}^{\infty}  \frac{d \omega}{2 \pi i \omega} e^{i \omega t} \tilde{q}_n(\omega) , \label{eq:def_Qn_v_a} \\
  \tilde{q}_n(\omega) &=  \prod_{\alpha=1}^n \int_0^t d v_\alpha e^{- i \omega v_\alpha} T_\alpha(v_\alpha) P_\alpha^{-1} \label{eq:def_Qn_v_b} .
\end{align}
\end{subequations}
Instead of the initial $n$-dimensional integral, the above equation for $\tilde{q}_n(\omega)$ is a product of $n$ one-dimensional integrals, which can be computed numerically.
Moreover, the function $\tilde{q}_n(\omega)$ can be precomputed once and the entire $Q_n(t)$ curve obtained \textit{a posteriori} by evaluating the remaining one-dimensional integral in \eqref{eq:def_Qn_v_a}  for different values of the  time $t$.
In practice, as the integrands in  (\ref{eq:def_Qn_v_a}) and
(\ref{eq:def_Qn_v_b}) decrease fast, the integrals are cut-off at a finite,
large enough values of $\omega$ and $v$.
Appropriate quadratures~\cite{quadpack} are used to compute the principal value
integral numerically around the pole at $\omega = 0$, as well as for the
oscillatory integrals in (\ref{eq:def_Qn_v_a}) and (\ref{eq:def_Qn_v_b}).  For
the precomputation, $\tilde{q}_n(\omega)$ is interpolated using piecewise
adaptive polynomials as in Ref.\ \cite{Gonnet10}.

\section{Efficient quadrature for $T_\alpha(v)$}
\label{app:tailored_quad}

In the case of the single impurity embedded in a two-dimensional lattice,
the tensors $T_\alpha(v)$ oscillate rapidly and
decay slowly with respect to $v$. We present a specialized quadrature
scheme to compute them efficiently.

Since the integrand $\tilde Q_n(u_i)$ is a sum of products of
non-interacting Green's functions, we are able to characterize the
behavior of $T_\alpha(v)$ as $v \to \infty$. For the single
impurity embedded in a two-dimensional lattice, we find empirically that we can
accurately approximate $T_\alpha(v)$ by an expansion of the type
\begin{equation} \label{eq:tasymp}
  T_\alpha(v) \approx \sum_{p=0}^{N_p} \sum_{n=2}^{N_n} a_{pn}
  \frac{\cos(4pv)}{v^n} + b_{pn} \frac{\sin(4pv)}{v^n}
\end{equation}
when $v > v_{\text{cut}}$, for $v_{\text{cut}}$ a sufficiently large
cutoff. The chosen frequencies originate from the bandwidth of the
non-interacting Hamiltonian, and the algebraic decay is observed empirically. $N_p$ and $N_n$ are used to control the precision
of the expansion, and in practice we observe rapid convergence in these
parameters. We therefore split the integral into two parts:
\begin{equation}
\int_0^\infty T_\alpha(v) \, dv = \int_0^{v_{\text{cut}}} T_\alpha(v) \, dv
+ \int_{v_{\text{cut}}}^\infty T_\alpha(v) \, dv.
\end{equation}
$T_\alpha(v)$ typically only contains only a few oscillations on
$[0,v_{\text{cut}}]$, so the first integral can
be computed efficiently using a standard Gauss-Legendre quadrature rule. For the second integral, we use
\eqref{eq:tasymp} to design a custom quadrature rule, as follows.

We describe the method for a generic collection of functions $f: (a,b) \to
\mathbb{C}$ defined as the span of $N$ basis functions $\phi_k$:
\begin{equation}
f(x) = \sum_{k=1}^N \widehat{f_k} \phi_k(x).
\end{equation}
The functions $T_\alpha$ form such a class approximately, with $(a,b) =
(v_{\text{cut}},\infty)$ and the basis functions $\phi_k$ given by
\eqref{eq:tasymp}. Given a collection $x_j$ of $N$ sampling points for
the functions $\phi_k$, we define the matrix $\Phi_{jk} \equiv
\phi_k(x_j)$, and have $f_j \equiv f(x_j) = \sum_{k=1}^N
\Phi_{jk} \widehat{f_k}$. If
$I_k \equiv \int_a^b \phi_k(x) \, dx$, then
\begin{equation}
\int_a^b f(x) \, dx = \sum_{k=1}^N I_k \widehat{f_k} = \sum_{j,k=1}^N
I_k \Phi_{kj}^{-1} f_j \equiv \sum_{j=1}^N w_j f_j,
\end{equation}
where we have defined the quadrature weights $w_j$.

The nodes $x_j$ must be chosen properly to ensure stability. To do so, we form the $M \times N$ matrix
$\phi_k(\bar x_j)$, where $\{\bar x_j\}_{j=1}^M$ is a fine grid
on $(a,b)$, sufficient to accurately discretize all of the functions
$\phi_k$. It can be shown that the nodes $x_j$ corresponding to the pivot
indices obtained by pivoted Gram-Schmidt orthogonalization on the
rows of this matrix yield a stable quadrature rule \cite{bremer10}.
Roughly speaking, this
procedure chooses the $N$ most linearly independent rows of the matrix,
yielding the $N$ most independent nodes in the fine grid.
Alternatively, we
have found in practice that the nodes corresponding to the pivots
of the cross interpolation of $\phi_k(\bar x_j)$ may be used as well.
%Although,
%this alternative method does not have the same guarantees as the Gram-Schmidt ("QR") decomposition, 
%we have found in practice that it provides a stable rule for this application.

We can follow this procedure to compute a quadrature rule for the functions $T_\alpha$ using
the expansion \eqref{eq:tasymp}. In this case, it is straightforward to write the
integrals $I_k$ in terms of the well-known $E_n$ functions:
\begin{equation}
E_n(z) \equiv \int_1^\infty \frac{e^{-zt}}{t^n} \, dt.
\end{equation}
$E_1$ is the exponential integral, which can be evaluated using standard
libraries \cite{gsl}, and $E_n(z)$ can then be obtained by a simple recurrence
\cite[Eq. 8.19.12]{NIST:DLMF}. In practice, we set $v_{\text{cut}} =
10$, use $63$ Gauss-Legendre nodes for the integral on
$[0,v_{\text{cut}}]$, and set $N_p = 2$, $N_n = 4$ to obtain $15$ nodes
for the integral on $[v_{\text{cut}},\infty)$. These $78$ quadrature
nodes yield 2-3 digits of accuracy in the final result.

\bibliography{refs}

\begin{thebibliography}{58}%
\makeatletter
\providecommand \@ifxundefined [1]{%
 \@ifx{#1\undefined}
}%
\providecommand \@ifnum [1]{%
 \ifnum #1\expandafter \@firstoftwo
 \else \expandafter \@secondoftwo
 \fi
}%
\providecommand \@ifx [1]{%
 \ifx #1\expandafter \@firstoftwo
 \else \expandafter \@secondoftwo
 \fi
}%
\providecommand \natexlab [1]{#1}%
\providecommand \enquote  [1]{``#1''}%
\providecommand \bibnamefont  [1]{#1}%
\providecommand \bibfnamefont [1]{#1}%
\providecommand \citenamefont [1]{#1}%
\providecommand \href@noop [0]{\@secondoftwo}%
\providecommand \href [0]{\begingroup \@sanitize@url \@href}%
\providecommand \@href[1]{\@@startlink{#1}\@@href}%
\providecommand \@@href[1]{\endgroup#1\@@endlink}%
\providecommand \@sanitize@url [0]{\catcode `\\12\catcode `\$12\catcode
  `\&12\catcode `\#12\catcode `\^12\catcode `\_12\catcode `\%12\relax}%
\providecommand \@@startlink[1]{}%
\providecommand \@@endlink[0]{}%
\providecommand \url  [0]{\begingroup\@sanitize@url \@url }%
\providecommand \@url [1]{\endgroup\@href {#1}{\urlprefix }}%
\providecommand \urlprefix  [0]{URL }%
\providecommand \Eprint [0]{\href }%
\providecommand \doibase [0]{https://doi.org/}%
\providecommand \selectlanguage [0]{\@gobble}%
\providecommand \bibinfo  [0]{\@secondoftwo}%
\providecommand \bibfield  [0]{\@secondoftwo}%
\providecommand \translation [1]{[#1]}%
\providecommand \BibitemOpen [0]{}%
\providecommand \bibitemStop [0]{}%
\providecommand \bibitemNoStop [0]{.\EOS\space}%
\providecommand \EOS [0]{\spacefactor3000\relax}%
\providecommand \BibitemShut  [1]{\csname bibitem#1\endcsname}%
\let\auto@bib@innerbib\@empty
%</preamble>
\bibitem [{\citenamefont {Blankenbecler}\ \emph {et~al.}(1981)\citenamefont
  {Blankenbecler}, \citenamefont {Scalapino},\ and\ \citenamefont
  {Sugar}}]{Sugar1981}%
  \BibitemOpen
  \bibfield  {author} {\bibinfo {author} {\bibfnamefont {R.}~\bibnamefont
  {Blankenbecler}}, \bibinfo {author} {\bibfnamefont {D.~J.}\ \bibnamefont
  {Scalapino}},\ and\ \bibinfo {author} {\bibfnamefont {R.~L.}\ \bibnamefont
  {Sugar}},\ }\bibfield  {title} {\bibinfo {title} {Monte carlo calculations of
  coupled boson-fermion systems. i},\ }\href
  {https://doi.org/10.1103/PhysRevD.24.2278} {\bibfield  {journal} {\bibinfo
  {journal} {Phys. Rev. D}\ }\textbf {\bibinfo {volume} {24}},\ \bibinfo
  {pages} {2278} (\bibinfo {year} {1981})}\BibitemShut {NoStop}%
\bibitem [{\citenamefont {Foulkes}\ \emph {et~al.}(2001)\citenamefont
  {Foulkes}, \citenamefont {Mitas}, \citenamefont {Needs},\ and\ \citenamefont
  {Rajagopal}}]{Foulkes2001}%
  \BibitemOpen
  \bibfield  {author} {\bibinfo {author} {\bibfnamefont {W.~M.~C.}\
  \bibnamefont {Foulkes}}, \bibinfo {author} {\bibfnamefont {L.}~\bibnamefont
  {Mitas}}, \bibinfo {author} {\bibfnamefont {R.~J.}\ \bibnamefont {Needs}},\
  and\ \bibinfo {author} {\bibfnamefont {G.}~\bibnamefont {Rajagopal}},\
  }\bibfield  {title} {\bibinfo {title} {Quantum monte carlo simulations of
  solids},\ }\href {https://doi.org/10.1103/RevModPhys.73.33} {\bibfield
  {journal} {\bibinfo  {journal} {Rev. Mod. Phys.}\ }\textbf {\bibinfo {volume}
  {73}},\ \bibinfo {pages} {33} (\bibinfo {year} {2001})}\BibitemShut {NoStop}%
\bibitem [{\citenamefont {Sandvik}(2010)}]{Sandvik2010}%
  \BibitemOpen
  \bibfield  {author} {\bibinfo {author} {\bibfnamefont {A.~W.}\ \bibnamefont
  {Sandvik}},\ }\bibfield  {title} {\bibinfo {title} {Computational studies of
  quantum spin systems},\ }\href {https://doi.org/10.1063/1.3518900} {\bibfield
   {journal} {\bibinfo  {journal} {AIP Conference Proceedings}\ }\textbf
  {\bibinfo {volume} {1297}},\ \bibinfo {pages} {135} (\bibinfo {year}
  {2010})},\ \Eprint
  {https://arxiv.org/abs/https://aip.scitation.org/doi/pdf/10.1063/1.3518900}
  {https://aip.scitation.org/doi/pdf/10.1063/1.3518900} \BibitemShut {NoStop}%
\bibitem [{\citenamefont {{Van Houcke}}\ \emph {et~al.}(2010)\citenamefont
  {{Van Houcke}}, \citenamefont {Kozik}, \citenamefont {Prokof’ev},\ and\
  \citenamefont {Svistunov}}]{Vanhoucke2010}%
  \BibitemOpen
  \bibfield  {author} {\bibinfo {author} {\bibfnamefont {K.}~\bibnamefont {{Van
  Houcke}}}, \bibinfo {author} {\bibfnamefont {E.}~\bibnamefont {Kozik}},
  \bibinfo {author} {\bibfnamefont {N.}~\bibnamefont {Prokof’ev}},\ and\
  \bibinfo {author} {\bibfnamefont {B.}~\bibnamefont {Svistunov}},\ }\bibfield
  {title} {\bibinfo {title} {Diagrammatic monte carlo},\ }\href
  {https://doi.org/https://doi.org/10.1016/j.phpro.2010.09.034} {\bibfield
  {journal} {\bibinfo  {journal} {Physics Procedia}\ }\textbf {\bibinfo
  {volume} {6}},\ \bibinfo {pages} {95} (\bibinfo {year} {2010})},\ \bibinfo
  {note} {computer Simulations Studies in Condensed Matter Physics
  XXI}\BibitemShut {NoStop}%
\bibitem [{\citenamefont {Carlson}\ \emph {et~al.}(2015)\citenamefont
  {Carlson}, \citenamefont {Gandolfi}, \citenamefont {Pederiva}, \citenamefont
  {Pieper}, \citenamefont {Schiavilla}, \citenamefont {Schmidt},\ and\
  \citenamefont {Wiringa}}]{Carlson2015}%
  \BibitemOpen
  \bibfield  {author} {\bibinfo {author} {\bibfnamefont {J.}~\bibnamefont
  {Carlson}}, \bibinfo {author} {\bibfnamefont {S.}~\bibnamefont {Gandolfi}},
  \bibinfo {author} {\bibfnamefont {F.}~\bibnamefont {Pederiva}}, \bibinfo
  {author} {\bibfnamefont {S.~C.}\ \bibnamefont {Pieper}}, \bibinfo {author}
  {\bibfnamefont {R.}~\bibnamefont {Schiavilla}}, \bibinfo {author}
  {\bibfnamefont {K.~E.}\ \bibnamefont {Schmidt}},\ and\ \bibinfo {author}
  {\bibfnamefont {R.~B.}\ \bibnamefont {Wiringa}},\ }\bibfield  {title}
  {\bibinfo {title} {Quantum monte carlo methods for nuclear physics},\ }\href
  {https://doi.org/10.1103/RevModPhys.87.1067} {\bibfield  {journal} {\bibinfo
  {journal} {Rev. Mod. Phys.}\ }\textbf {\bibinfo {volume} {87}},\ \bibinfo
  {pages} {1067} (\bibinfo {year} {2015})}\BibitemShut {NoStop}%
\bibitem [{\citenamefont {Oseledets}(2011)}]{oseledets2011ttd}%
  \BibitemOpen
  \bibfield  {author} {\bibinfo {author} {\bibfnamefont {I.~V.}\ \bibnamefont
  {Oseledets}},\ }\bibfield  {title} {\bibinfo {title} {Tensor-train
  decomposition},\ }\href@noop {} {\bibfield  {journal} {\bibinfo  {journal}
  {SIAM Journal on Scientific Computing}\ }\textbf {\bibinfo {volume} {33}},\
  \bibinfo {pages} {2295} (\bibinfo {year} {2011})}\BibitemShut {NoStop}%
\bibitem [{\citenamefont {Dolgov}\ and\ \citenamefont
  {Savostyanov}(2020)}]{dolgov2020integralRn}%
  \BibitemOpen
  \bibfield  {author} {\bibinfo {author} {\bibfnamefont {S.}~\bibnamefont
  {Dolgov}}\ and\ \bibinfo {author} {\bibfnamefont {D.}~\bibnamefont
  {Savostyanov}},\ }\bibfield  {title} {\bibinfo {title} {Parallel cross
  interpolation for high-precision calculation of high-dimensional integrals},\
  }\href@noop {} {\bibfield  {journal} {\bibinfo  {journal} {Computer Physics
  Communications}\ }\textbf {\bibinfo {volume} {246}},\ \bibinfo {pages}
  {106869} (\bibinfo {year} {2020})}\BibitemShut {NoStop}%
\bibitem [{\citenamefont {Dolgov}\ \emph {et~al.}(2020)\citenamefont {Dolgov},
  \citenamefont {Anaya-Izquierdo}, \citenamefont {Fox},\ and\ \citenamefont
  {Scheichl}}]{tt-integral1}%
  \BibitemOpen
  \bibfield  {author} {\bibinfo {author} {\bibfnamefont {S.}~\bibnamefont
  {Dolgov}}, \bibinfo {author} {\bibfnamefont {K.}~\bibnamefont
  {Anaya-Izquierdo}}, \bibinfo {author} {\bibfnamefont {C.}~\bibnamefont
  {Fox}},\ and\ \bibinfo {author} {\bibfnamefont {R.}~\bibnamefont
  {Scheichl}},\ }\bibfield  {title} {\bibinfo {title} {Approximation and
  sampling of multivariate probability distributions in the tensor train
  decomposition},\ }\href@noop {} {\bibfield  {journal} {\bibinfo  {journal}
  {Statistics and Computing}\ }\textbf {\bibinfo {volume} {30}},\ \bibinfo
  {pages} {603} (\bibinfo {year} {2020})}\BibitemShut {NoStop}%
\bibitem [{\citenamefont {Vysotsky}\ \emph {et~al.}(2021)\citenamefont
  {Vysotsky}, \citenamefont {Smirnov},\ and\ \citenamefont
  {Tyrtyshnikov}}]{tt-integral2}%
  \BibitemOpen
  \bibfield  {author} {\bibinfo {author} {\bibfnamefont {L.~I.}\ \bibnamefont
  {Vysotsky}}, \bibinfo {author} {\bibfnamefont {A.~V.}\ \bibnamefont
  {Smirnov}},\ and\ \bibinfo {author} {\bibfnamefont {E.~E.}\ \bibnamefont
  {Tyrtyshnikov}},\ }\bibfield  {title} {\bibinfo {title} {Tensor-train
  numerical integration of multivariate functions with singularities},\
  }\href@noop {} {\bibfield  {journal} {\bibinfo  {journal} {Lobachevskii
  Journal of Mathematics}\ }\textbf {\bibinfo {volume} {42}},\ \bibinfo {pages}
  {1608} (\bibinfo {year} {2021})}\BibitemShut {NoStop}%
\bibitem [{\citenamefont {Chertkov}\ and\ \citenamefont
  {Oseledets}(2021)}]{tt-example-FP}%
  \BibitemOpen
  \bibfield  {author} {\bibinfo {author} {\bibfnamefont {A.}~\bibnamefont
  {Chertkov}}\ and\ \bibinfo {author} {\bibfnamefont {I.}~\bibnamefont
  {Oseledets}},\ }\bibfield  {title} {\bibinfo {title} {Solution of the
  fokker--planck equation by cross approximation method in the tensor train
  format},\ }\href@noop {} {\bibfield  {journal} {\bibinfo  {journal}
  {Frontiers in Artificial Intelligence}\ }\textbf {\bibinfo {volume} {4}}
  (\bibinfo {year} {2021})}\BibitemShut {NoStop}%
\bibitem [{\citenamefont {Smirnov}\ \emph {et~al.}(2022)\citenamefont
  {Smirnov}, \citenamefont {Shapurov},\ and\ \citenamefont
  {Vysotsky}}]{fiesta5}%
  \BibitemOpen
  \bibfield  {author} {\bibinfo {author} {\bibfnamefont {A.}~\bibnamefont
  {Smirnov}}, \bibinfo {author} {\bibfnamefont {N.}~\bibnamefont {Shapurov}},\
  and\ \bibinfo {author} {\bibfnamefont {L.}~\bibnamefont {Vysotsky}},\
  }\bibfield  {title} {\bibinfo {title} {Fiesta5: Numerical high-performance
  feynman integral evaluation},\ }\href
  {https://doi.org/https://doi.org/10.1016/j.cpc.2022.108386} {\bibfield
  {journal} {\bibinfo  {journal} {Computer Physics Communications}\ }\textbf
  {\bibinfo {volume} {277}},\ \bibinfo {pages} {108386} (\bibinfo {year}
  {2022})}\BibitemShut {NoStop}%
\bibitem [{\citenamefont {Oseledets}\ and\ \citenamefont
  {Tyrtyshnikov}(2010)}]{tci-dmrg1-2010}%
  \BibitemOpen
  \bibfield  {author} {\bibinfo {author} {\bibfnamefont {I.}~\bibnamefont
  {Oseledets}}\ and\ \bibinfo {author} {\bibfnamefont {E.}~\bibnamefont
  {Tyrtyshnikov}},\ }\bibfield  {title} {\bibinfo {title} {Tt-cross
  approximation for multidimensional arrays},\ }\href
  {https://doi.org/https://doi.org/10.1016/j.laa.2009.07.024} {\bibfield
  {journal} {\bibinfo  {journal} {Linear Algebra and its Applications}\
  }\textbf {\bibinfo {volume} {432}},\ \bibinfo {pages} {70} (\bibinfo {year}
  {2010})}\BibitemShut {NoStop}%
\bibitem [{\citenamefont {Savostyanov}\ and\ \citenamefont
  {Oseledets}(2011)}]{tci-dmrg2-2011}%
  \BibitemOpen
  \bibfield  {author} {\bibinfo {author} {\bibfnamefont {D.}~\bibnamefont
  {Savostyanov}}\ and\ \bibinfo {author} {\bibfnamefont {I.}~\bibnamefont
  {Oseledets}},\ }\bibfield  {title} {\bibinfo {title} {Fast adaptive
  interpolation of multi-dimensional arrays in tensor train format},\ }in\
  \href {https://doi.org/10.1109/nDS.2011.6076873} {\emph {\bibinfo {booktitle}
  {The 2011 International Workshop on Multidimensional (nD) Systems}}}\
  (\bibinfo {year} {2011})\ pp.\ \bibinfo {pages} {1--8}\BibitemShut {NoStop}%
\bibitem [{\citenamefont {Savostyanov}(2014)}]{ACA-tci-2014}%
  \BibitemOpen
  \bibfield  {author} {\bibinfo {author} {\bibfnamefont {D.~V.}\ \bibnamefont
  {Savostyanov}},\ }\bibfield  {title} {\bibinfo {title} {Quasioptimality of
  maximum-volume cross interpolation of tensors},\ }\href
  {https://doi.org/https://doi.org/10.1016/j.laa.2014.06.006} {\bibfield
  {journal} {\bibinfo  {journal} {Linear Algebra and its Applications}\
  }\textbf {\bibinfo {volume} {458}},\ \bibinfo {pages} {217} (\bibinfo {year}
  {2014})}\BibitemShut {NoStop}%
\bibitem [{\citenamefont {Goreinov}\ \emph {et~al.}(1997)\citenamefont
  {Goreinov}, \citenamefont {Zamarashkin},\ and\ \citenamefont
  {Tyrtyshnikov}}]{maxvol1997}%
  \BibitemOpen
  \bibfield  {author} {\bibinfo {author} {\bibfnamefont {S.~A.}\ \bibnamefont
  {Goreinov}}, \bibinfo {author} {\bibfnamefont {N.~L.}\ \bibnamefont
  {Zamarashkin}},\ and\ \bibinfo {author} {\bibfnamefont {E.~E.}\ \bibnamefont
  {Tyrtyshnikov}},\ }\bibfield  {title} {\bibinfo {title} {Pseudo-skeleton
  approximations by matrices of maximal volume},\ }\href@noop {} {\bibfield
  {journal} {\bibinfo  {journal} {Mathematical Notes}\ }\textbf {\bibinfo
  {volume} {62}},\ \bibinfo {pages} {515} (\bibinfo {year} {1997})}\BibitemShut
  {NoStop}%
\bibitem [{\citenamefont {Bebendorf}(2000)}]{ACA2000}%
  \BibitemOpen
  \bibfield  {author} {\bibinfo {author} {\bibfnamefont {M.}~\bibnamefont
  {Bebendorf}},\ }\bibfield  {title} {\bibinfo {title} {Approximation of
  boundary element matrices},\ }\href
  {https://doi.org/https://doi.org/10.1007/PL00005410} {\bibfield  {journal}
  {\bibinfo  {journal} {Numerische Mathematik}\ }\textbf {\bibinfo {volume}
  {86}},\ \bibinfo {pages} {565} (\bibinfo {year} {2000})}\BibitemShut
  {NoStop}%
\bibitem [{\citenamefont {Goreinov}\ \emph {et~al.}(2010)\citenamefont
  {Goreinov}, \citenamefont {Oseledets}, \citenamefont {Savostyanov},
  \citenamefont {Tyrtyshnikov},\ and\ \citenamefont
  {Zamarashkin}}]{maxvol2010}%
  \BibitemOpen
  \bibfield  {author} {\bibinfo {author} {\bibfnamefont {S.~A.}\ \bibnamefont
  {Goreinov}}, \bibinfo {author} {\bibfnamefont {I.~V.}\ \bibnamefont
  {Oseledets}}, \bibinfo {author} {\bibfnamefont {D.~V.}\ \bibnamefont
  {Savostyanov}}, \bibinfo {author} {\bibfnamefont {E.~E.}\ \bibnamefont
  {Tyrtyshnikov}},\ and\ \bibinfo {author} {\bibfnamefont {N.~L.}\ \bibnamefont
  {Zamarashkin}},\ }\bibinfo {title} {How to find a good submatrix},\ in\ \href
  {https://doi.org/10.1142/9789812836021_0015} {\emph {\bibinfo {booktitle}
  {Matrix Methods: Theory, Algorithms and Applications}}}\ (\bibinfo
  {publisher} {World Scientific},\ \bibinfo {year} {2010})\ pp.\ \bibinfo
  {pages} {247--256}\BibitemShut {NoStop}%
\bibitem [{\citenamefont {Kishore~Kumar}\ and\ \citenamefont
  {Schneider}(2017)}]{matDecomp2017survey}%
  \BibitemOpen
  \bibfield  {author} {\bibinfo {author} {\bibfnamefont {N.}~\bibnamefont
  {Kishore~Kumar}}\ and\ \bibinfo {author} {\bibfnamefont {J.}~\bibnamefont
  {Schneider}},\ }\bibfield  {title} {\bibinfo {title} {Literature survey on
  low rank approximation of matrices},\ }\href@noop {} {\bibfield  {journal}
  {\bibinfo  {journal} {Linear and Multilinear Algebra}\ }\textbf {\bibinfo
  {volume} {65}},\ \bibinfo {pages} {2212} (\bibinfo {year}
  {2017})}\BibitemShut {NoStop}%
\bibitem [{\citenamefont {White}(1992)}]{dmrg-White-1992}%
  \BibitemOpen
  \bibfield  {author} {\bibinfo {author} {\bibfnamefont {S.~R.}\ \bibnamefont
  {White}},\ }\bibfield  {title} {\bibinfo {title} {Density matrix formulation
  for quantum renormalization groups},\ }\href
  {https://doi.org/10.1103/PhysRevLett.69.2863} {\bibfield  {journal} {\bibinfo
   {journal} {Phys. Rev. Lett.}\ }\textbf {\bibinfo {volume} {69}},\ \bibinfo
  {pages} {2863} (\bibinfo {year} {1992})}\BibitemShut {NoStop}%
\bibitem [{\citenamefont {Huggins}\ \emph {et~al.}(2019)\citenamefont
  {Huggins}, \citenamefont {Patil}, \citenamefont {Mitchell}, \citenamefont
  {Whaley},\ and\ \citenamefont {Stoudenmire}}]{Huggins2019}%
  \BibitemOpen
  \bibfield  {author} {\bibinfo {author} {\bibfnamefont {W.}~\bibnamefont
  {Huggins}}, \bibinfo {author} {\bibfnamefont {P.}~\bibnamefont {Patil}},
  \bibinfo {author} {\bibfnamefont {B.}~\bibnamefont {Mitchell}}, \bibinfo
  {author} {\bibfnamefont {K.~B.}\ \bibnamefont {Whaley}},\ and\ \bibinfo
  {author} {\bibfnamefont {E.~M.}\ \bibnamefont {Stoudenmire}},\ }\bibfield
  {title} {\bibinfo {title} {Towards quantum machine learning with tensor
  networks},\ }\href {https://doi.org/10.1088/2058-9565/aaea94} {\bibfield
  {journal} {\bibinfo  {journal} {Quantum Science and Technology}\ }\textbf
  {\bibinfo {volume} {4}},\ \bibinfo {pages} {024001} (\bibinfo {year}
  {2019})}\BibitemShut {NoStop}%
\bibitem [{\citenamefont {Prokof'ev}\ and\ \citenamefont
  {Svistunov}(1998)}]{Prokofev_9804}%
  \BibitemOpen
  \bibfield  {author} {\bibinfo {author} {\bibfnamefont {N.~V.}\ \bibnamefont
  {Prokof'ev}}\ and\ \bibinfo {author} {\bibfnamefont {B.~V.}\ \bibnamefont
  {Svistunov}},\ }\bibfield  {title} {\bibinfo {title} {Polaron problem by
  diagrammatic quantum {Monte} {Carlo}},\ }\href
  {https://doi.org/10.1103/PhysRevLett.81.2514} {\bibfield  {journal} {\bibinfo
   {journal} {Phys. Rev. Lett.}\ }\textbf {\bibinfo {volume} {81}},\ \bibinfo
  {pages} {2514} (\bibinfo {year} {1998})},\ \Eprint
  {https://arxiv.org/abs/cond-mat/9804097} {arXiv:cond-mat/9804097}
  \BibitemShut {NoStop}%
\bibitem [{\citenamefont {Prokof’ev}\ and\ \citenamefont
  {Svistunov}(2008)}]{Prokofev_0801}%
  \BibitemOpen
  \bibfield  {author} {\bibinfo {author} {\bibfnamefont {N.~V.}\ \bibnamefont
  {Prokof’ev}}\ and\ \bibinfo {author} {\bibfnamefont {B.~V.}\ \bibnamefont
  {Svistunov}},\ }\bibfield  {title} {\bibinfo {title} {Bold diagrammatic
  {Monte} {Carlo}: A generic sign-problem tolerant technique for polaron models
  and possibly interacting many-body problems},\ }\href
  {https://doi.org/10.1103/PhysRevB.77.125101} {\bibfield  {journal} {\bibinfo
  {journal} {Phys. Rev. B}\ }\textbf {\bibinfo {volume} {77}},\ \bibinfo
  {pages} {125101} (\bibinfo {year} {2008})},\ \Eprint
  {https://arxiv.org/abs/0801.0911} {arXiv:0801.0911} \BibitemShut {NoStop}%
\bibitem [{\citenamefont {Mishchenko}\ \emph {et~al.}(2001)\citenamefont
  {Mishchenko}, \citenamefont {Prokof'ev}, \citenamefont {Svistunov},\ and\
  \citenamefont {Sakamoto}}]{Mishchenko_9910}%
  \BibitemOpen
  \bibfield  {author} {\bibinfo {author} {\bibfnamefont {A.~S.}\ \bibnamefont
  {Mishchenko}}, \bibinfo {author} {\bibfnamefont {N.~V.}\ \bibnamefont
  {Prokof'ev}}, \bibinfo {author} {\bibfnamefont {B.~V.}\ \bibnamefont
  {Svistunov}},\ and\ \bibinfo {author} {\bibfnamefont {A.}~\bibnamefont
  {Sakamoto}},\ }\bibfield  {title} {\bibinfo {title} {Comprehensive study of
  {Fröhlich} polaron},\ }\href {https://doi.org/10.1142/S0217979201009050}
  {\bibfield  {journal} {\bibinfo  {journal} {Int. J. Mod. Phys. B}\ }\textbf
  {\bibinfo {volume} {15}},\ \bibinfo {pages} {3940} (\bibinfo {year}
  {2001})}\BibitemShut {NoStop}%
\bibitem [{\citenamefont {Van~Houcke}\ \emph {et~al.}(2012)\citenamefont
  {Van~Houcke}, \citenamefont {Werner}, \citenamefont {Kozik}, \citenamefont
  {Prokof’ev}, \citenamefont {Svistunov}, \citenamefont {Ku}, \citenamefont
  {Sommer}, \citenamefont {Cheuk}, \citenamefont {Schirotzek},\ and\
  \citenamefont {Zwierlein}}]{VanHoucke_1110}%
  \BibitemOpen
  \bibfield  {author} {\bibinfo {author} {\bibfnamefont {K.}~\bibnamefont
  {Van~Houcke}}, \bibinfo {author} {\bibfnamefont {F.}~\bibnamefont {Werner}},
  \bibinfo {author} {\bibfnamefont {E.}~\bibnamefont {Kozik}}, \bibinfo
  {author} {\bibfnamefont {N.}~\bibnamefont {Prokof’ev}}, \bibinfo {author}
  {\bibfnamefont {B.}~\bibnamefont {Svistunov}}, \bibinfo {author}
  {\bibfnamefont {M.~J.~H.}\ \bibnamefont {Ku}}, \bibinfo {author}
  {\bibfnamefont {A.~T.}\ \bibnamefont {Sommer}}, \bibinfo {author}
  {\bibfnamefont {L.~W.}\ \bibnamefont {Cheuk}}, \bibinfo {author}
  {\bibfnamefont {A.}~\bibnamefont {Schirotzek}},\ and\ \bibinfo {author}
  {\bibfnamefont {M.~W.}\ \bibnamefont {Zwierlein}},\ }\bibfield  {title}
  {\bibinfo {title} {Feynman diagrams versus {Fermi}-gas {Feynman} emulator},\
  }\href {https://doi.org/10.1038/NPHYS2273} {\bibfield  {journal} {\bibinfo
  {journal} {Nature Phys}\ }\textbf {\bibinfo {volume} {8}},\ \bibinfo {pages}
  {366} (\bibinfo {year} {2012})},\ \Eprint {https://arxiv.org/abs/1110.3747}
  {arXiv:1110.3747} \BibitemShut {NoStop}%
\bibitem [{\citenamefont {Profumo}\ \emph {et~al.}(2015)\citenamefont
  {Profumo}, \citenamefont {Groth}, \citenamefont {Messio}, \citenamefont
  {Parcollet},\ and\ \citenamefont {Waintal}}]{profumo2015}%
  \BibitemOpen
  \bibfield  {author} {\bibinfo {author} {\bibfnamefont {R.~E.~V.}\
  \bibnamefont {Profumo}}, \bibinfo {author} {\bibfnamefont {C.}~\bibnamefont
  {Groth}}, \bibinfo {author} {\bibfnamefont {L.}~\bibnamefont {Messio}},
  \bibinfo {author} {\bibfnamefont {O.}~\bibnamefont {Parcollet}},\ and\
  \bibinfo {author} {\bibfnamefont {X.}~\bibnamefont {Waintal}},\ }\bibfield
  {title} {\bibinfo {title} {Quantum {Monte} {Carlo} for correlated
  out-of-equilibrium nanoelectronic devices},\ }\href
  {https://doi.org/10.1103/PhysRevB.91.245154} {\bibfield  {journal} {\bibinfo
  {journal} {Phys. Rev. B}\ }\textbf {\bibinfo {volume} {91}},\ \bibinfo
  {pages} {245154} (\bibinfo {year} {2015})},\ \Eprint
  {https://arxiv.org/abs/1504.02132} {arXiv:1504.02132} \BibitemShut {NoStop}%
\bibitem [{\citenamefont {Wu}\ \emph {et~al.}(2017)\citenamefont {Wu},
  \citenamefont {Ferrero}, \citenamefont {Georges},\ and\ \citenamefont
  {Kozik}}]{Wu_1608}%
  \BibitemOpen
  \bibfield  {author} {\bibinfo {author} {\bibfnamefont {W.}~\bibnamefont
  {Wu}}, \bibinfo {author} {\bibfnamefont {M.}~\bibnamefont {Ferrero}},
  \bibinfo {author} {\bibfnamefont {A.}~\bibnamefont {Georges}},\ and\ \bibinfo
  {author} {\bibfnamefont {E.}~\bibnamefont {Kozik}},\ }\bibfield  {title}
  {\bibinfo {title} {Controlling {Feynman} diagrammatic expansions: Physical
  nature of the pseudogap in the two-dimensional {Hubbard} model},\ }\href
  {https://doi.org/10.1103/PhysRevB.96.041105} {\bibfield  {journal} {\bibinfo
  {journal} {Phys. Rev. B}\ }\textbf {\bibinfo {volume} {96}},\ \bibinfo
  {pages} {041105} (\bibinfo {year} {2017})},\ \Eprint
  {https://arxiv.org/abs/1608.08402} {arXiv:1608.08402} \BibitemShut {NoStop}%
\bibitem [{\citenamefont {Rossi}(2017)}]{Rossi_1612}%
  \BibitemOpen
  \bibfield  {author} {\bibinfo {author} {\bibfnamefont {R.}~\bibnamefont
  {Rossi}},\ }\bibfield  {title} {\bibinfo {title} {Determinant diagrammatic
  {Monte} {Carlo} algorithm in the thermodynamic limit},\ }\href
  {https://doi.org/10.1103/PhysRevLett.119.045701} {\bibfield  {journal}
  {\bibinfo  {journal} {Phys. Rev. Lett.}\ }\textbf {\bibinfo {volume} {119}},\
  \bibinfo {pages} {045701} (\bibinfo {year} {2017})},\ \Eprint
  {https://arxiv.org/abs/1612.05184} {arXiv:1612.05184} \BibitemShut {NoStop}%
\bibitem [{\citenamefont {Chen}\ and\ \citenamefont {Haule}(2019)}]{Chen_1809}%
  \BibitemOpen
  \bibfield  {author} {\bibinfo {author} {\bibfnamefont {K.}~\bibnamefont
  {Chen}}\ and\ \bibinfo {author} {\bibfnamefont {K.}~\bibnamefont {Haule}},\
  }\bibfield  {title} {\bibinfo {title} {A combined variational and
  diagrammatic quantum {Monte} {Carlo} approach to the many-electron problem},\
  }\href {https://doi.org/10.1038/s41467-019-11708-6} {\bibfield  {journal}
  {\bibinfo  {journal} {Nat Commun}\ }\textbf {\bibinfo {volume} {10}},\
  \bibinfo {pages} {3725} (\bibinfo {year} {2019})},\ \Eprint
  {https://arxiv.org/abs/1809.04651} {arXiv:1809.04651} \BibitemShut {NoStop}%
\bibitem [{\citenamefont {Bertrand}\ \emph
  {et~al.}(2019{\natexlab{a}})\citenamefont {Bertrand}, \citenamefont
  {Florens}, \citenamefont {Parcollet},\ and\ \citenamefont
  {Waintal}}]{Bertrand_1903_series}%
  \BibitemOpen
  \bibfield  {author} {\bibinfo {author} {\bibfnamefont {C.}~\bibnamefont
  {Bertrand}}, \bibinfo {author} {\bibfnamefont {S.}~\bibnamefont {Florens}},
  \bibinfo {author} {\bibfnamefont {O.}~\bibnamefont {Parcollet}},\ and\
  \bibinfo {author} {\bibfnamefont {X.}~\bibnamefont {Waintal}},\ }\bibfield
  {title} {\bibinfo {title} {Reconstructing nonequilibrium regimes of quantum
  many-body systems from the analytical structure of perturbative expansions},\
  }\href {https://doi.org/10.1103/PhysRevX.9.041008} {\bibfield  {journal}
  {\bibinfo  {journal} {Phys. Rev. X}\ }\textbf {\bibinfo {volume} {9}},\
  \bibinfo {pages} {041008} (\bibinfo {year} {2019}{\natexlab{a}})},\ \Eprint
  {https://arxiv.org/abs/1903.11646} {arXiv:1903.11646} \BibitemShut {NoStop}%
\bibitem [{\citenamefont {Bertrand}\ \emph
  {et~al.}(2019{\natexlab{b}})\citenamefont {Bertrand}, \citenamefont
  {Parcollet}, \citenamefont {Maillard},\ and\ \citenamefont
  {Waintal}}]{Bertrand_1903_kernel}%
  \BibitemOpen
  \bibfield  {author} {\bibinfo {author} {\bibfnamefont {C.}~\bibnamefont
  {Bertrand}}, \bibinfo {author} {\bibfnamefont {O.}~\bibnamefont {Parcollet}},
  \bibinfo {author} {\bibfnamefont {A.}~\bibnamefont {Maillard}},\ and\
  \bibinfo {author} {\bibfnamefont {X.}~\bibnamefont {Waintal}},\ }\bibfield
  {title} {\bibinfo {title} {Quantum {Monte} {Carlo} algorithm for
  out-of-equilibrium {Green}'s functions at long times},\ }\href
  {https://doi.org/10.1103/PhysRevB.100.125129} {\bibfield  {journal} {\bibinfo
   {journal} {Phys. Rev. B}\ }\textbf {\bibinfo {volume} {100}},\ \bibinfo
  {pages} {125129} (\bibinfo {year} {2019}{\natexlab{b}})},\ \Eprint
  {https://arxiv.org/abs/1903.11636} {arXiv:1903.11636} \BibitemShut {NoStop}%
\bibitem [{\citenamefont {Moutenet}\ \emph {et~al.}(2019)\citenamefont
  {Moutenet}, \citenamefont {Seth}, \citenamefont {Ferrero},\ and\
  \citenamefont {Parcollet}}]{Moutenet_1904}%
  \BibitemOpen
  \bibfield  {author} {\bibinfo {author} {\bibfnamefont {A.}~\bibnamefont
  {Moutenet}}, \bibinfo {author} {\bibfnamefont {P.}~\bibnamefont {Seth}},
  \bibinfo {author} {\bibfnamefont {M.}~\bibnamefont {Ferrero}},\ and\ \bibinfo
  {author} {\bibfnamefont {O.}~\bibnamefont {Parcollet}},\ }\bibfield  {title}
  {\bibinfo {title} {Cancellation of vacuum diagrams and the long-time limit in
  out-of-equilibrium diagrammatic quantum {Monte} {Carlo}},\ }\href
  {https://doi.org/10.1103/PhysRevB.100.085125} {\bibfield  {journal} {\bibinfo
   {journal} {Phys. Rev. B}\ }\textbf {\bibinfo {volume} {100}},\ \bibinfo
  {pages} {085125} (\bibinfo {year} {2019})},\ \Eprint
  {https://arxiv.org/abs/1904.11969} {arXiv:1904.11969} \BibitemShut {NoStop}%
\bibitem [{\citenamefont {Rossi}\ \emph {et~al.}(2020)\citenamefont {Rossi},
  \citenamefont {Simkovic},\ and\ \citenamefont {Ferrero}}]{Rossi_2001}%
  \BibitemOpen
  \bibfield  {author} {\bibinfo {author} {\bibfnamefont {R.}~\bibnamefont
  {Rossi}}, \bibinfo {author} {\bibfnamefont {F.}~\bibnamefont {Simkovic}},\
  and\ \bibinfo {author} {\bibfnamefont {M.}~\bibnamefont {Ferrero}},\
  }\bibfield  {title} {\bibinfo {title} {Renormalized perturbation theory at
  large expansion orders},\ }\href
  {https://doi.org/10.1209/0295-5075/132/11001} {\bibfield  {journal} {\bibinfo
   {journal} {Europhysics Letters}\ }\textbf {\bibinfo {volume} {132}},\
  \bibinfo {pages} {11001} (\bibinfo {year} {2020})}\BibitemShut {NoStop}%
\bibitem [{\citenamefont {Ma{\v{c}}ek}\ \emph {et~al.}(2020)\citenamefont
  {Ma{\v{c}}ek}, \citenamefont {Dumitrescu}, \citenamefont {Bertrand},
  \citenamefont {Triggs}, \citenamefont {Parcollet},\ and\ \citenamefont
  {Waintal}}]{macek2020qqmc}%
  \BibitemOpen
  \bibfield  {author} {\bibinfo {author} {\bibfnamefont {M.}~\bibnamefont
  {Ma{\v{c}}ek}}, \bibinfo {author} {\bibfnamefont {P.~T.}\ \bibnamefont
  {Dumitrescu}}, \bibinfo {author} {\bibfnamefont {C.}~\bibnamefont
  {Bertrand}}, \bibinfo {author} {\bibfnamefont {B.}~\bibnamefont {Triggs}},
  \bibinfo {author} {\bibfnamefont {O.}~\bibnamefont {Parcollet}},\ and\
  \bibinfo {author} {\bibfnamefont {X.}~\bibnamefont {Waintal}},\ }\bibfield
  {title} {\bibinfo {title} {Quantum quasi-monte carlo technique for many-body
  perturbative expansions},\ }\href@noop {} {\bibfield  {journal} {\bibinfo
  {journal} {Physical Review Letters}\ }\textbf {\bibinfo {volume} {125}},\
  \bibinfo {pages} {047702} (\bibinfo {year} {2020})}\BibitemShut {NoStop}%
\bibitem [{\citenamefont {Haule}\ and\ \citenamefont
  {Chen}(2020)}]{HauleChen2020}%
  \BibitemOpen
  \bibfield  {author} {\bibinfo {author} {\bibfnamefont {K.}~\bibnamefont
  {Haule}}\ and\ \bibinfo {author} {\bibfnamefont {K.}~\bibnamefont {Chen}},\
  }\href {https://doi.org/10.48550/ARXIV.2012.03146} {\bibinfo {title}
  {Single-particle excitations in the uniform electron gas by diagrammatic
  monte carlo}} (\bibinfo {year} {2020})\BibitemShut {NoStop}%
\bibitem [{\citenamefont {Bertrand}\ \emph {et~al.}(2021)\citenamefont
  {Bertrand}, \citenamefont {Bauernfeind}, \citenamefont {Dumitrescu},
  \citenamefont {Macek}, \citenamefont {Waintal},\ and\ \citenamefont
  {Parcollet}}]{Bertrand_2021}%
  \BibitemOpen
  \bibfield  {author} {\bibinfo {author} {\bibfnamefont {C.}~\bibnamefont
  {Bertrand}}, \bibinfo {author} {\bibfnamefont {D.}~\bibnamefont
  {Bauernfeind}}, \bibinfo {author} {\bibfnamefont {P.~T.}\ \bibnamefont
  {Dumitrescu}}, \bibinfo {author} {\bibfnamefont {M.}~\bibnamefont {Macek}},
  \bibinfo {author} {\bibfnamefont {X.}~\bibnamefont {Waintal}},\ and\ \bibinfo
  {author} {\bibfnamefont {O.}~\bibnamefont {Parcollet}},\ }\bibfield  {title}
  {\bibinfo {title} {Quantum quasi monte carlo algorithm for out-of-equilibrium
  green functions at long times},\ }\bibfield  {journal} {\bibinfo  {journal}
  {Physical Review B}\ }\textbf {\bibinfo {volume} {103}},\ \href
  {https://doi.org/10.1103/physrevb.103.155104} {10.1103/physrevb.103.155104}
  (\bibinfo {year} {2021})\BibitemShut {NoStop}%
\bibitem [{\citenamefont {Simkovic}\ \emph {et~al.}(2021)\citenamefont
  {Simkovic}, \citenamefont {Rossi},\ and\ \citenamefont
  {Ferrero}}]{SimkovicRossiFerrero2021}%
  \BibitemOpen
  \bibfield  {author} {\bibinfo {author} {\bibfnamefont {F.}~\bibnamefont
  {Simkovic}}, \bibinfo {author} {\bibfnamefont {R.}~\bibnamefont {Rossi}},\
  and\ \bibinfo {author} {\bibfnamefont {M.}~\bibnamefont {Ferrero}},\ }\href
  {https://doi.org/10.48550/ARXIV.2110.05863} {\bibinfo {title} {The weak, the
  strong and the long correlation regimes of the two-dimensional hubbard model
  at finite temperature}} (\bibinfo {year} {2021})\BibitemShut {NoStop}%
\bibitem [{\citenamefont {Schneider}(2010)}]{matCIerror2010}%
  \BibitemOpen
  \bibfield  {author} {\bibinfo {author} {\bibfnamefont {J.}~\bibnamefont
  {Schneider}},\ }\bibfield  {title} {\bibinfo {title} {Error estimates for
  two-dimensional cross approximation},\ }\href
  {https://doi.org/https://doi.org/10.1016/j.jat.2010.04.012} {\bibfield
  {journal} {\bibinfo  {journal} {Journal of Approximation Theory}\ }\textbf
  {\bibinfo {volume} {162}},\ \bibinfo {pages} {1685} (\bibinfo {year}
  {2010})}\BibitemShut {NoStop}%
\bibitem [{\citenamefont {Goreinov}\ and\ \citenamefont
  {Tyrtyshnikov}(2011)}]{matCIquasioptimality2011}%
  \BibitemOpen
  \bibfield  {author} {\bibinfo {author} {\bibfnamefont {S.~A.}\ \bibnamefont
  {Goreinov}}\ and\ \bibinfo {author} {\bibfnamefont {E.~E.}\ \bibnamefont
  {Tyrtyshnikov}},\ }\bibfield  {title} {\bibinfo {title} {Quasioptimality of
  skeleton approximation of a matrix in the chebyshev norm},\ }in\ \href
  {https://doi.org/https://doi.org/10.1134/S1064562411030355} {\emph {\bibinfo
  {booktitle} {Doklady Mathematics}}},\ Vol.~\bibinfo {volume} {83}\ (\bibinfo
  {organization} {Springer},\ \bibinfo {year} {2011})\ pp.\ \bibinfo {pages}
  {374--375}\BibitemShut {NoStop}%
\bibitem [{\citenamefont {Kronrod}(1965)}]{kronrod1965nodes}%
  \BibitemOpen
  \bibfield  {author} {\bibinfo {author} {\bibfnamefont {A.}~\bibnamefont
  {Kronrod}},\ }\href@noop {} {\bibinfo {title} {Nodes and weights of
  quadrature formulas: sixteen-place tables. consultants bureau}} (\bibinfo
  {year} {1965})\BibitemShut {NoStop}%
\bibitem [{\citenamefont {Rubtsov}\ and\ \citenamefont
  {Lichtenstein}(2004)}]{Rubtsov2004}%
  \BibitemOpen
  \bibfield  {author} {\bibinfo {author} {\bibfnamefont {A.~N.}\ \bibnamefont
  {Rubtsov}}\ and\ \bibinfo {author} {\bibfnamefont {A.~I.}\ \bibnamefont
  {Lichtenstein}},\ }\bibfield  {title} {\bibinfo {title} {Continuous-time
  quantum {Monte} {Carlo} method for fermions: Beyond auxiliary field
  framework},\ }\href {https://doi.org/10.1134/1.1800216} {\bibfield  {journal}
  {\bibinfo  {journal} {Journal of Experimental and Theoretical Physics
  Letters}\ }\textbf {\bibinfo {volume} {80}},\ \bibinfo {pages} {61} (\bibinfo
  {year} {2004})}\BibitemShut {NoStop}%
\bibitem [{\citenamefont {Kloss}\ \emph {et~al.}(2021)\citenamefont {Kloss},
  \citenamefont {Weston}, \citenamefont {Gaury}, \citenamefont {Rossignol},
  \citenamefont {Groth},\ and\ \citenamefont {Waintal}}]{Kloss2021}%
  \BibitemOpen
  \bibfield  {author} {\bibinfo {author} {\bibfnamefont {T.}~\bibnamefont
  {Kloss}}, \bibinfo {author} {\bibfnamefont {J.}~\bibnamefont {Weston}},
  \bibinfo {author} {\bibfnamefont {B.}~\bibnamefont {Gaury}}, \bibinfo
  {author} {\bibfnamefont {B.}~\bibnamefont {Rossignol}}, \bibinfo {author}
  {\bibfnamefont {C.}~\bibnamefont {Groth}},\ and\ \bibinfo {author}
  {\bibfnamefont {X.}~\bibnamefont {Waintal}},\ }\bibfield  {title} {\bibinfo
  {title} {Tkwant: a software package for time-dependent quantum transport},\
  }\href {https://doi.org/10.1088/1367-2630/abddf7} {\bibfield  {journal}
  {\bibinfo  {journal} {New Journal of Physics}\ }\textbf {\bibinfo {volume}
  {23}},\ \bibinfo {pages} {023025} (\bibinfo {year} {2021})}\BibitemShut
  {NoStop}%
\bibitem [{\citenamefont {Griffin}\ and\ \citenamefont
  {Tsatsomeros}(2006)}]{Griffin_2006_Pminors}%
  \BibitemOpen
  \bibfield  {author} {\bibinfo {author} {\bibfnamefont {K.}~\bibnamefont
  {Griffin}}\ and\ \bibinfo {author} {\bibfnamefont {M.~J.}\ \bibnamefont
  {Tsatsomeros}},\ }\bibfield  {title} {\bibinfo {title} {Principal minors,
  part i: A method for computing all the principal minors of a matrix},\ }\href
  {https://doi.org/https://doi.org/10.1016/j.laa.2006.04.008} {\bibfield
  {journal} {\bibinfo  {journal} {Linear Algebra and its Applications}\
  }\textbf {\bibinfo {volume} {419}},\ \bibinfo {pages} {107} (\bibinfo {year}
  {2006})}\BibitemShut {NoStop}%
\bibitem [{\citenamefont {Simkovic}\ and\ \citenamefont
  {Ferrero}(2022)}]{Simkovic_2022_Pminors}%
  \BibitemOpen
  \bibfield  {author} {\bibinfo {author} {\bibfnamefont {F.}~\bibnamefont
  {Simkovic}}\ and\ \bibinfo {author} {\bibfnamefont {M.}~\bibnamefont
  {Ferrero}},\ }\bibfield  {title} {\bibinfo {title} {Fast principal minor
  algorithms for diagrammatic monte carlo},\ }\href
  {https://doi.org/10.1103/PhysRevB.105.125104} {\bibfield  {journal} {\bibinfo
   {journal} {Phys. Rev. B}\ }\textbf {\bibinfo {volume} {105}},\ \bibinfo
  {pages} {125104} (\bibinfo {year} {2022})}\BibitemShut {NoStop}%
\bibitem [{\citenamefont {Wiegmann}\ and\ \citenamefont
  {Tsvelick}(1983)}]{Wiegmann_1983_a}%
  \BibitemOpen
  \bibfield  {author} {\bibinfo {author} {\bibfnamefont {P.~B.}\ \bibnamefont
  {Wiegmann}}\ and\ \bibinfo {author} {\bibfnamefont {A.~M.}\ \bibnamefont
  {Tsvelick}},\ }\bibfield  {title} {\bibinfo {title} {Exact solution of the
  {Anderson} model: I},\ }\href {https://doi.org/10.1088/0022-3719/16/12/017}
  {\bibfield  {journal} {\bibinfo  {journal} {J. Phys. C: Solid State Phys.}\
  }\textbf {\bibinfo {volume} {16}},\ \bibinfo {pages} {2281} (\bibinfo {year}
  {1983})}\BibitemShut {NoStop}%
\bibitem [{\citenamefont {Cohen}\ \emph {et~al.}(2013)\citenamefont {Cohen},
  \citenamefont {Gull}, \citenamefont {Reichman}, \citenamefont {Millis},\ and\
  \citenamefont {Rabani}}]{Cohen2013}%
  \BibitemOpen
  \bibfield  {author} {\bibinfo {author} {\bibfnamefont {G.}~\bibnamefont
  {Cohen}}, \bibinfo {author} {\bibfnamefont {E.}~\bibnamefont {Gull}},
  \bibinfo {author} {\bibfnamefont {D.~R.}\ \bibnamefont {Reichman}}, \bibinfo
  {author} {\bibfnamefont {A.~J.}\ \bibnamefont {Millis}},\ and\ \bibinfo
  {author} {\bibfnamefont {E.}~\bibnamefont {Rabani}},\ }\bibfield  {title}
  {\bibinfo {title} {Numerically exact long-time magnetization dynamics at the
  nonequilibrium {Kondo} crossover of the {Anderson} impurity model},\ }\href
  {https://doi.org/10.1103/PhysRevB.87.195108} {\bibfield  {journal} {\bibinfo
  {journal} {Phys. Rev. B}\ }\textbf {\bibinfo {volume} {87}},\ \bibinfo
  {pages} {195108} (\bibinfo {year} {2013})}\BibitemShut {NoStop}%
\bibitem [{\citenamefont {Cohen}\ \emph
  {et~al.}(2014{\natexlab{a}})\citenamefont {Cohen}, \citenamefont {Reichman},
  \citenamefont {Millis},\ and\ \citenamefont {Gull}}]{Cohen2014a}%
  \BibitemOpen
  \bibfield  {author} {\bibinfo {author} {\bibfnamefont {G.}~\bibnamefont
  {Cohen}}, \bibinfo {author} {\bibfnamefont {D.~R.}\ \bibnamefont {Reichman}},
  \bibinfo {author} {\bibfnamefont {A.~J.}\ \bibnamefont {Millis}},\ and\
  \bibinfo {author} {\bibfnamefont {E.}~\bibnamefont {Gull}},\ }\bibfield
  {title} {\bibinfo {title} {Green's functions from real-time bold-line {Monte}
  {Carlo}},\ }\href {https://doi.org/10.1103/PhysRevB.89.115139} {\bibfield
  {journal} {\bibinfo  {journal} {Phys. Rev. B}\ }\textbf {\bibinfo {volume}
  {89}},\ \bibinfo {pages} {115139} (\bibinfo {year}
  {2014}{\natexlab{a}})}\BibitemShut {NoStop}%
\bibitem [{\citenamefont {Cohen}\ \emph
  {et~al.}(2014{\natexlab{b}})\citenamefont {Cohen}, \citenamefont {Gull},
  \citenamefont {Reichman},\ and\ \citenamefont {Millis}}]{Cohen2014b}%
  \BibitemOpen
  \bibfield  {author} {\bibinfo {author} {\bibfnamefont {G.}~\bibnamefont
  {Cohen}}, \bibinfo {author} {\bibfnamefont {E.}~\bibnamefont {Gull}},
  \bibinfo {author} {\bibfnamefont {D.~R.}\ \bibnamefont {Reichman}},\ and\
  \bibinfo {author} {\bibfnamefont {A.~J.}\ \bibnamefont {Millis}},\ }\bibfield
   {title} {\bibinfo {title} {Green's functions from real-time bold-line
  {Monte} {Carlo} calculations: Spectral properties of the nonequilibrium
  {Anderson} impurity model},\ }\href
  {https://doi.org/10.1103/PhysRevLett.112.146802} {\bibfield  {journal}
  {\bibinfo  {journal} {Phys. Rev. Lett.}\ }\textbf {\bibinfo {volume} {112}},\
  \bibinfo {pages} {146802} (\bibinfo {year} {2014}{\natexlab{b}})}\BibitemShut
  {NoStop}%
\bibitem [{\citenamefont {Cohen}\ \emph {et~al.}(2015)\citenamefont {Cohen},
  \citenamefont {Gull}, \citenamefont {Reichman},\ and\ \citenamefont
  {Millis}}]{Cohen2015}%
  \BibitemOpen
  \bibfield  {author} {\bibinfo {author} {\bibfnamefont {G.}~\bibnamefont
  {Cohen}}, \bibinfo {author} {\bibfnamefont {E.}~\bibnamefont {Gull}},
  \bibinfo {author} {\bibfnamefont {D.~R.}\ \bibnamefont {Reichman}},\ and\
  \bibinfo {author} {\bibfnamefont {A.~J.}\ \bibnamefont {Millis}},\ }\bibfield
   {title} {\bibinfo {title} {Taming the dynamical sign problem in real-time
  evolution of quantum many-body problems},\ }\href
  {https://doi.org/10.1103/PhysRevLett.115.266802} {\bibfield  {journal}
  {\bibinfo  {journal} {Phys. Rev. Lett.}\ }\textbf {\bibinfo {volume} {115}},\
  \bibinfo {pages} {266802} (\bibinfo {year} {2015})}\BibitemShut {NoStop}%
\bibitem [{\citenamefont {Eidelstein}\ \emph {et~al.}(2020)\citenamefont
  {Eidelstein}, \citenamefont {Gull},\ and\ \citenamefont
  {Cohen}}]{Eidelstein_2020}%
  \BibitemOpen
  \bibfield  {author} {\bibinfo {author} {\bibfnamefont {E.}~\bibnamefont
  {Eidelstein}}, \bibinfo {author} {\bibfnamefont {E.}~\bibnamefont {Gull}},\
  and\ \bibinfo {author} {\bibfnamefont {G.}~\bibnamefont {Cohen}},\ }\bibfield
   {title} {\bibinfo {title} {Multiorbital quantum impurity solver for general
  interactions and hybridizations},\ }\bibfield  {journal} {\bibinfo  {journal}
  {Physical Review Letters}\ }\textbf {\bibinfo {volume} {124}},\ \href
  {https://doi.org/10.1103/physrevlett.124.206405}
  {10.1103/physrevlett.124.206405} (\bibinfo {year} {2020})\BibitemShut
  {NoStop}%
\bibitem [{\citenamefont {Li}\ \emph {et~al.}(2022)\citenamefont {Li},
  \citenamefont {Yu}, \citenamefont {Gull},\ and\ \citenamefont
  {Cohen}}]{Li_2022}%
  \BibitemOpen
  \bibfield  {author} {\bibinfo {author} {\bibfnamefont {J.}~\bibnamefont
  {Li}}, \bibinfo {author} {\bibfnamefont {Y.}~\bibnamefont {Yu}}, \bibinfo
  {author} {\bibfnamefont {E.}~\bibnamefont {Gull}},\ and\ \bibinfo {author}
  {\bibfnamefont {G.}~\bibnamefont {Cohen}},\ }\bibfield  {title} {\bibinfo
  {title} {Interaction-expansion inchworm monte carlo solver for lattice and
  impurity models},\ }\bibfield  {journal} {\bibinfo  {journal} {Physical
  Review B}\ }\textbf {\bibinfo {volume} {105}},\ \href
  {https://doi.org/10.1103/physrevb.105.165133} {10.1103/physrevb.105.165133}
  (\bibinfo {year} {2022})\BibitemShut {NoStop}%
\bibitem [{\citenamefont {Golub}\ and\ \citenamefont
  {Van~Loan}(1996)}]{GoluVanl96}%
  \BibitemOpen
  \bibfield  {author} {\bibinfo {author} {\bibfnamefont {G.~H.}\ \bibnamefont
  {Golub}}\ and\ \bibinfo {author} {\bibfnamefont {C.~F.}\ \bibnamefont
  {Van~Loan}},\ }\href@noop {} {\emph {\bibinfo {title} {Matrix
  Computations}}},\ \bibinfo {edition} {3rd}\ ed.\ (\bibinfo  {publisher} {The
  Johns Hopkins University Press},\ \bibinfo {year} {1996})\BibitemShut
  {NoStop}%
\bibitem [{\citenamefont {Schollwöck}(2011)}]{Schollwoeck_1008}%
  \BibitemOpen
  \bibfield  {author} {\bibinfo {author} {\bibfnamefont {U.}~\bibnamefont
  {Schollwöck}},\ }\bibfield  {title} {\bibinfo {title} {The density-matrix
  renormalization group in the age of matrix product states},\ }\href
  {https://doi.org/10.1016/j.aop.2010.09.012} {\bibfield  {journal} {\bibinfo
  {journal} {Annals of Physics}\ }\textbf {\bibinfo {volume} {326}},\ \bibinfo
  {pages} {96} (\bibinfo {year} {2011})},\ \Eprint
  {https://arxiv.org/abs/1008.3477} {arXiv:1008.3477} \BibitemShut {NoStop}%
\bibitem [{\citenamefont {Sanderson}\ and\ \citenamefont
  {Curtin}(2016)}]{armaSanderson2016}%
  \BibitemOpen
  \bibfield  {author} {\bibinfo {author} {\bibfnamefont {C.}~\bibnamefont
  {Sanderson}}\ and\ \bibinfo {author} {\bibfnamefont {R.}~\bibnamefont
  {Curtin}},\ }\bibfield  {title} {\bibinfo {title} {Armadillo: a
  template-based c++ library for linear algebra},\ }\href
  {https://doi.org/10.21105/joss.00026} {\bibfield  {journal} {\bibinfo
  {journal} {Journal of Open Source Software}\ }\textbf {\bibinfo {volume}
  {1}},\ \bibinfo {pages} {26} (\bibinfo {year} {2016})}\BibitemShut {NoStop}%
\bibitem [{{\relax DLMF}()}]{NIST:DLMF}%
  \BibitemOpen
  {\relax DLMF},\ \href {http://dlmf.nist.gov/} {\bibinfo {title} {{\it NIST
  Digital Library of Mathematical Functions}}},\ \bibinfo {howpublished}
  {http://dlmf.nist.gov/, Release 1.1.5 of 2022-03-15},\ \bibinfo {note}
  {f.~W.~J. Olver, A.~B. {Olde Daalhuis}, D.~W. Lozier, B.~I. Schneider, R.~F.
  Boisvert, C.~W. Clark, B.~R. Miller, B.~V. Saunders, H.~S. Cohl, and M.~A.
  McClain, eds.}\BibitemShut {Stop}%
\bibitem [{\citenamefont {Piessens}\ \emph {et~al.}(1983)\citenamefont
  {Piessens}, \citenamefont {de~Doncker-Kapenga}, \citenamefont {\"Uberhuber},\
  and\ \citenamefont {Kahaner}}]{quadpack}%
  \BibitemOpen
  \bibfield  {author} {\bibinfo {author} {\bibfnamefont {R.}~\bibnamefont
  {Piessens}}, \bibinfo {author} {\bibfnamefont {E.}~\bibnamefont
  {de~Doncker-Kapenga}}, \bibinfo {author} {\bibfnamefont {C.~W.}\ \bibnamefont
  {\"Uberhuber}},\ and\ \bibinfo {author} {\bibfnamefont {D.~K.}\ \bibnamefont
  {Kahaner}},\ }\bibfield  {title} {\bibinfo {title} {{QUADPACK A Subroutine
  Package for Automatic Integration.}},\ }\href
  {https://doi.org/10.1007/978-3-642-61786-7} {\bibfield  {journal} {\bibinfo
  {journal} {Springer Series in Comput. Math.}\ } (\bibinfo {year}
  {1983})}\BibitemShut {NoStop}%
\bibitem [{\citenamefont {Gonnet}(2010)}]{Gonnet10}%
  \BibitemOpen
  \bibfield  {author} {\bibinfo {author} {\bibfnamefont {P.}~\bibnamefont
  {Gonnet}},\ }\bibfield  {title} {\bibinfo {title} {{Increasing the
  Reliability of Adaptive Quadrature Using Explicit Interpolants}},\ }\href
  {https://doi.org/10.1145/1824801.1824804} {\bibfield  {journal} {\bibinfo
  {journal} {ACM Trans. Math. Softw.}\ }\textbf {\bibinfo {volume} {37}},\
  \bibinfo {pages} {26} (\bibinfo {year} {2010})}\BibitemShut {NoStop}%
\bibitem [{\citenamefont {Bremer}\ \emph {et~al.}(2010)\citenamefont {Bremer},
  \citenamefont {Gimbutas},\ and\ \citenamefont {Rokhlin}}]{bremer10}%
  \BibitemOpen
  \bibfield  {author} {\bibinfo {author} {\bibfnamefont {J.}~\bibnamefont
  {Bremer}}, \bibinfo {author} {\bibfnamefont {Z.}~\bibnamefont {Gimbutas}},\
  and\ \bibinfo {author} {\bibfnamefont {V.}~\bibnamefont {Rokhlin}},\
  }\bibfield  {title} {\bibinfo {title} {A nonlinear optimization procedure for
  generalized {G}aussian quadratures},\ }\href@noop {} {\bibfield  {journal}
  {\bibinfo  {journal} {SIAM J. Sci. Comput.}\ }\textbf {\bibinfo {volume}
  {32}},\ \bibinfo {pages} {1761} (\bibinfo {year} {2010})}\BibitemShut
  {NoStop}%
\bibitem [{\citenamefont {Gough}(2009)}]{gsl}%
  \BibitemOpen
  \bibfield  {author} {\bibinfo {author} {\bibfnamefont {B.}~\bibnamefont
  {Gough}},\ }\href {http://www.gnu.org/software/gsl/} {\emph {\bibinfo {title}
  {GNU scientific library reference manual}}}\ (\bibinfo  {publisher} {Network
  Theory Ltd.},\ \bibinfo {year} {2009})\BibitemShut {NoStop}%
\end{thebibliography}%

\end{document}